\documentclass{emulateapj}
  \usepackage{epsfig}
  \received{}
  \accepted{}
\slugcomment{Accepted for publication in ApJS}

\shortauthors{Ptak et al.}
\shorttitle{Optical Counterparts of ULXs}

\newcommand{\etal}{{et al.~}}

\def\setxsize{\epsfxsize=0.32\hsize}
\def\doyskip{\vspace{0.175truecm}}
\def\doxskip{\hspace{0.175truecm}}
\def\rowoffigs#1#2#3{\centerline{\setxsize\epsfbox{#1}\doxskip\setxsize\epsfbox{#2}\doxskip\setxsize\epsfbox{#3}}}

\def\setxtsize{\epsfxsize=0.22\hsize}
\def\doytskip{\vspace{0.2truecm}}
\def\doxtskip{\hspace{0.3truecm}}
\def\fouroffigs#1#2#3#4{\centerline{\setxtsize\epsfbox{#1}\doxtskip\setxtsize\epsfbox{#2}\doxtskip\doxtskip\doxtskip\setxtsize\epsfbox{#3}\doxtskip\setxtsize\epsfbox{#4}}}

\begin{document}

\title{Optical Counterparts of Ultra-Luminous X-ray Sources identified 
from Archival Hubble Space Telescope/WFPC2 Images\altaffilmark{1}}

\author{Andrew Ptak, Ed Colbert}
\affil{Department of Physics \& Astronomy, Johns Hopkins University, 
       3400 North Charles Street, Baltimore, MD 21218}

\author{Roeland P.~van der Marel, Erin Roye}
\affil{Space Telescope Science Institute, 3700 San Martin Drive, 
       Baltimore, MD 21218}

%
\author{Tim Heckman, Brian Towne}
\affil{Department of Physics \& Astronomy, Johns Hopkins University, 
       3400 North Charles Street, Baltimore, MD 21218}


\altaffiltext{1}{Based on observations made with the NASA/ESA Hubble Space 
Telescope, obtained from the Data Archive at the Space Telescope
Science Institute, which is operated by the Association of
Universities for Research in Astronomy, Inc., under NASA contract NAS
5-26555. This project is associated with Archival proposal \#9545.}






\begin{abstract}
We present a systematic analysis of archival HST WFPC2 ``Association'' data
sets that correlate with the Chandra positions of a set of 44
ultra-luminous X-ray sources (ULXs) of nearby galaxies.  The main
motivation is to address the nature of ULXs by searching for optical
counterparts. Sixteen of the ULXs are found in early-type galaxies
(RC3 Hubble type $<$ 3). We have improved the Chandra-HST 
relative astrometry whenever possible, resulting in errors circles of
0.3-1.7'' in size.  Disparate numbers of potential ULX counterparts
are found, and in some cases none are found.  The lack of or low
number of counterparts in some cases may be due to insufficient depth
in the WFPC2 images.
Particularly in
late-type galaxies, the HST image in the ULX region was often complex
or crowded, requiring source detection to be performed manually.  We
therefore address various scenarios for the nature of the ULX since it
is not known which, if any, of the sources found are true
counterparts.  The  optical luminosities  
of the sources are typically in the range  $10^{4-6} L_{\odot}$,
with (effective) V magnitudes typically in the range 22-24.  In
several cases color information is available, with the colors roughly
tending to be more red in early-type galaxies.  This suggests that, in
general, the (potential) counterparts found in early-type galaxies are
likely to be older stellar populations, and are probably globular
clusters. 
Several early-type galaxy counterparts have blue colors, which may be
due to younger stellar populations in the host galaxies, however these
could also be background sources.  
In spiral galaxies the sources may also be due to localized
structure in the disks rather than bound stellar systems.
Alternatively some of the counterparts in late-type galaxies may be
isolated supergiant stars.  The observed X-ray/optical flux ratio is
diluted by the optical emission of the cluster in cases where the system
is an X-ray binary in a cluster, particularly in the case of a low-mass
X-ray binaries in old cluster.
If any of the  counterparts are bound
systems with $\sim 10^{4-6}$ stars {\it and} are the true counterparts
to the ULX sources, then the X-ray luminosities of the 
ULX are generally well below the Eddington limit for a 
black hole with mass $\sim  0.1\%$ of the cluster mass.
Finally, we find that the optical flux of the counterparts is consistent
with being dominated by emission from an accretion disk around an
intermediate-mass black hole if the black hole happens to have a mass
$\ga 10^{2} M_{\odot}$ and is accreting at close to
the Eddington rate, unless the accretion disk is irradiated (which
would result in high optical disk luminosities at lower black hole
masses).  
\end{abstract}



\clearpage


\section{Introduction}
\label{s:intro}

In the early 1980s, surveys of normal galaxies with the Einstein X-ray
satellite revealed intermediate-luminosity (L$_X$ $\sim$
10$^{39}$$-$10$^{40}$ erg~s$^{-1}$) X-ray sources which were seemingly
located in the centers of spiral galaxies (Fabbiano 1989).  This was
very interesting, since Seyfert nuclei (active galactic nuclei [AGNs]
in nearby spirals) are typically much more luminous (L$_X$ $\sim$
10$^{42}$$-$10$^{44}$ erg~s$^{-1}$) and black hole X-ray binaries (BH
XRBs) are much less luminous (typically L$_X$ $\lesssim$ 10$^{38}$
erg~s$^{-1}$).  It was not clear whether these intriguing sources were
underluminous accreting supermassive BHs, overluminous XRBs located
near the galactic nucleus, or an entirely new type of astrophysical
object altogether (e.g., Colbert \etal 1995).  In the 1990s, ROSAT
High Resolution Imager (HRI) observations showed that these
Ultraluminous X-ray objects (ULXs; also known as
Intermediate-luminosity X-ray Objects, or IXOs) are compact X-ray
sources, and are quite common in the local Universe (Colbert \&
Mushotzky 1999; Roberts \& Warwick 2000).  Many ULXs were found to be
offset from the optical nucleus (e.g., Colbert \& Mushotzky 1999).
More complete surveys from ROSAT (e.g., Roberts \& Warwick 2000;
Colbert \& Ptak 2002) and Chandra imaging data place their frequency
of occurrence at about one in every five galaxies (Ptak \& Colbert 2004). 
Most of the ULXs found
in these surveys are clearly significantly displaced from the galactic
nuclei, implying that their masses must be $\lesssim$10$^5$
M$_{\odot}$ since dynamical friction would otherwise cause them to
sink to the center of the galaxy in a Hubble time (Tremaine, Ostriker
\& Spitzer 1975).

ULXs are generally unresolved at the high spatial resolution ($\sim
0.5''$) of Chandra, and many show variability (Colbert \& Ptak 2002;
Fabbiano \etal 2003), which rules out the hypothesis that ULXs are
closely-spaced aggregates of lower-luminosity X-ray sources. Bondi
accretion from a dense ISM may not be sufficient to explain the observed
luminosities (King \etal 2001), although an intermediate mass black
hole ($M \gtrsim 300 M_{\odot}$) in a dense molecular cloud may
achieve luminosities in excess of $10^{39} \rm \ ergs \ s^{-1}$
(Krolik 2004) . They are therefore most likely powered
by accretion in binary systems. This is supported by evidence for
periodicity in a few individual objects (Sugiho \etal 2001; Liu
\etal 2002b). If one assumes that ULXs have L$_X$ below the Eddington
luminosity and are not beamed, then the mass of the central object is
required to be in the range $\sim 15-500 M_\odot$. This is too massive
to be a neutron star or a black hole formed through normal stellar
evolution, thus we would have to be dealing with intermediate-mass
black holes (IMBHs). This interpretation is quite fascinating, as
reviewed by van der Marel (2004) and Miller \& Colbert (2004). For
some individual ULXs it has been argued that the presence of an IMBH
is plausible (Portegies Zwart \etal 2004), but in general it poses
many challenges from a theoretical point of view (see King
\etal 2001; King 2002). Alternatively, ULXs may instead be
stellar-mass black hole systems that are emitting anisotropically with
either mild (King \etal 2001) or relativistic beaming (K\"ording,
Falcke \& Markoff 2002), or are actually emitting at super-Eddington
rates (Begelman 2002; Grimm, Gilfanov \& Sunyaev 2002). For example,
the galactic microquasar GRS 1915 has a known black hole mass of $14
M_{\odot}$ and is often observed with X-ray luminosities exceeding
$10^{39}\rm \ ergs \ s^{-1}$, qualifying it as an ULX. In this case
the high luminosity is due to both relativistic beaming and
super-Eddington emission (King 2002; Mirabel \& Rodriguez 1999).

Phenomenologically there appear to be several breeds of ULXs.  They
may be preferentially found in star-forming galaxies (Kilgard \etal
2002; Grimm \etal 2003), with starburst galaxies such as the Antennae
(Zezas \etal 2002; Zezas \& Fabbiano 2002) and the Cartwheel (Gao
\etal 2003) having large numbers of ULXs. This suggests an
association between the incidence of ULXs and star formation (although
Ptak \& Colbert [2004] have found that this is not necessarily the
case in general). Indeed, optical counterparts reported for ULXs
suggest a young star cluster in one case (Goad \etal 2002) and a
single O-star in another case (Liu, Bregman \& Seitzer
2002a). \citet{tera06} discuss a study of optical counterparts to ULXs
in M51 and varyingly find no, one or several candidate counterparts.
The colors and absolute magnitudes of counterparts in several cases
(where there were only one or two counterparts) were consistent with
stars of mass 7-9 $M_{\odot}$ and 10-15 $M_{\odot}$, implying
that these ULXs may be high-mass X-ray binaries.
 A
radio counterpart was found for a ULX in the dwarf galaxy NGC 5408
(Kaaret \etal 2003), with two possible HST counterparts in the X-ray
error circle.  The brighter of these counterparts has $M_V
\sim -6.2$, consistent with a supergiant O star.  

ULXs have also been
found in elliptical galaxies and galactic bulges (Colbert \& Ptak
2002; Angelini \etal 2001; Wu \etal 2002; Jeltema \etal 2003). This
suggests that some ULXs are associated with old systems such as
globular clusters.  However, ULXs with $L_X > 2 \times 10^{39} \rm \
ergs \ s^{-1}$ found in ellipticals are often not associated with
globular clusters and are observed in numbers consistent with the
expected number of background sources (Irwin \etal 2004).  A study of
optical counterparts to four ULXs by Gutierrez (2006) found that three
are background AGN and one is likely to be foreground star.  Three of
these ULXs were found in early-type galaxies, again showing a tendency
for ULXs associated with early-type galaxies to be background AGN.

X-ray spectroscopy can give clues about accretion disks and coronae
that may be present in ULXs (Colbert \& Mushotzky 1999; Makishima 
\etal 2000; Roberts \etal 2001; Strickland \etal 2001; Zezas \etal 2002; 
Foschini \etal 2002; Miller \etal 2003). However, there are important
questions about the nature of ULXs that can only be answered using the
multi-wavelength properties of as large a sample of ULXs as
possible. In the Antennae, ULXs are often observed close to star
clusters, but not coincident with them. This may suggest a scenario in
which ULXs are X-ray binaries that can be ejected out of clusters
through recoil (Portegies Zwart \& McMillan 2000). If this is
generally true then it would preclude an IMBH because the mass would
be too large for the binary to be ejected (Miller \& Hamilton 2002).
In cases where individual counterparts are not detected learning about
the environment of the ULXs can also be instructive. For example, it
has been suggested that ULXs in late-type galaxies are preferentially
found in or near HII regions (Pakull \& Mirioni 2002).

We have established a pipeline to continuously
analyze Chandra data from galaxies as it becomes publicly available,
with emphasis on producing a catalog of ULXs. The
present paper presents a study of archival Hubble Space Telescope
(HST) images obtained with the Second Wide Field and Planetary Camera
(WFPC2) of fields that contain ULXs in our Chandra catalog. The goal
is to search for optical counterparts to the ULXs, and to characterize
their environment. While a few individual optical ULX counterparts
have been reported by other groups, the present work represents the
first large-scale Chandra/HST ULX cross-comparison and cataloging
effort. The use of HST data is critical, given that ground-based
observations do not have the spatial resolution to resolve individual
stars (unless the star is unusually isolated) and often cannot reach
the apparent magnitudes of even supergiant stars except for the
nearest galaxies. The results provide much improved statistics with
fewer biases than previous work, as well as a catalog of sources for
use in follow-up studies.

The paper is structured as follows. Section~\ref{s:sample} discusses
the sample selection and the Chandra and HST data that we have used
for the study. Section~\ref{s:astrocor} discusses the astrometric
registration of the Chandra and HST data. Section~\ref{s:photometry}
discusses the results from photometric analysis on the potential
counterparts that were located. Section~\ref{s:discuss} discusses the
main results of the study, and the implications for our understanding
of ULXs. Section~\ref{s:conc} summarizes the main conclusions.

\section{Sample Selection}
\label{s:sample}

\subsection{Catalog of Chandra ULXs}
\label{ss:chadracat}

We searched for Chandra X-ray sources within the R$_{25}$ radius of
all galaxies in the Third Reference Catalog of Bright Galaxies (RC3;
de Vaucouleurs \etal 1991) with recessional velocities $cz <$ 5000
km~s$^{-1}$.  Here R$_{25}$ is the 25 mag~arcsec$^{-2}$ isophotal
diameter, as listed in RC3.  An X-ray point source database was
created with the XAssist (Ptak \& Griffith 2003) Chandra ACIS
pipeline, using all public ACIS data as of 17~July~2002.  ULXs were
selected by requiring a 2$-$10 keV X-ray luminosity L$_X \ge$
10$^{39.0}$ erg~s$^{-1}$, assuming an absorbed power-law model with
Galactic absorption (Dickey \& Lockman 1990) and photon index $\Gamma
=$ 1.7.  For galaxies with RC3 recessional velocities $cz <$ 1000
km~s$^{-1}$, distances were taken from Tully (1988), if available,
otherwise from Kraan-Kortewag \& Tamman (1979).  Distances to galaxies
with $cz \ge$ 1000 km~s$^{-1}$ were computed using H$_0 =$ 75
km~s$^{-1}$~Mpc$^{-1}$ (for consistency with Tully 1988).  H~I line
velocities from RC3 were used when available, otherwise ``optical'' RC3
velocities were used. 

The preliminary source list thus obtained was tapered by merging
duplicate sources and examining the individual X-ray images to verify
that the source was not an X-ray instrumental artifact and that it was
indeed point-like.  We next cross-correlated the X-ray positions with
known sources listed in the NASA Extragalactic Database (NED) to
identify known AGNs, galaxy nuclei, and X-ray
supernovae.  Any objects thus identified were omitted, and questionable
objects were flagged as 
such. The remaining list contained 125 ULXs, which were used as a
starting point for searching the HST WFPC2 archive using the selection
criteria discussed in Section~\ref{ss:HSTdata}. For those ULXs with
WFPC2 data that passed our search criteria, we performed an
astrometric registration as described in
Section~\ref{s:astrocor}. With more accurate absolute positions
available, we reexamined all of the questionable objects by checking
their X-ray position with respect to the galaxy nucleus in the WFPC2
images, and by rejecting sources when spatially coincident with known
AGNs or SNe. This ultimately yielded 44 ULXs in 25 different galaxies,
that make up our sample for this paper. Table~\ref{t:sample} lists the
basic properties of the sample. 

\begin{deluxetable*}{lrrcrrrr}
\tablecolumns{9}
\tablewidth{0pc}
\tabletypesize{\scriptsize}
\tablecaption{ULX Sample}
\tablehead{
\colhead{ULX} &
\colhead{RA} & \colhead{DEC} & \colhead{Galaxy} & \colhead{T-type} & \colhead{Dist.} & 
\colhead{Chandra} & \colhead{L$_X$} \\
\colhead{ID} & \multicolumn{2}{c}{(J2000)} & \colhead{} & \colhead{(RC3)} & \colhead{(Mpc)} & \colhead{ObsID} & 
\colhead{(10$^{39}$ erg~s$^{-1}$)} \\
\colhead{(1)} & \colhead{(2)} & \colhead{(3)} & \colhead{(4)} & \colhead{(5)} & \colhead{(6)} & \colhead{(7)} & \colhead{(8)} \\
}
\startdata
  1       &    0:52:02.72   &    47:33:07.7   &                NGC 278   &   3.0   &   11.8   &   2056  &  2.0 \\ 
  2  	 &    2:22:31.36   &	42:20:24.4   &  			  NGC 891   &   3.0	&	9.6   &    794  &  1.8 \\ 
  3  	 &    2:42:38.89   &   -00:00:55.1   &  			  NGC 1068   &   3.0	&	15.2   &	344  &  2.9 \\ 
  4  	 &    2:42:39.71   &   -00:01:01.4   &  			  NGC 1068   &   3.0	&	15.2   &	344  &  1.0 \\ 
  5  	 &    2:46:19.82   &   -30:16:02.6   &  			  NGC 1097   &   3.0	&	17.0   &   1611  &  2.9 \\ 
    	 &  			   &				 &  					  & 		&		   &   2339  &  2.8 \\ 
  6 	 &    3:38:27.64   &   -35:26:48.2   &  			  NGC 1399   &   -5.0   &   19.3	&	 319  &  1.0 \\ 
  7  	 &    3:38:31.81   &   -35:26:03.8   &  			  NGC 1399   &   -5.0   &   19.3	&	 319  &  1.3 \\ 
  8  	 &    3:38:32.60   &   -35:27:05.1   &  			  NGC 1399   &   -5.0   &   19.3	&	 319  &  1.6 \\ 
 9  	&	 4:56:52.21   &    -4:52:17.9	&				 NGC 1700   &   -5.0	&	52.2   &   2069  &  2.3 \\ 
 10   	&	 4:56:56.03   &    -4:51:59.5	&				 NGC 1700   &   -5.0	&	52.2   &   2069  &  3.0 \\ 
 11  	&	 4:57:01.82   &    -4:51:15.9	&				 NGC 1700   &   -5.0	&	52.2   &   2069  &  6.4 \\ 
 12  	&	 9:55:46.44   &    69:40:40.5	&				 NGC 3034   &   90.0	&	5.2   &    379  &  1.0 \\ 
 13  	&	 9:55:50.01   &    69:40:46.0	&				 NGC 3034   &   90.0	&	5.2   &    361  &  2.1 \\ 
    	&				  & 				&						 &  		&		  &   1302  &  2.2 \\ 
    	&				  & 				&						 &  		&		  &    378  &  6.9 \\ 
    	&				  & 				&						 &  		&		  &    379  &  9.0 \\ 
 14  	&	 9:55:50.91   &    69:40:46.6	&				 NGC 3034   &   90.0	&	5.2   &    378  &  2.8 \\ 
 15  	&	10:27:52.54   &   -43:53:50.2	&				 NGC 3256   &   99.0	&	37.1   &	835  &  1.3  \\ 
 16  	&	10:27:55.15   &   -43:54:47.2	&				 NGC 3256   &   99.0	&	37.1   &	835  &  2.3  \\ 
 17  	&	10:36:42.66   &   -27:31:40.8	&				 NGC 3311   &   -4.0	&	50.5   &   2220  &  4.5  \\ 
 18  	&	10:36:42.75   &   -27:31:43.5	&				 NGC 3311   &   -4.0	&	50.5   &   2220  &  9.2  \\ 
 19  	&	11:20:15.76   &    13:35:13.7	&				 NGC 3628   &   3.0   &   7.7   &   2039  &  1.2  \\ 
    	&				  & 				&						 &  	   &		 &    395  &  2.3  \\ 
 20 	&	11:20:20.90   &    12:58:46.0	&				 NGC 3627   &   3.0   &   6.6   &    394  &  1.2  \\ 
 21  	&	12:01:51.32   &   -18:52:25.4	&			   NGC 4038/9   &   9.0   &   21.7   &    315  &  2.0  \\ 
 22  	&	12:01:52.08   &   -18:51:33.6	&				 NGC 4038   &   9.0   &   21.7   &    315  &  5.5  \\ 
 23  	&	12:01:54.27   &   -18:52:01.9	&			   NGC 4038/9   &   9.0   &   21.7   &    315  &  1.0  \\ 
 24  	&	12:01:54.35   &   -18:52:10.3	&				 NGC 4039   &   9.0   &   22.1   &    315  &  1.0  \\ 
 25  	&	12:01:54.97   &   -18:53:15.1	&			   NGC 4038/9   &   9.0   &   21.7   &    315  &  2.8  \\ 
 26  	&	12:01:55.65   &   -18:52:15.1	&			   NGC 4038/9   &   9.0   &   21.7   &    315  &  3.8  \\ 
 27  	&	12:01:56.43   &   -18:51:57.9	&			   NGC 4038/9   &   9.0   &   21.7   &    315  &  3.6  \\ 
 28  	&	12:01:58.22   &   -18:52:04.5	&				 NGC 4038/9   &   9.0   &   21.7	&	 315  &  1.0  \\ 
 29  	&	12:10:33.77   &    30:23:57.9	&				 NGC 4150   &   -2.0	&	9.7   &   1638  &  1.1  \\ 
 30  	&	12:30:43.26   &    41:38:18.4	&				 NGC 4490   &   7.0   &   7.8   &   1579  &  1.1  \\ 
 31  	&	12:35:58.56   &    27:57:41.9	&				 NGC 4559   &   6.0   &   9.7   &   2026  &  2.2  \\ 
    	&				  & 				&						 &  	   &		 &   2027  &  4.2  \\ 
 32  	&	12:36:17.40   &    25:58:55.5	&				 NGC 4565   &   3.0   &   16.4   &    404  &  9.8  \\ 
 33  	&	12:40:00.35   &   -11:37:24.0	&				 NGC 4594   &   1.0   &   14.5   &    407  &  1.4  \\ 
 34  	&	12:50:25.68   &    25:31:29.2	&				 NGC 4725   &   2.0   &   16.1   &    409  &  1.6  \\ 
 35  	&	13:05:21.94   &   -49:28:26.6	&				 NGC 4945   &   6.0  &   5.2	&	 864  &  1.0  \\   
 36     &   13:12:55.59   &   -19:30:39.7   &			   NGC 5018   &   -5.0   &   37.3   &   2070  &  5.3  \\ 
 37     &   13:13:02.03   &   -19:31:05.5   &			   NGC 5018   &   -5.0   &   37.3   &   2070  &  2.5  \\ 
 38     &   13:13:29.66   &    36:35:23.1   &			   NGC 5033   &   5.0   &   18.7	&	 412  &  2.2  \\ 
 39     &   13:13:29.46   &    36:35:17.4   &			   NGC 5033   &   5.0   &   18.7	&	 412  &  1.5  \\ 
 40     &   13:25:19.84   &   -43:03:17.2   &			   NGC 5128   &   -2.0   &   4.9	&	 316  &  1.0  \\ 
 41 	&	14:13:10.03   &   -65:20:45.1	&			 Circinus	&	3.0   &   3.7	&	 355  &  1.6  \\ 
 42  	&	14:13:12.24   &   -65:20:14.0	&			 Circinus	&	3.0   &   3.7	&	 356  &  1.8  \\ 
       &				 &  			   &						&		  & 		&	 365  &  4.9  \\ 
 43  	&	17:56:01.59   &    18:20:22.6	&				NGC 6500	&	1.7   &   40.0   &    416  &  2.5  \\ 
 44  	&	22:02:07.92   &   -31:59:19.7	&				NGC 7174	&	1.8   &   37.0   &    905  &  1.5  \\ 
\enddata

\tablecomments{The table lists basic properties for the ULXs in our sample,
selected as described in Section~\ref{ss:chadracat}. Column~(1) lists
the running ID number of the ULX used throughout this
paper. Columns~(2) and~(3) list the X-ray source position from the
Chandra ACIS data. Column~(4) lists the host galaxy name, and
column~(5) lists the galaxy morphology T-type from the Third Reference
Catalog of Bright Galaxies (RC3, de Vaucouleurs
\etal 1991). Column~(6) lists the galaxy distance, obtained as
described in Section~\ref{ss:chadracat}. Column~(7) lists the Chandra
ACIS Observation ID. Column~(8) lists the observed X-ray luminosity in
the 2--10 keV band, estimated as described in
Section~\ref{ss:chadracat}.\label{t:sample}}
\end{deluxetable*}

\subsection{Selection of HST/WFPC2 Data}
\label{ss:HSTdata}

For the HST parts of our project we chose to work with WFPC2
associations, instead of individual WFPC2 images. Associations are
logically related sets of individual WFPC2 images (e.g. the same
field, filter, and orientation) with accompanying information on
positional offsets between the images. The vast majority of WFPC2
science data (93\%) has been made into associations (Micol \& Durand
2002). The WFPC2 Associations Science Products Pipeline\footnote{WASPP
was developed by the Canadian Astronomy Data Centre (CADC) in
conjunction with the Space Telescope European Coordinating Facility
(ST-ECF) and the Multimission Archive at STScI (MAST).} 
(WASPP)
produces data products from the images in an association. To this end
it starts by applying the regular WFPC2 pipeline calibration steps
(bias subtraction, dark subtraction, flat-fielding, etc.) to the
individual images of the association. The reduced images are then
registered and co-added with cosmic-ray rejection. If reference stars
from the USNO-A2.0 Catalog (Monet \etal 1998) are present in the
field, then these are identified by an automated algorithm and the
final combined image is registered to their reference frame. For the
present project we used the data made available in the first release
(November 6, 2002) of the WASPP association data. This release
contained data products for 66\% of all WFPC2 associations.

We searched the STScI HST Data Archive for WFPC2 images with pointings
that might include the ULXs that were identified with Chandra
(Section~\ref{ss:chadracat}). Each WFPC2 field was then visually
inspected to ensure that the ULX position fell on one of the WFPC2
chips. Fields in which the ULX position was located off the edge of
the WFPC2 detectors were eliminated. 
Of these images, 90\% was part of
an association. Each association typically contains 2--8 images. Of
these associations, 74\% had been processed by the WASPP at the time
of the first data release. This implies that two-thirds of all the
images that we could have used for the present study were part of the
first WASPP release. Since manual registration and co-addition of the
remaining one-third would have been a substantial effort while
providing only a modest increases in overall sample size, we decided
to ignore these data for the present study. The final WFPC2 data
sample thus obtained consists of 76 WFPC2 images that include the 44
ULXs in our Chandra catalog. Each WFPC2 image is a fully calibrated
association data product. For a detailed description of the WASPP data
reduction steps we refer the reader to Schade \etal (2002). The WFPC2
association products used in our study are listed in
Table~2, together with basic properties such as the
filter used and the total exposure time. For some of our ULXs we have
images in multiple filters, or even multiple images in the same filter
(the WASPP pipeline does not combine images taken at different epochs
and telescope orientation, even when taken with the same filter).

\section{Astrometric Registration}
\label{s:astrocor}

It is important for our study that the HST and Chandra data are
astrometrically aligned as well as possible. The absolute astrometric
accuracy of WFPC2 data is dominated by two components. The first is
the $\sim 0.5''$ RMS astrometric accuracy of the Guide Star Catalog,
which is used for pointing and guiding. The second is the accuracy
with which the relative positions of the WFPC2 and the Fine Guidance
Sensors in the focal plane are known. The latter effect dominates the
overall error budget, although its actual size has varied over the
WFPC2 lifetime (Brammer \etal 2002). By comparing the positions of
USNO-A2.0 stars in large numbers of WFPC2 images to their known
astrometric coordinates, Schade \etal (2002) found an overall RMS
astrometric accuracy for WFPC2 data of $1.6''$. By contrast, Chandra
has nominal 90\% and 99\% confidence astrometric accuracies of $0.6''$ and $0.8''$ (Aldcroft 2002).

The WASPP association products have the advantage that they have
significantly enhanced astrometric accuracy as compared to the nominal
WFPC2 accuracy. The WASPP pipeline identifies all stars in the
USNO-A2.0 Catalog that might be on the image, and searches for
correspondence between the catalog and observed bright sources. The
results are used to improve the astrometry of the association data
products. In principle the final astrometric accuracy could be as good
as that of the USNO-A2.0 Catalog, which has a nominal RMS accuracy of
$0.25''$. In practice, however, there are several caveats to
this. First, the USNO-A2.0 Catalog has epochs in the 1950s, and
accumulated proper motions bring the current RMS accuracy into the
range $0.3''$--$0.4''$ (Schade \etal 2002). Second, not all fields
have (identifiable) USNO-A2.0 stars, and for these fields the
association data products have the same $1.6''$ RMS astrometric
accuracy as regular pipeline-calibrated WFPC2 data. And third, the
nearby galaxies of interest for the present study generally fill the
field of view and have complicated morphology. In this case the
automated source detection software in WASPP may identify the wrong
source with a USNO-A2.0 star, and thereby significantly degrade the
astrometry of the image. In our sample we found several cases for
which this appeared to have happened. Many galaxy nuclei are
themselves contained in the USNO-A2.0 Catalog, and due to their
complicated nuclear morphologies these are particularly tricky to use
in any astrometric registration.

Because of these partial shortcomings of the WASPP astrometric
registration algorithm, we did not use its results. Instead, we
visually identified stars from the USNO-B1.0 Catalog (Monet \etal
2003) in the association data products, and used them to improve the
astrometry.  While USNO-A was a two-color, one-epoch catalog, USNO-B
instead is a three-color, two-epoch catalog. The main advantage in the
present context is that proper motions are included in the catalog,
and can hence be corrected for. The nominal RMS accuracy of the
catalog is $0.2''$.  Note that mosaics produced using the STSDAS task
wmosaic were used, which removes known relative rotations and
geometric distortions of the individual WPFC2 chips.  The relative
astrometry between WFPC2 positions should then be good to better than
0.1'' \footnote{\tiny see
http://www.stsci.edu/instruments/wfpc2/Wfpc2\_faq/wfpc2\_ast\_faq.html}.
The advantage of our visual identification of the
stars is that it avoids the potential errors that an automated
algorithm could make. This approach allowed us to correct the
astrometry of 34 of our 76 images. For the remaining 30 images there
were no usable USNO-B1.0 stars in the field (and in general, the WASPP
pipeline had not identified any USNO-A2.0 stars either). In these
cases we left the astrometry as contained in the header of the
original (pre-association) lead image which came out of the STScI
on-the-fly-recalibration (OTFR) pipeline.

For the purpose of the present study we are not really interested in
the absolute astrometry of either the Chandra or the WFPC2 data
itself, but rather in the relative astrometric accuracy between the
two. It is possible to increase this accuracy in those cases where the
WFPC2 field contains a known X-ray source that also emits at optical
wavelengths (other than the ULX itself). This is the case if the
galaxy nucleus is a point-like AGN, and if the nucleus is not obscured
by dust in the WFPC2 image. We found this to be the case in 27 of the
images, corresponding to half of all the galaxies in the sample. This
high fraction reflects the fact that the Chandra and/or HST observers
tend to bias their observations towards active galaxies.

With the aforementioned approaches we obtained three types of WFPC2
header coordinates: Method 1 coordinates (available for 27 images)
were shifted to align the galaxy nucleus with the AGN coordinates in
Chandra data; Method 2 coordinates (available for 46 images) were
shifted to align observed USNO-B1.0 stars with their catalog
coordinates; and Method 3 coordinates (available for all 76 images)
contain the original astrometry from the telescope. These methods are
listed in order of decreasing accuracy. For our final analysis, we
adopt for each image the coordinates from the lowest method number
available, as listed in Table~2. Out of 76 images total,
27 images have coordinates from method 1, 34 images have coordinates
from method 2, and 15 images have coordinates from method
3. Table~2 also lists for each image the final pixel
position of each ULX. Together with the RA and DEC values of the ULXs
listed in Table~\ref{t:sample} this uniquely defines the astrometry of
each image. Note that our overall approach focuses on relative
astrometry and not absolute astrometry. In particular, when we
register WFPC2 data to the Chandra frame in method 1 we improve the
relative astrometry, but not necessarily the absolute astrometry.

The relative accuracy of the coordinates from the different methods
can be analyzed with some simple statistics. Let $d_{ij}$ be the RMS
(two-dimensional) distance between the coordinates from methods $i$
and $j$, for those images that have coordinates available from both
methods. We find for our sample: $d_{12} = 0.82'' \pm 0.11''$, $d_{13}
= 1.67'' \pm 0.16''$ and $d_{23} = 1.40'' \pm 0.13''$. We can safely
assume that method 1 yields much better relative Chandra/HST alignment
than method 2. This implies that $d_{12}$ is a measure of the relative
RMS Chandra/HST alignment for those images for which the best
coordinates are available from method 2. Similarly, $d_{13}$ is a
measure of the relative RMS Chandra/HST alignment for those images for
which the the best coordinates are available from method 3. In the
following we use $d_{12}$ and $d_{13}$ to set the size of the error
circles in our search for ULX counterparts. For method 1 the relative
RMS Chandra/HST alignment is more difficult to assess, because it is
dominated by the accuracy and stability of the geometric distortion
solutions for the two instruments. For WFPC2 the relative positions of
the chips have shifted by $\sim 0.2''$ since launch (Anderson \& King
2003). Furthermore, this alignment depends on the accuracy to which a
centroid of a source can be determined in both the X-ray and the
optical data. The complexity of the morphology can affect this
determination. In the following we conservatively, and somewhat
arbitrarily, assume a RMS Chandra/HST alignment accuracy of $0.3''$
for method 1.

We note that our results are more-or-less consistent with
expectation. For example, if the absolute RMS accuracy of USNO-B1.0
registered coordinates is $0.4''$, the absolute RMS accuracy of Chandra
pipeline coordinates is $0.7''$, and the absolute accuracy of WFPC2
pipeline coordinates is $1.4''$, then we would have expected that
$d_{12} = 0.81''$, $d_{13} = 1.57''$ and $d_{23} = 1.46''$. This is in
statistically acceptable agreement with our sample
statistics. Absolute accuracies of $0.4''$ for USNO-B1.0 registered
coordinates and $0.7''$ for Chandra pipeline coordinates are slightly
larger than the nominally expected values. However, it is not
unreasonable to expect that a variety of small systematic errors could
increase the achieved accuracies so as to exceed the nominal values.

\section{Images}
\label{s:images}

Figure 1 shows large scale views of each of the galaxies
and ULXs in our sample. These images are intended to illustrate the
overall morphologies of the sample galaxies and the relative locations
of the ULXs with respect to the galaxy nucleus. We therefore show only
data for a single filter per galaxy. When images in multiple filters
were available we preferentially chose to show a broad- or medium-band
filter in the red or visual part of the spectrum, such as F814W,
F791W, F606W, F555W, or F547M (see Biretta \etal (2001) for a
discussion of the properties of WFPC2 filters). However, there are
some exceptions to this. For NGC 1097, NGC 5033, and NGC 7174 there is
only data in the ultra-violet (F218W, F300W) part of the spectrum. For
the Antennae, one of its ULXs (\#25) is included only in a U-band (F336W)
image. For NGC 4594 only a narrow-band filter image is available
(F658N). For Circinus, one of its ULXs (\#41) is included only in a
narrow-band image (F656N).

The greyscale panels in Figure 2 show $10 \times 10$
arcsecond regions centered around the individual ULXs in the sample. A
circle around each ULX indicates the 1$\sigma$ accuracy of the
astrometric registration, i.e., $0.30''$ for images aligned by
method~1, $0.82''$ for images aligned by method~2, or $1.67''$ for
images aligned by method~3. These images show the general galaxy
morphology around each ULX. They form the basis for the search for
optical counterparts described in
Section~\ref{s:photometry}. Figure 2 shows each ULX, but
not each dataset for each ULX. For conciseness we generally only show
the highest quality image for each ULX. In most cases the morphology
is visually very similar in the different filters for which data are
available. However, in some cases the images in different filters do
show different morphologies, and for these ULXs we show images of the
same region in multiple filters in Figure 2.

The CADC WASPP pipeline performs a full reduction and image
combination for the datasets in each association. We therefore did
not attempt any additional reduction or calibration of the images.
However, we did discover occasional peculiarities in the combined
images from the CADC WASPP pipeline. In particular, some of the images
show evidence of improperly removed cosmic rays. Most of these are
cases in which only two exposures are available, and for which the
data may simply having been inadequate to fully remove all cosmic rays
and their coincidences. Examples of galaxies for which some datasets
appear to have residual cosmic rays are those of, NGC 4594, NGC 5018,
and NGC 5033. All counterparts were visually inspected to screen
questionable cases. 

Another peculiarity in some of the images is a sprinkling of
negative pixels that seems inconsistent with the regular tail of the
noise distribution. We learned (CADC, priv.~comm.) that these are
generally the result of combining two or more images with
significantly different exposure times. We made sure that these pixels
generally do not affect our photometry of potential counterparts
(unless in a few rare cases, which we explicitly identify as such
below). Figure 2 also shows some peculiarities that are
intrinsic to the data and the WFPC2 instrument. Most notably, the
diagonal bright stripes visible in the images for ULXs 3, 4, and 31
are chip boundaries in the mosaiced data.

\newcommand{\figcapmosaics}{WFPC2 mosaics showing the large scale galaxy 
morphology and ULX position for all of the ULXs in our sample. Each
ULX is shown as a cross, marked by its ID number in
Table~\ref{t:sample}. Each image is $2.58 \times 2.58$ arcmin, and has
North pointing towards the top of the page. The galaxy name, dataset,
and filter are indicated at the top of each panel. For a few galaxies
we show multiple images for different pointings, because not all ULXs
fall on a single image. Only first page of figures are shown here, for full
version see http://xassist.pha.jhu.edu/ptak/hst\_ulx\_paper. \label{f:mosaics}}

\newcommand{\figcapstamps}{WFPC2 mosaics showing the 10"x10" regions around
the ULX positions with well-defined measurable counterparts.  North is
pointing up.  The picture on the left shows the region of the galaxy
with a one sigma error circle drawn on top.  The picture on the right
shows both the one and two sigma error circles, and the positions and
numbers of the optical counterparts overlayed and marked with
crosses. For discussion of the error circle sizes, see
Section~\ref{s:astrocor}. Details about the photometry for the
counterparts are shown in Table 3, and discussed in
Section~\ref{s:photometry}. Only first page of figures are shown here, for full
version see http://xassist.pha.jhu.edu/ptak/hst\_ulx\_paper\label{f:stamps}}


\begin{figure*}
\rowoffigs{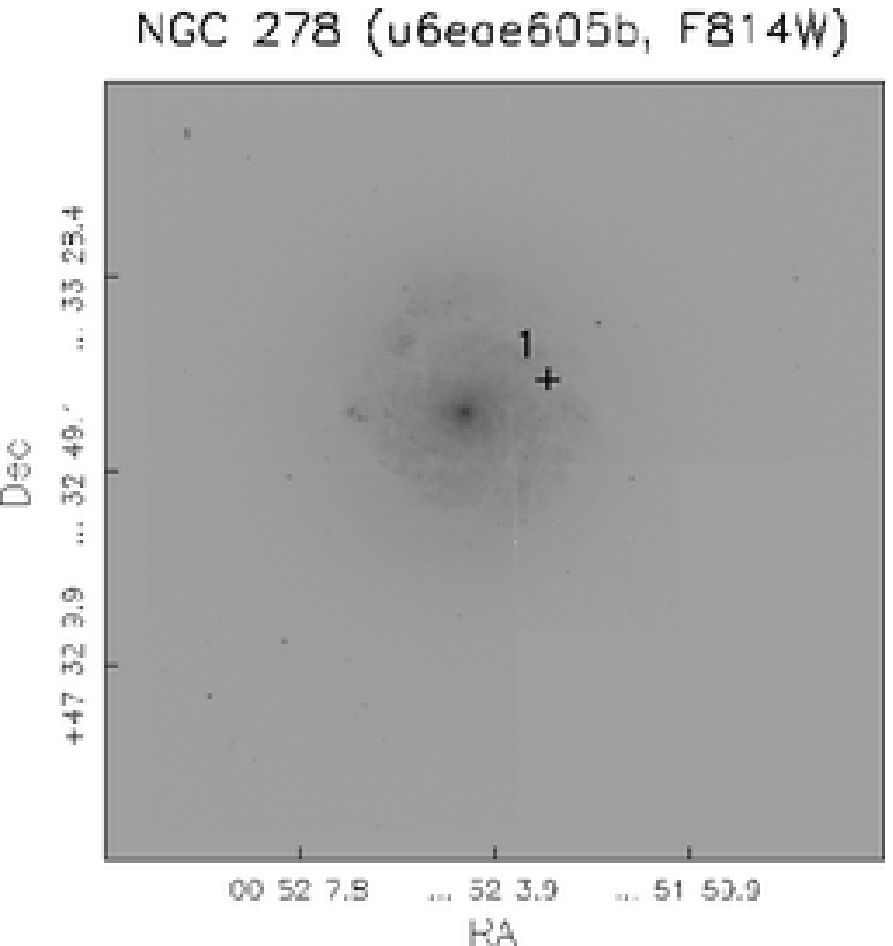}{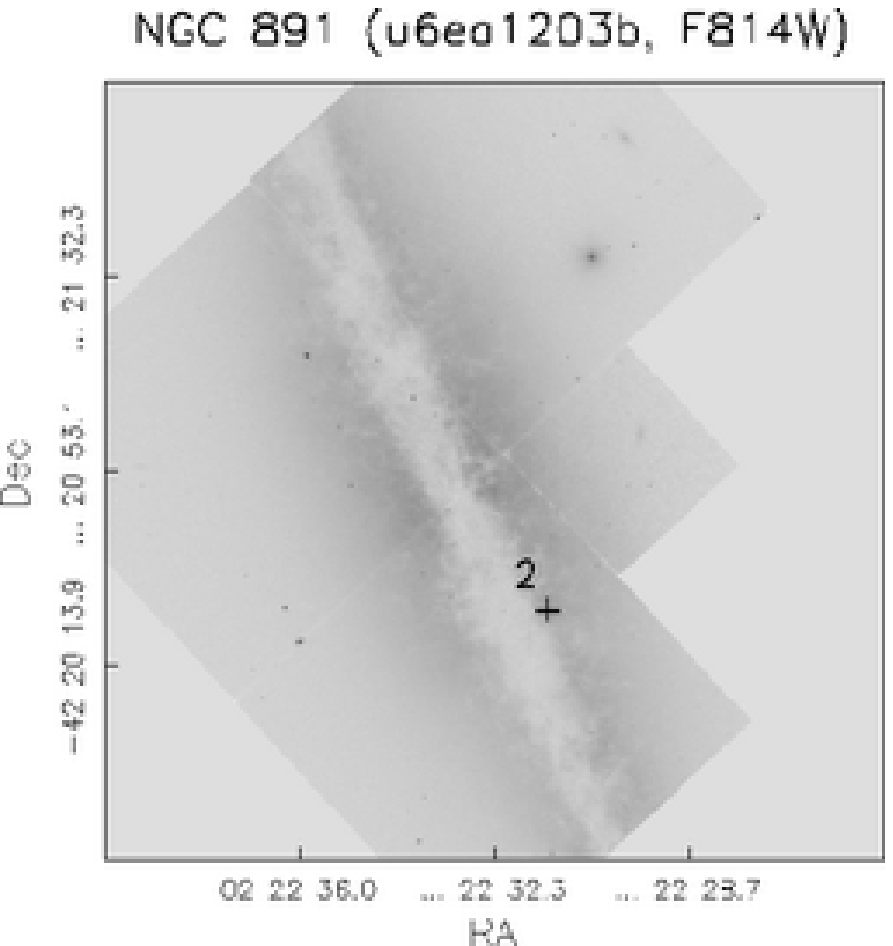}{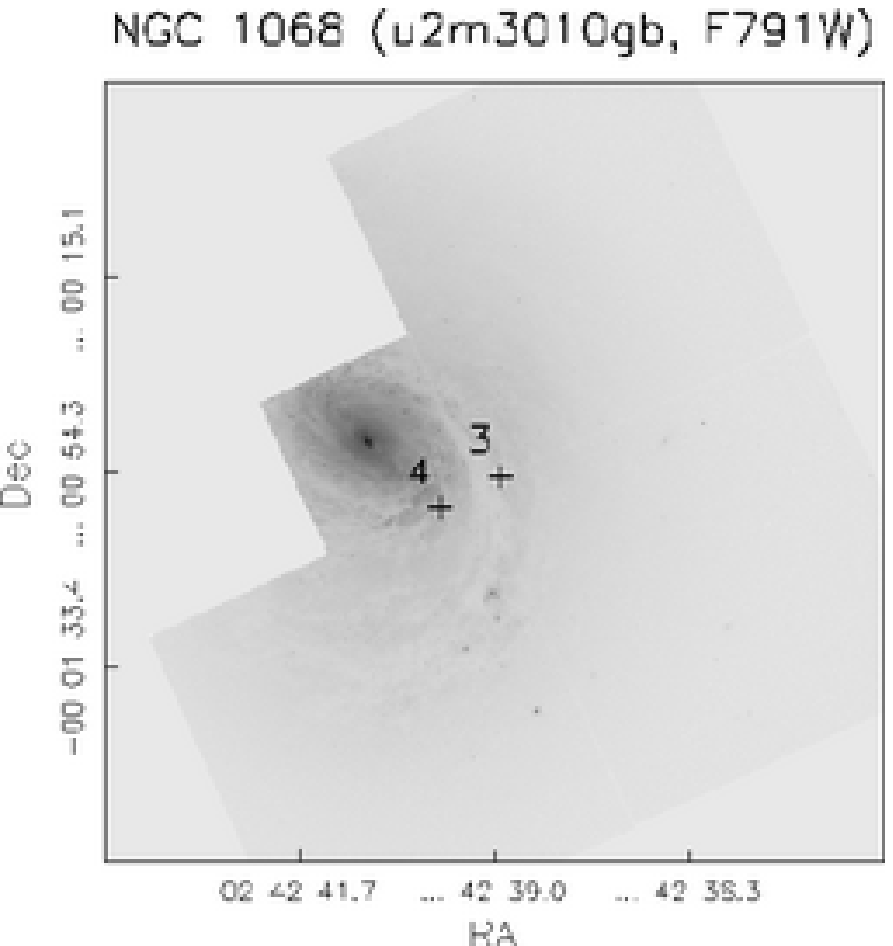}
\doyskip
\rowoffigs{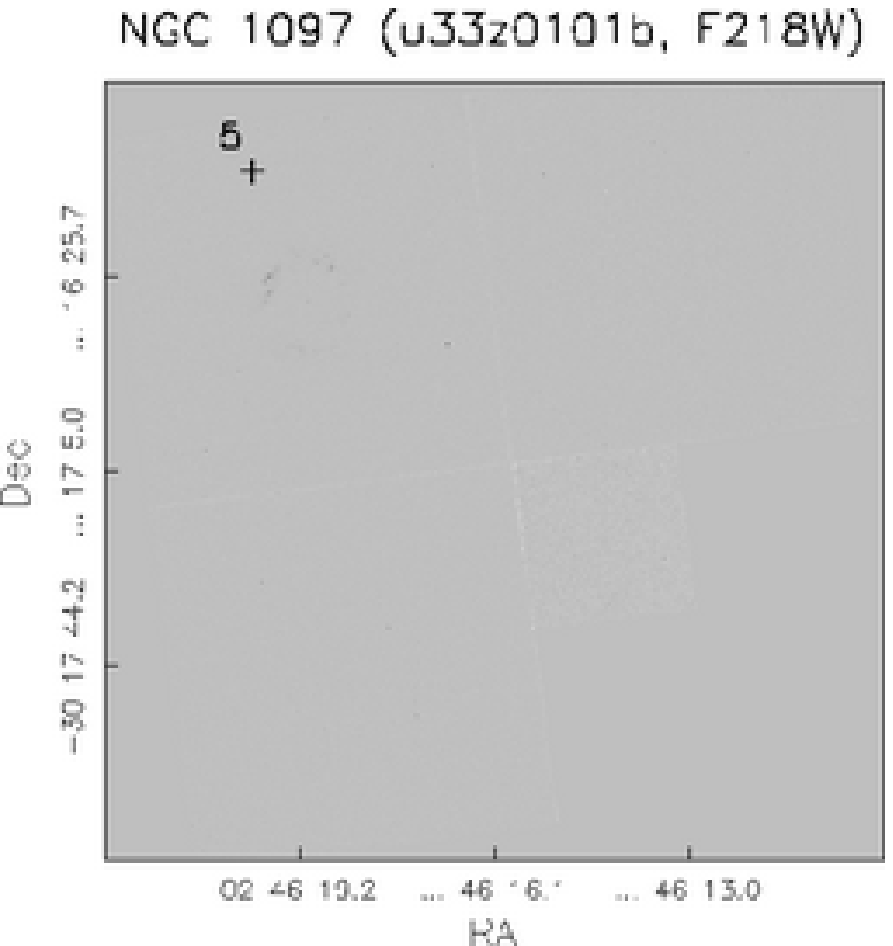}{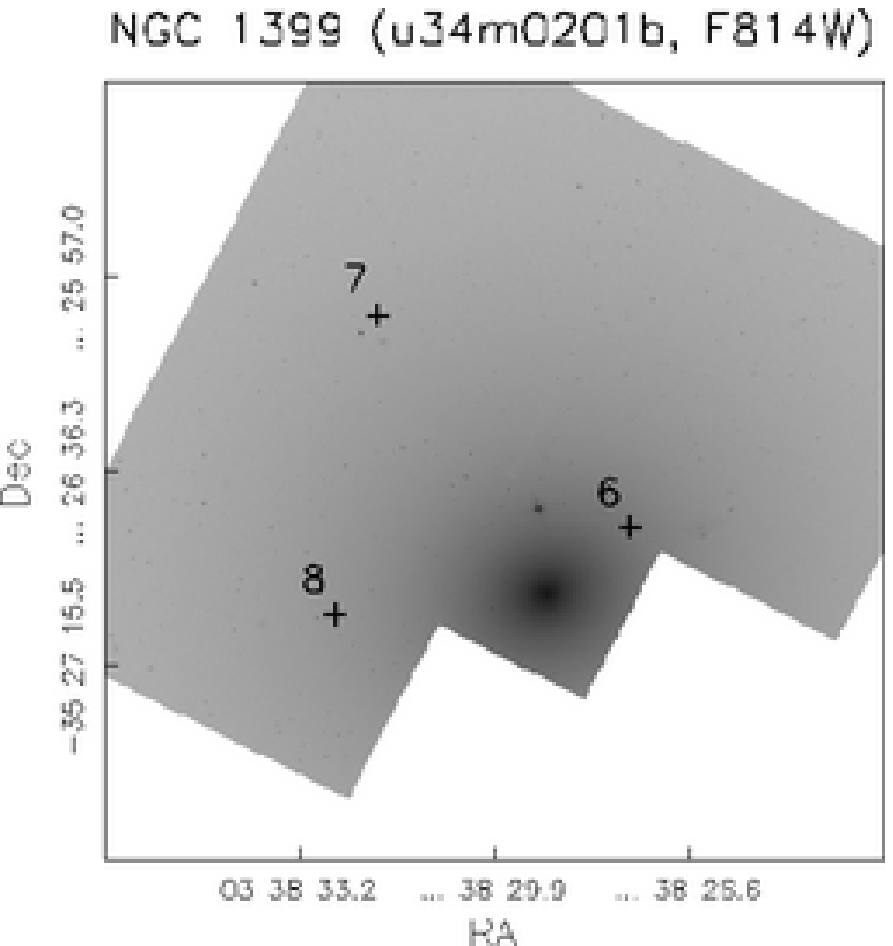}{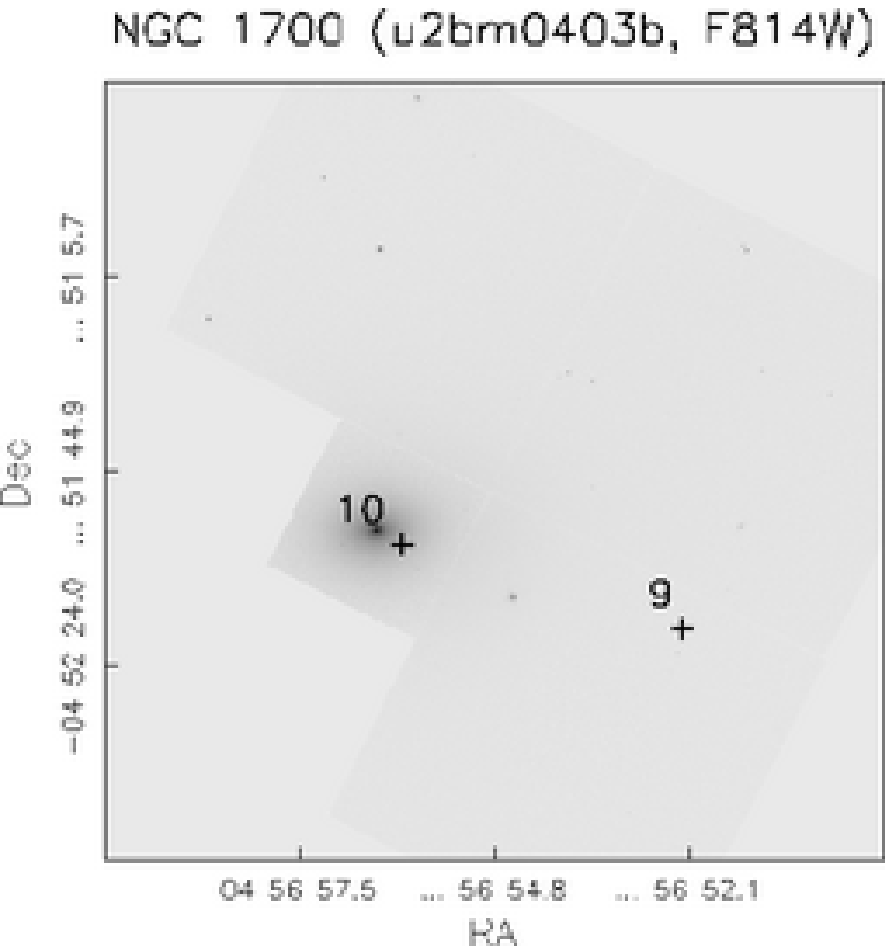}
\doyskip
\rowoffigs{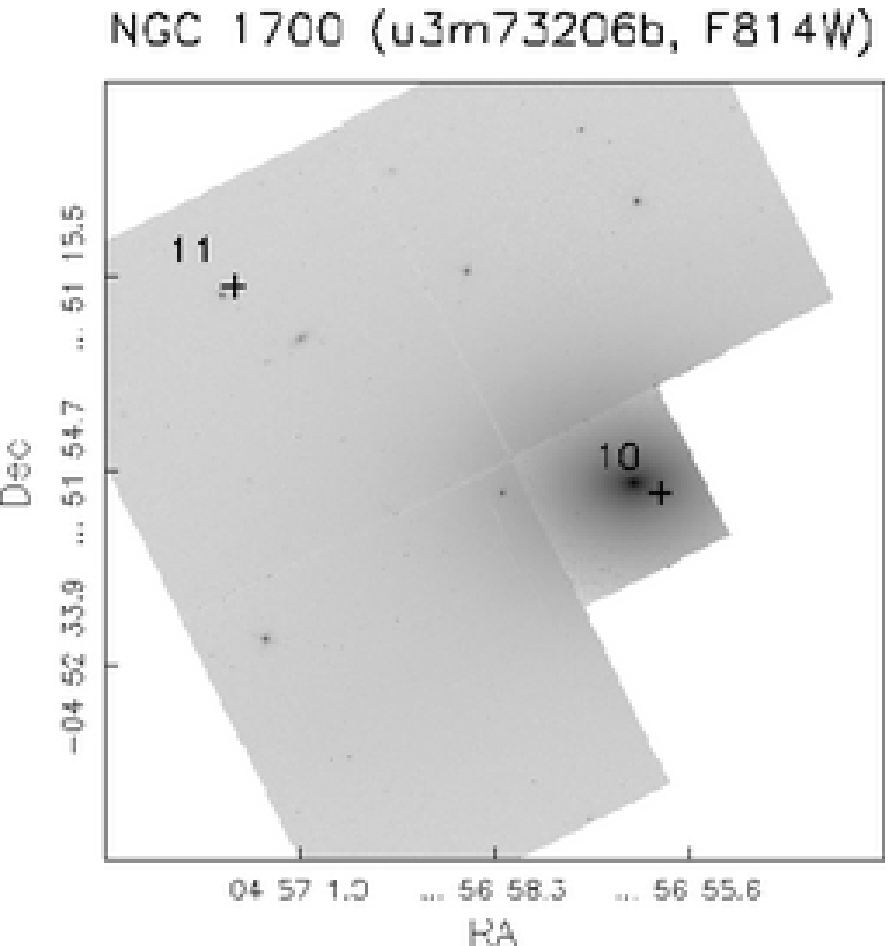}{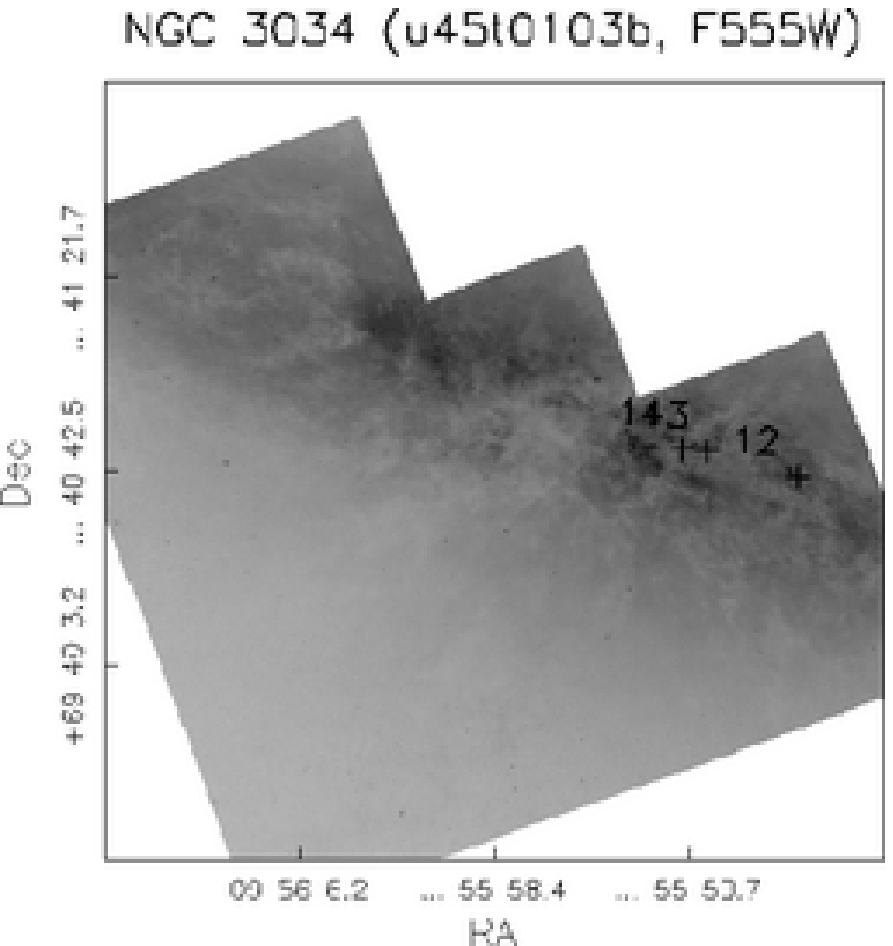}{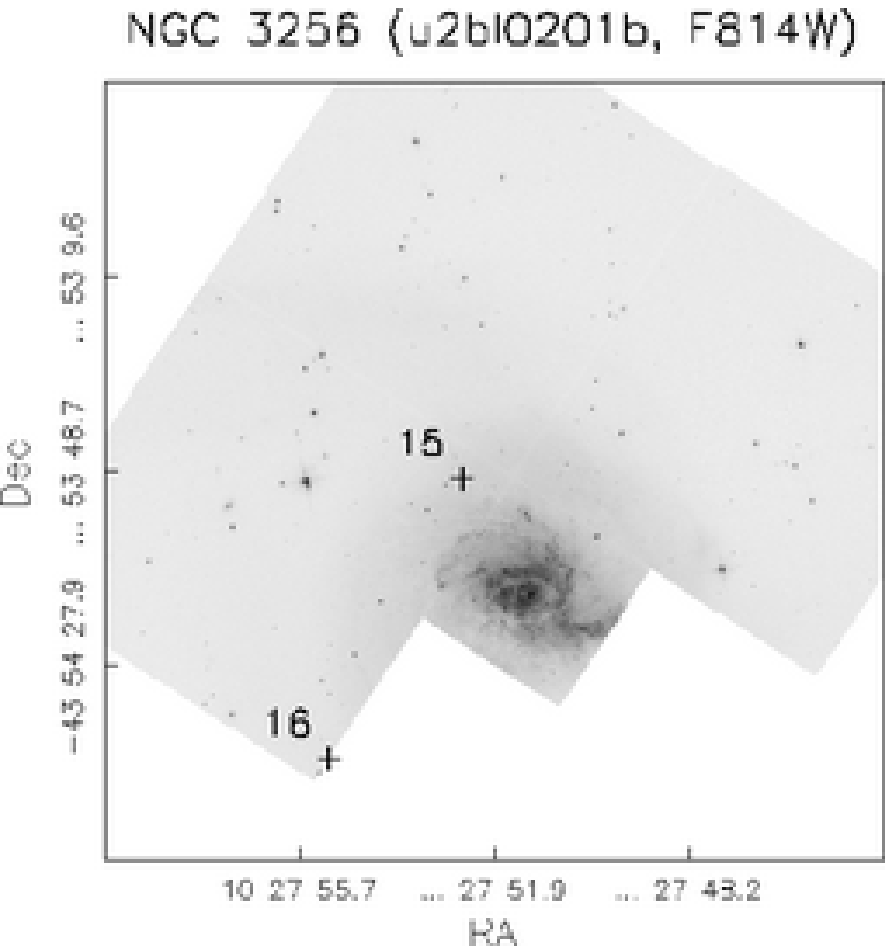}
\vskip3.0truecm
\setcounter{figure}{0}
\figcaption{\figcapmosaics}
\end{figure*}


\begin{figure*}
\fouroffigs{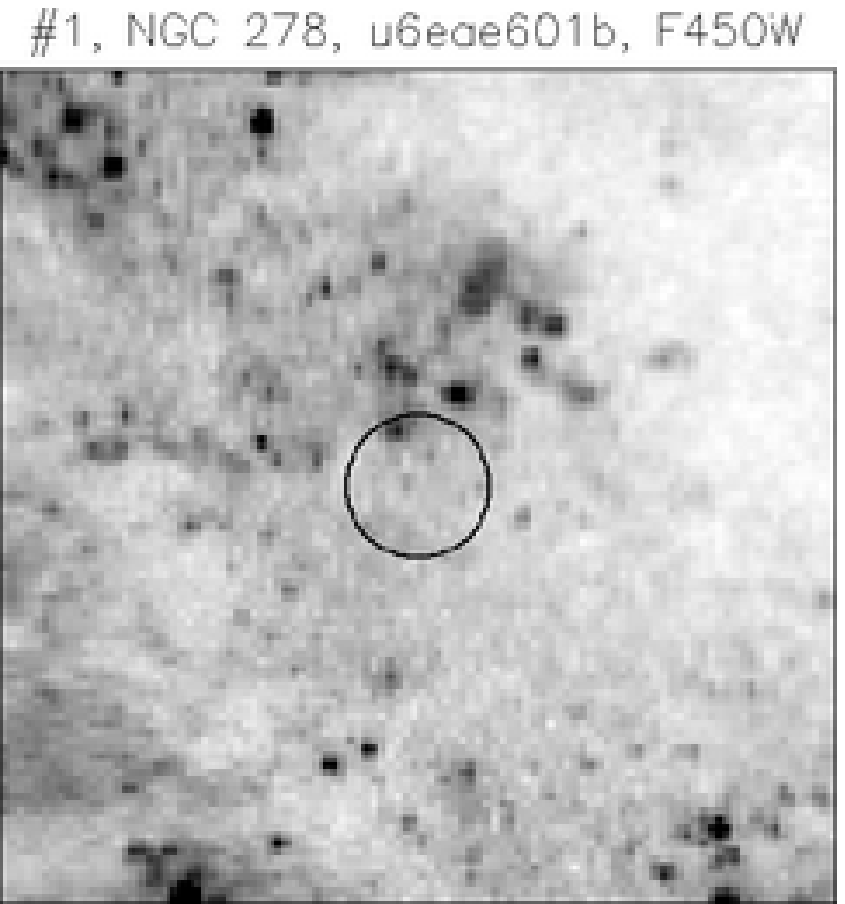}{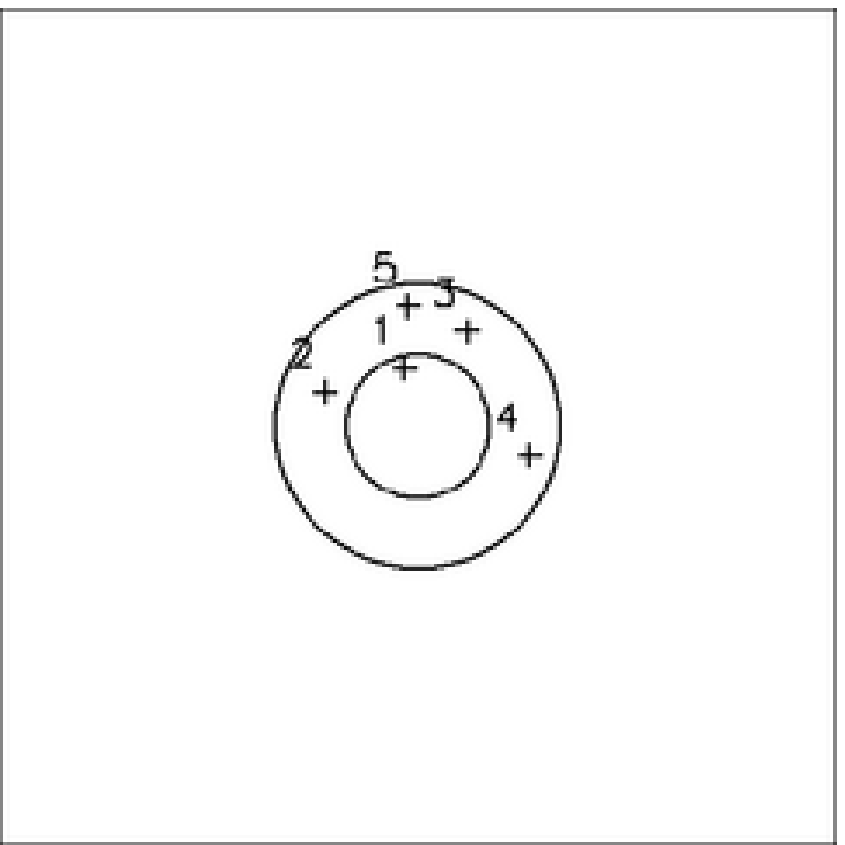}{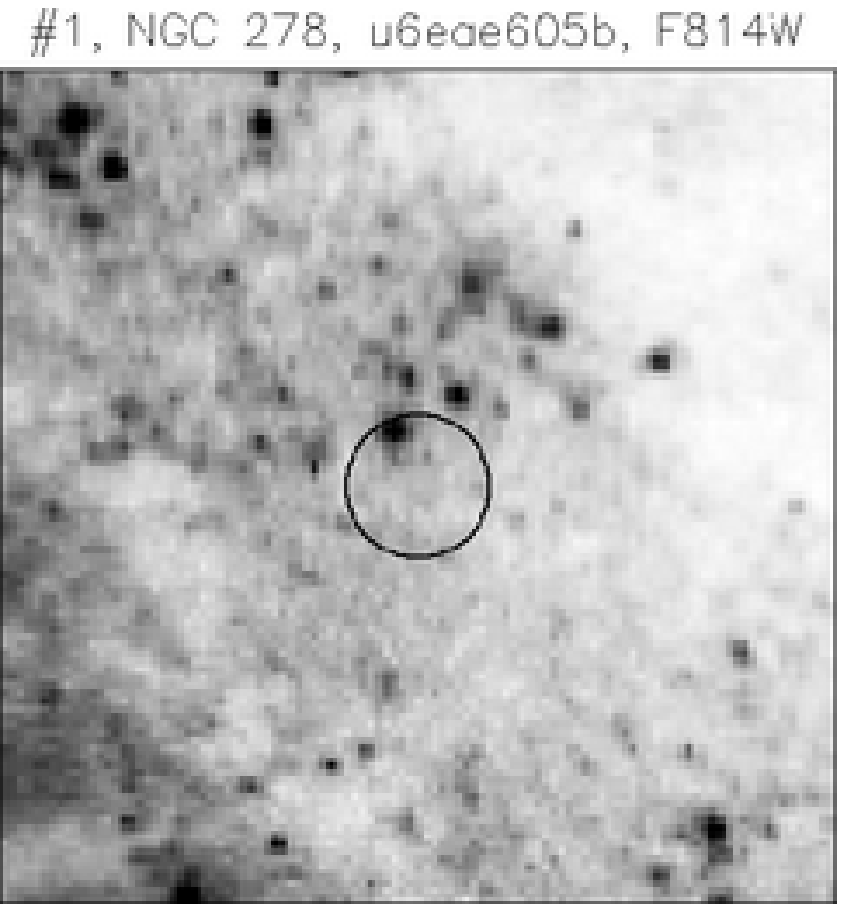}{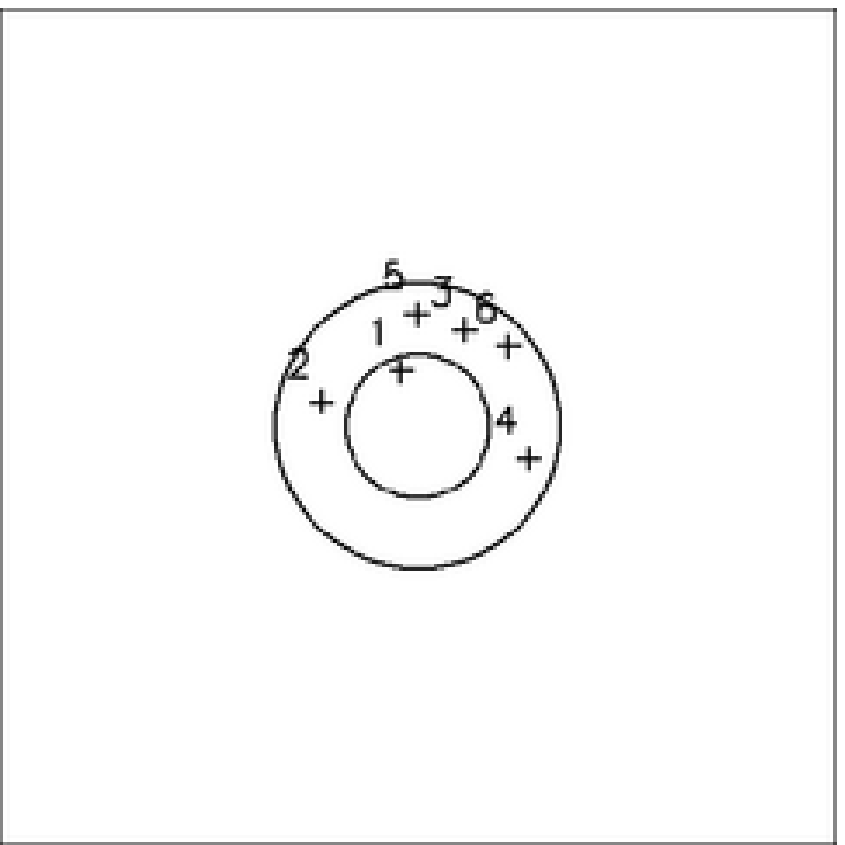}
\doytskip
\fouroffigs{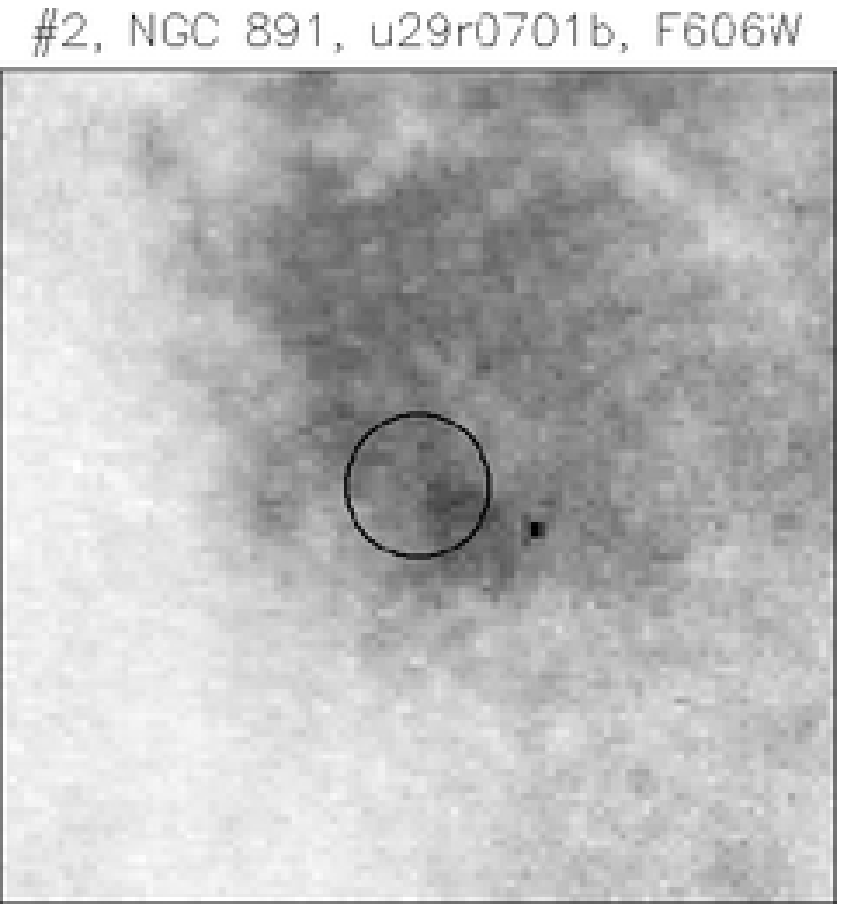}{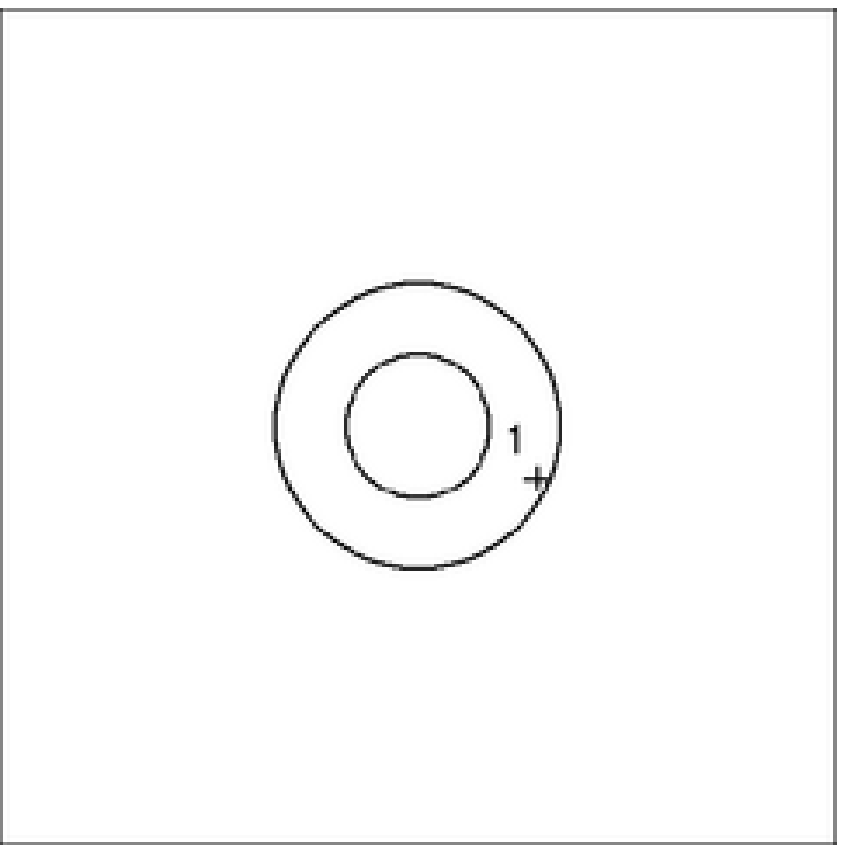}{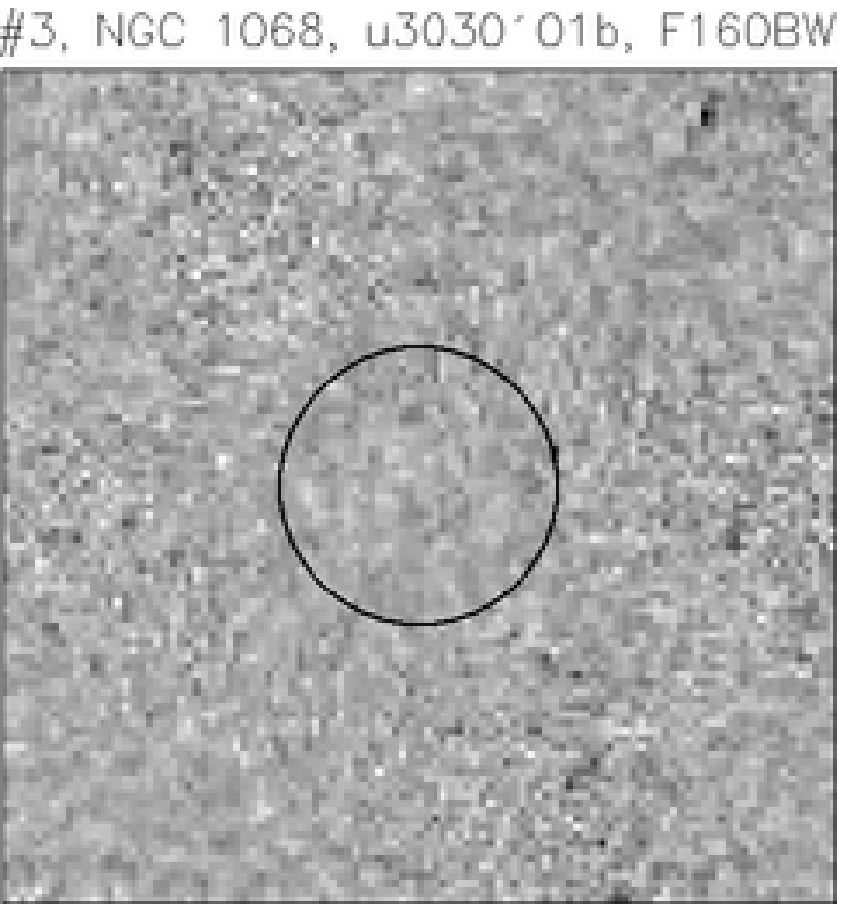}{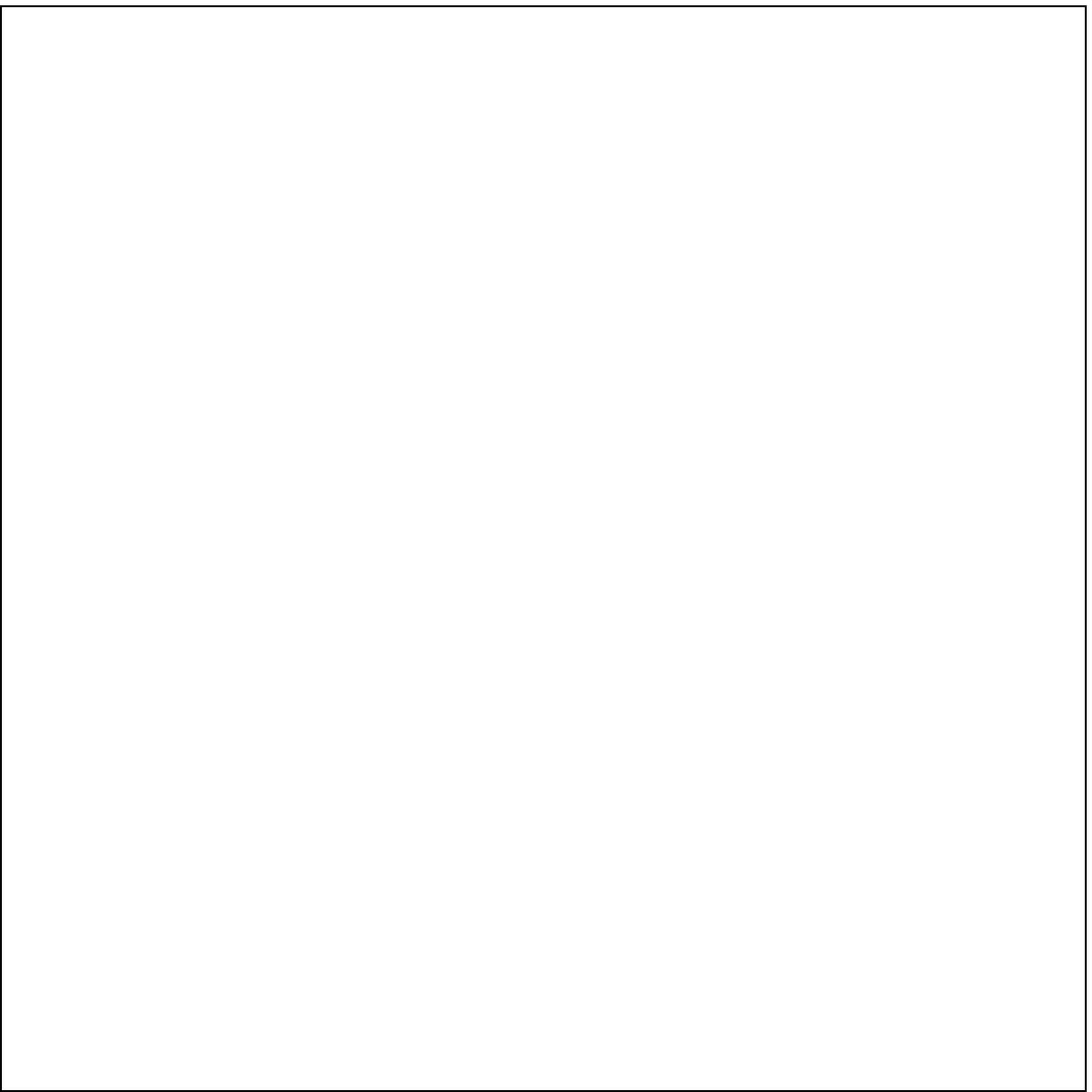}
\doytskip
\fouroffigs{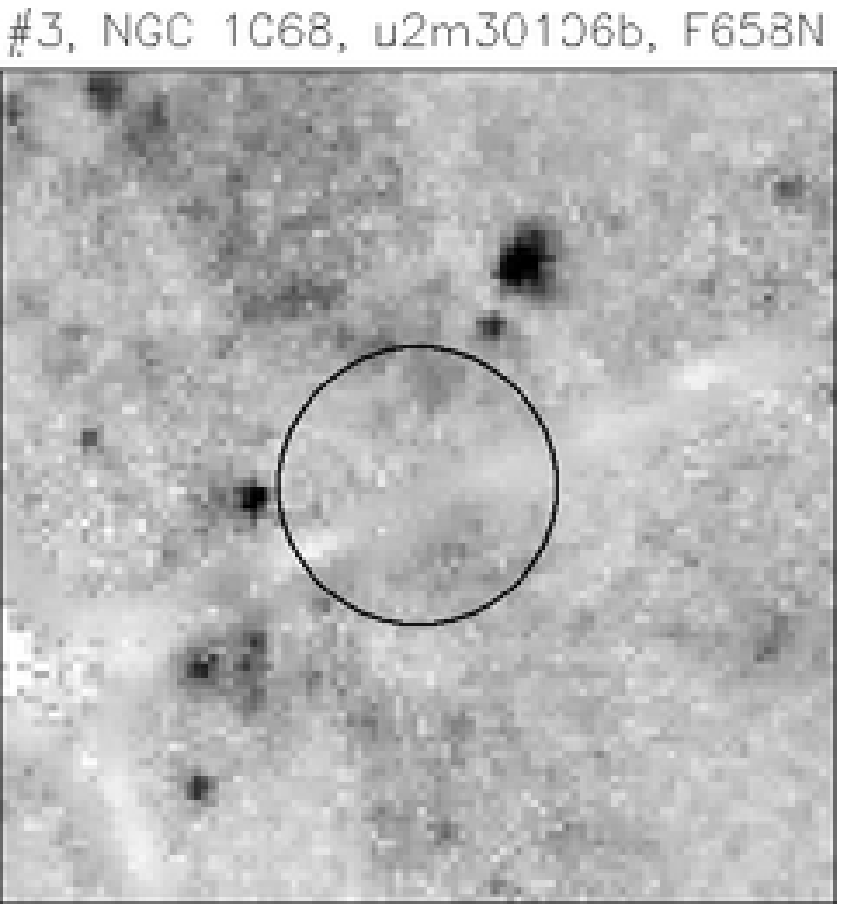}{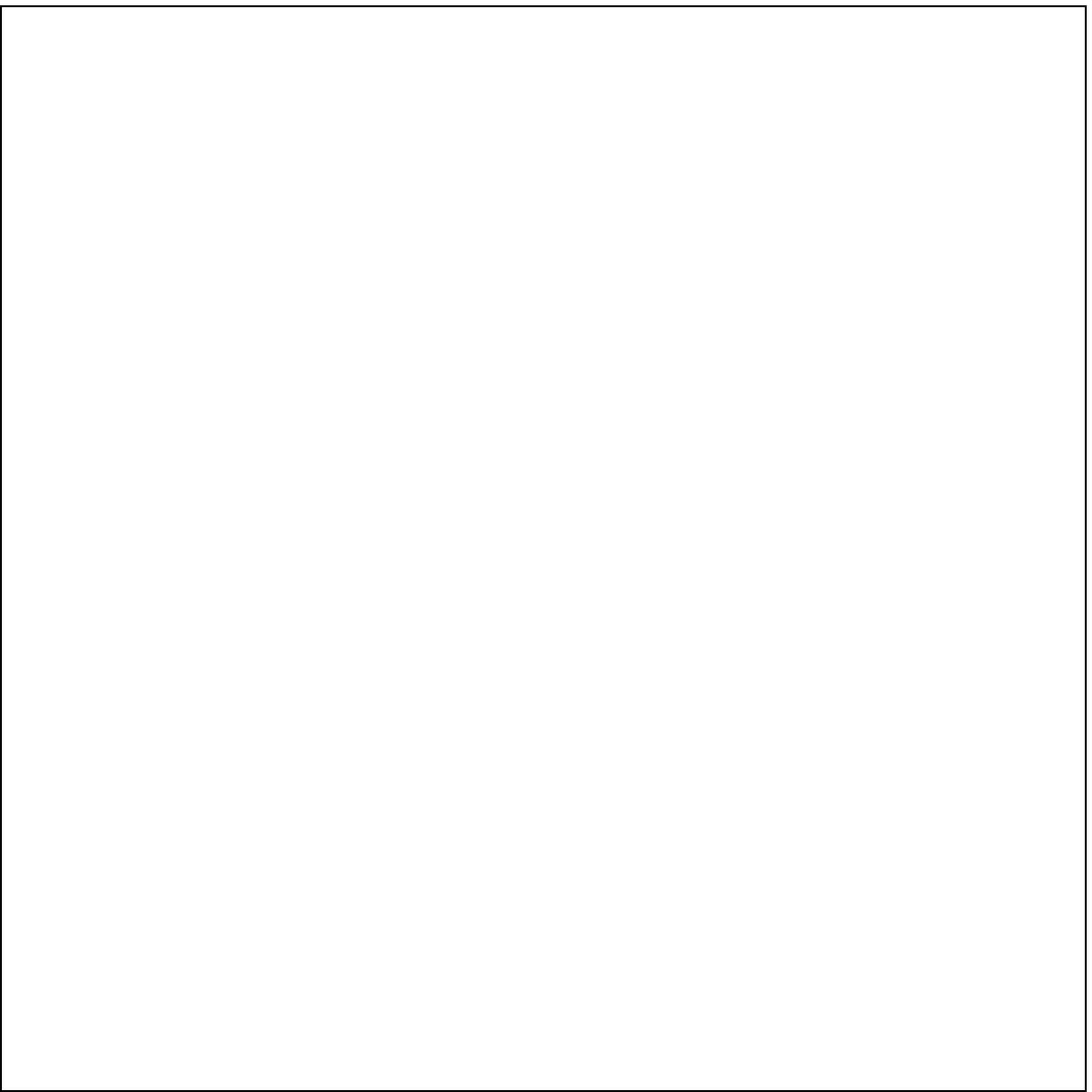}{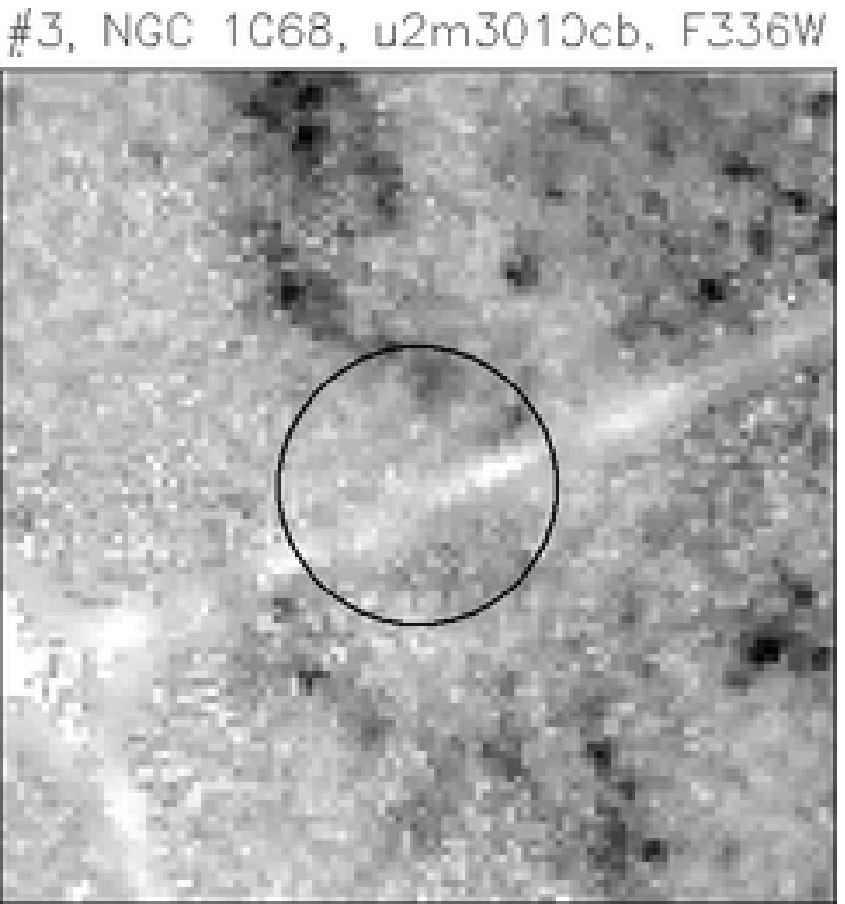}{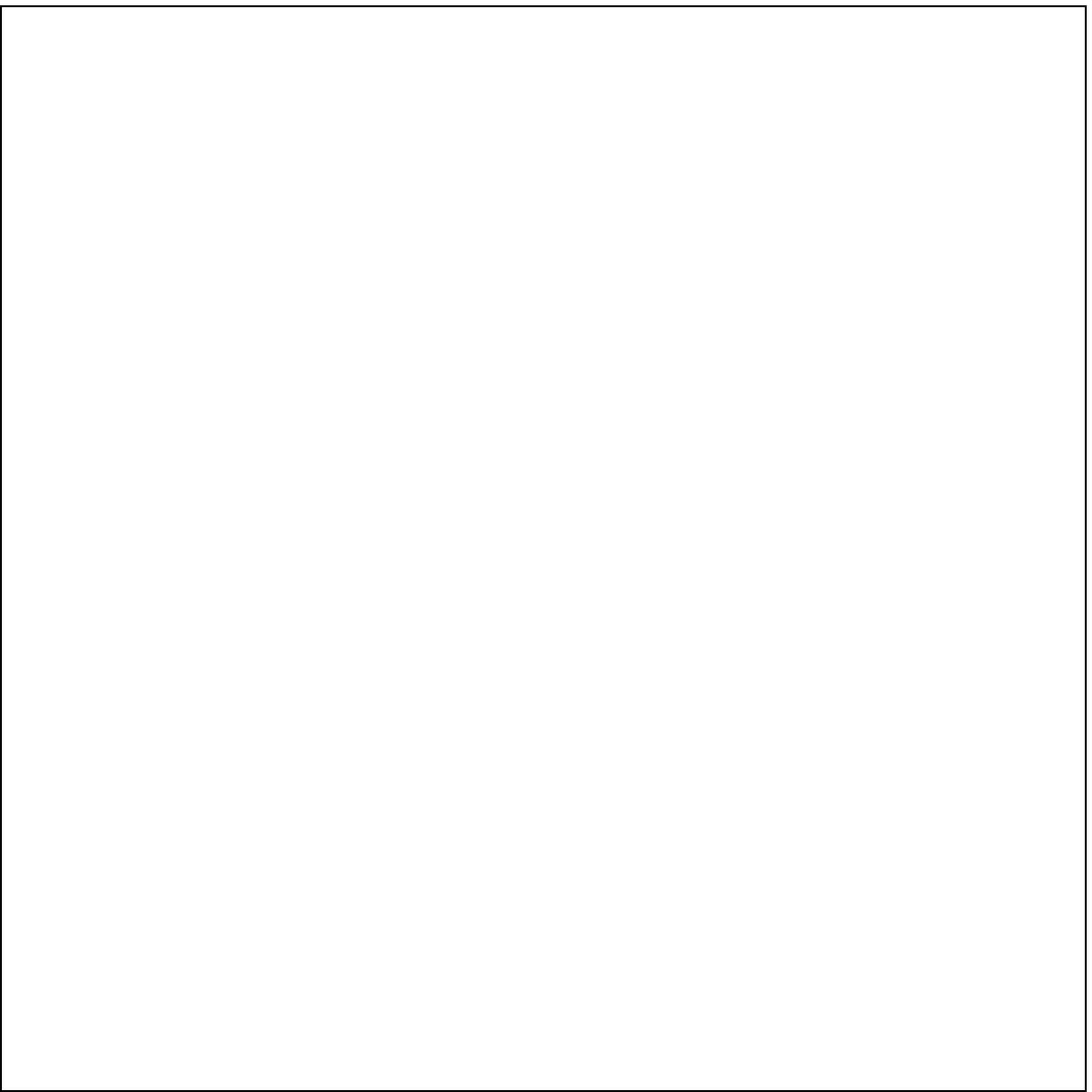}
\doytskip
\fouroffigs{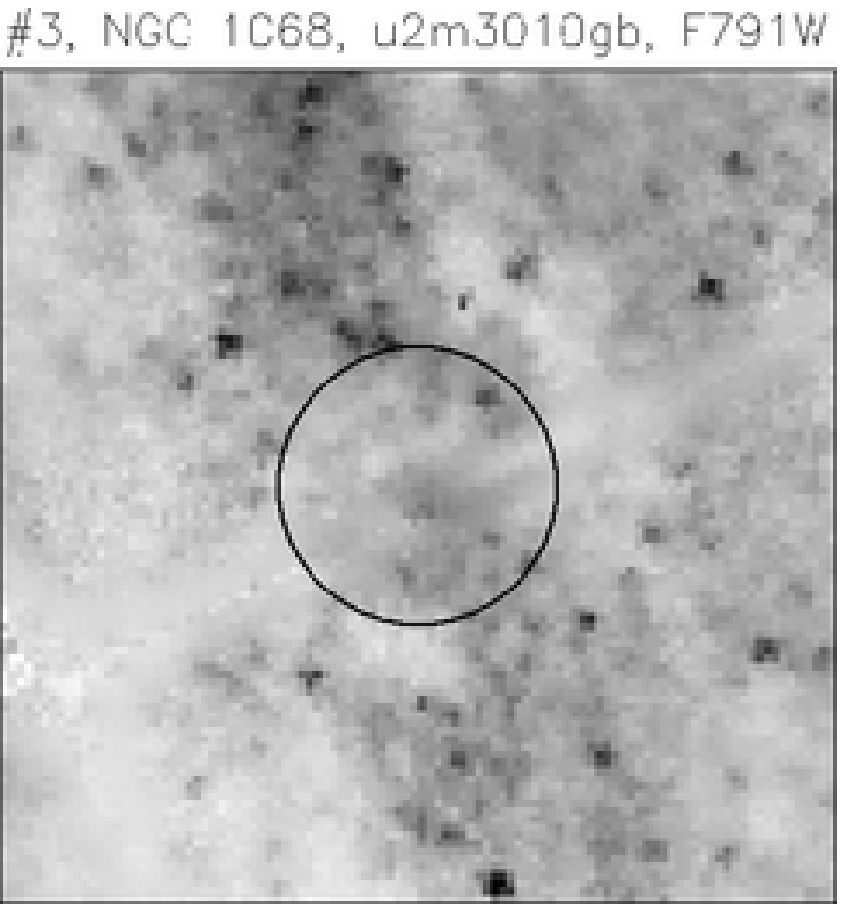}{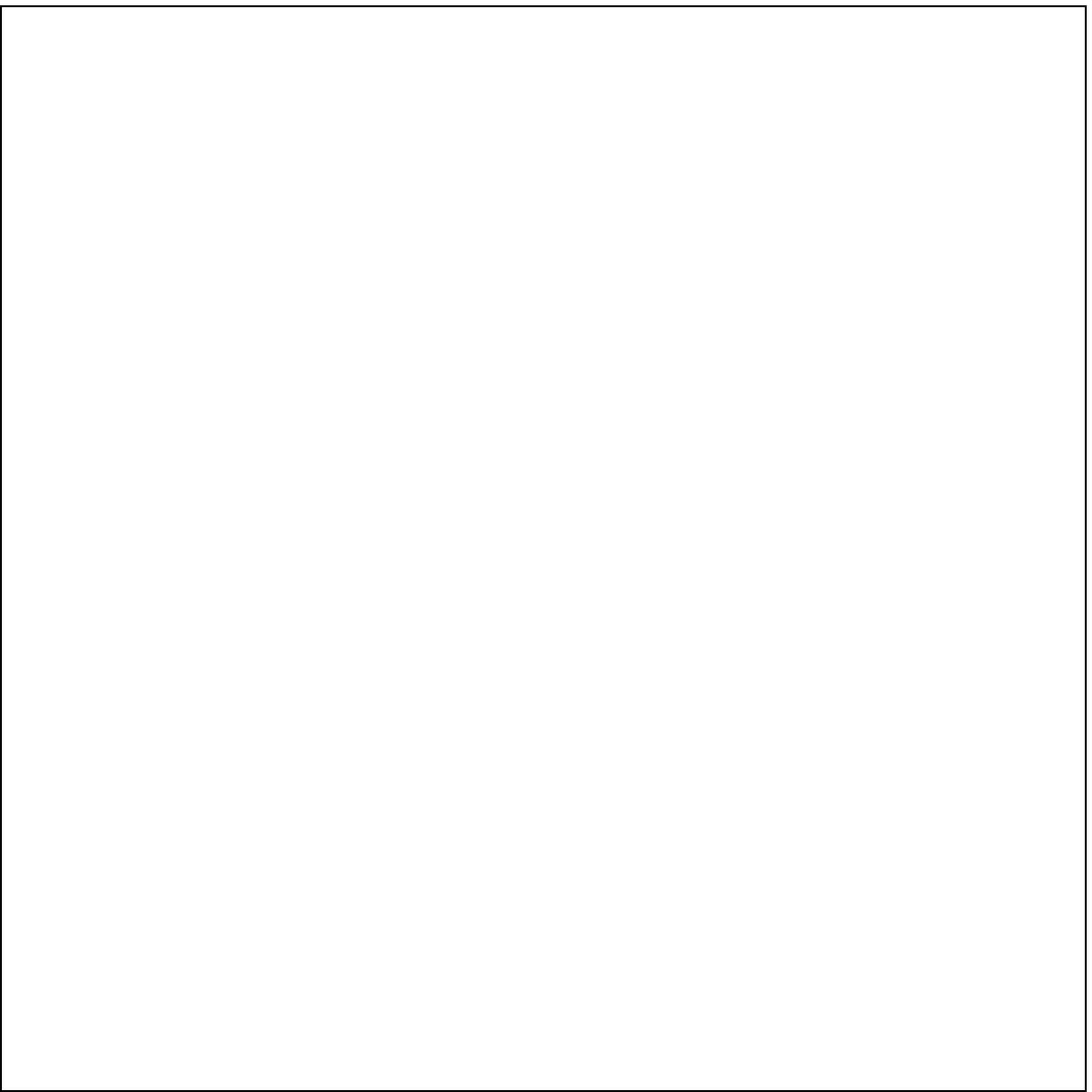}{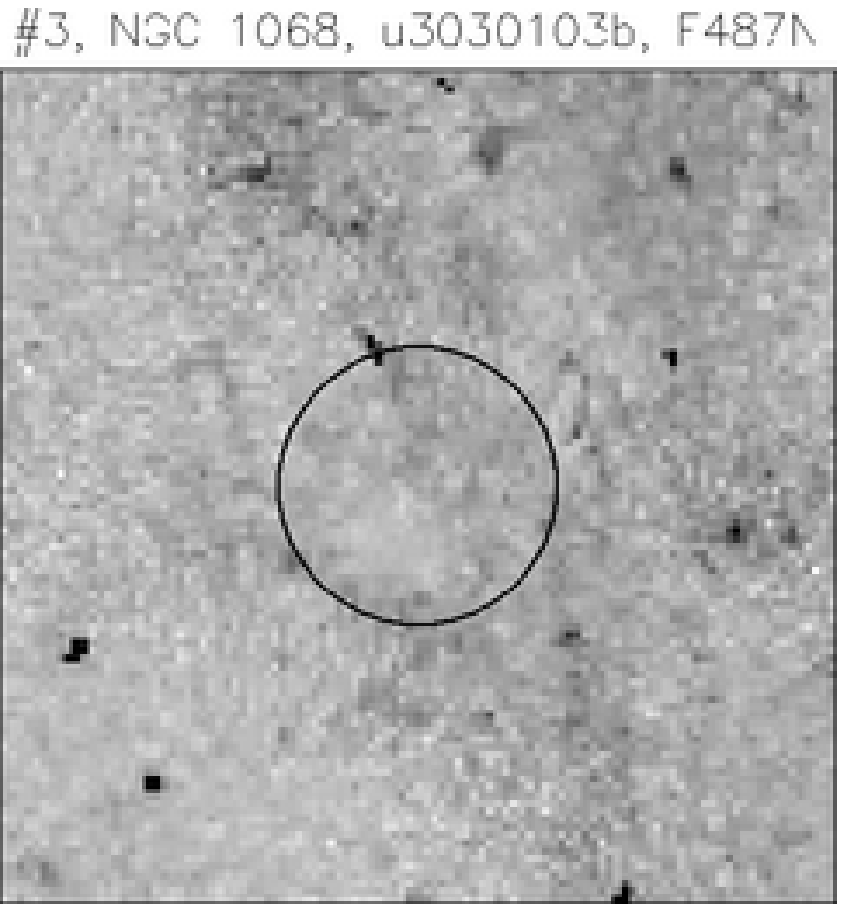}{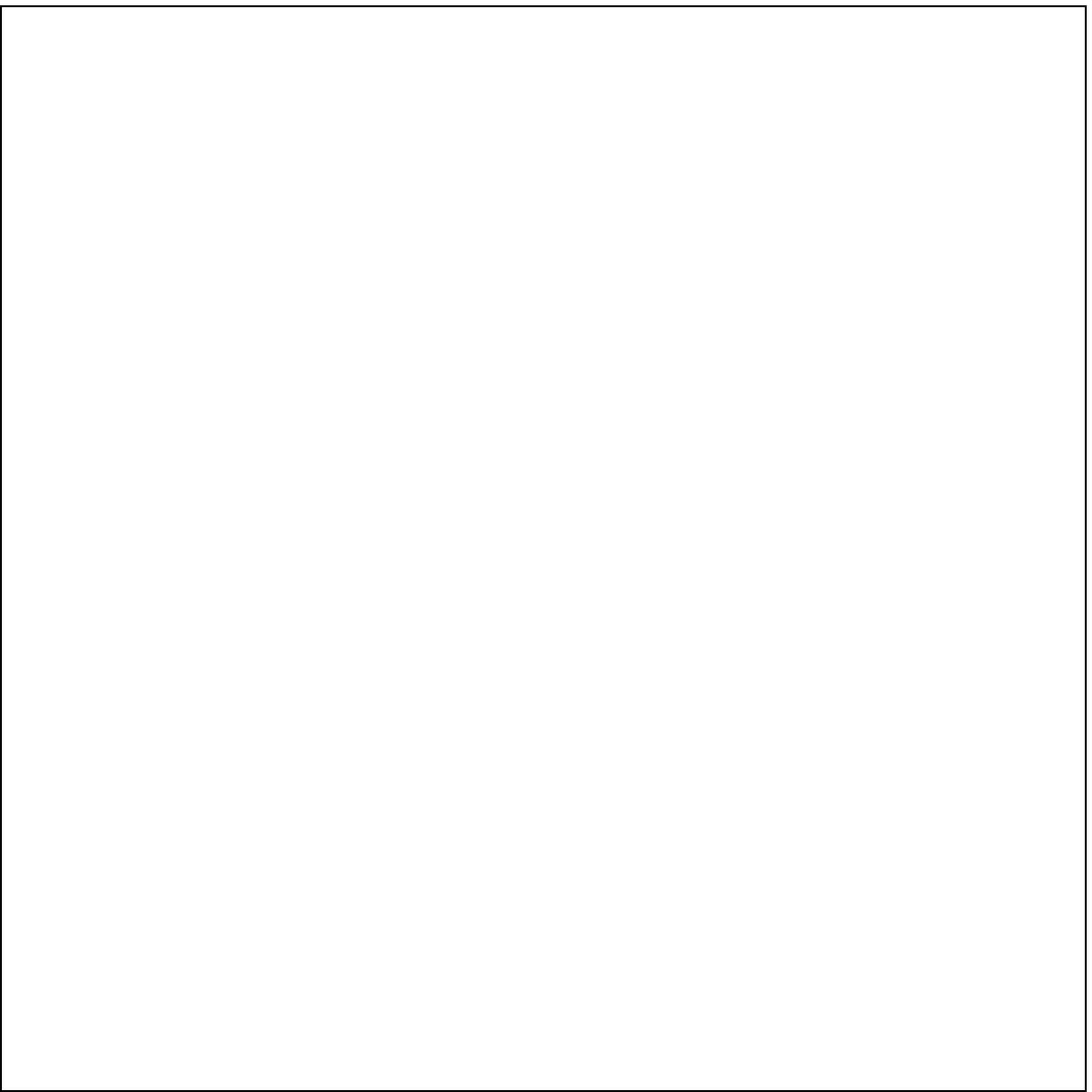}
\doytskip
\fouroffigs{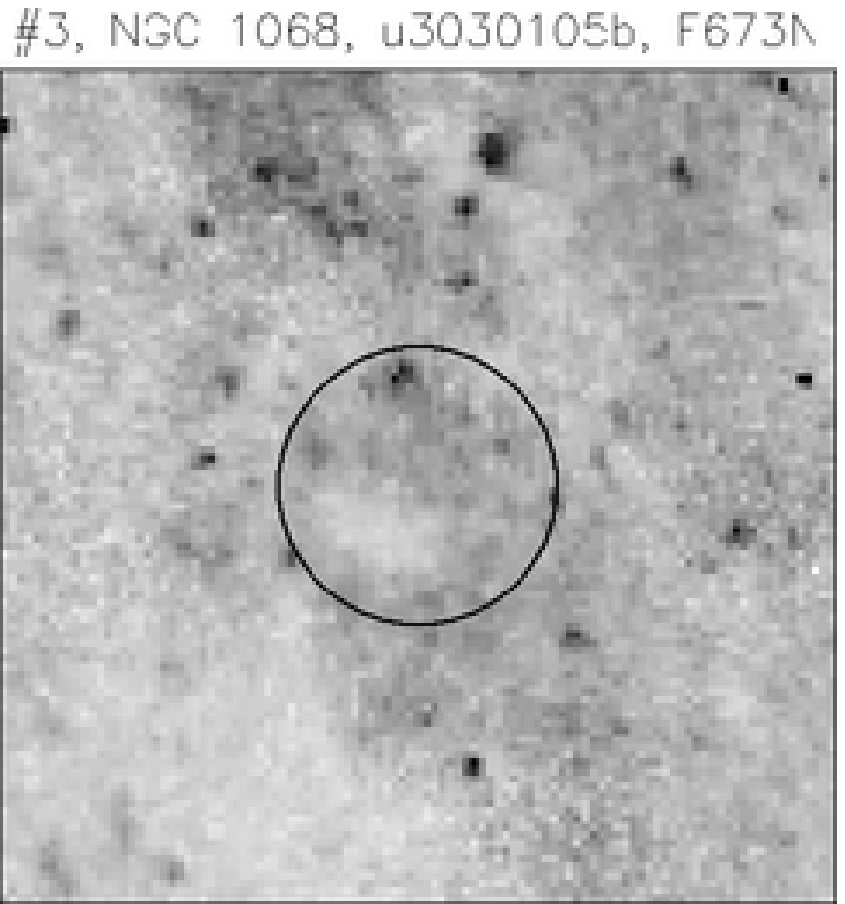}{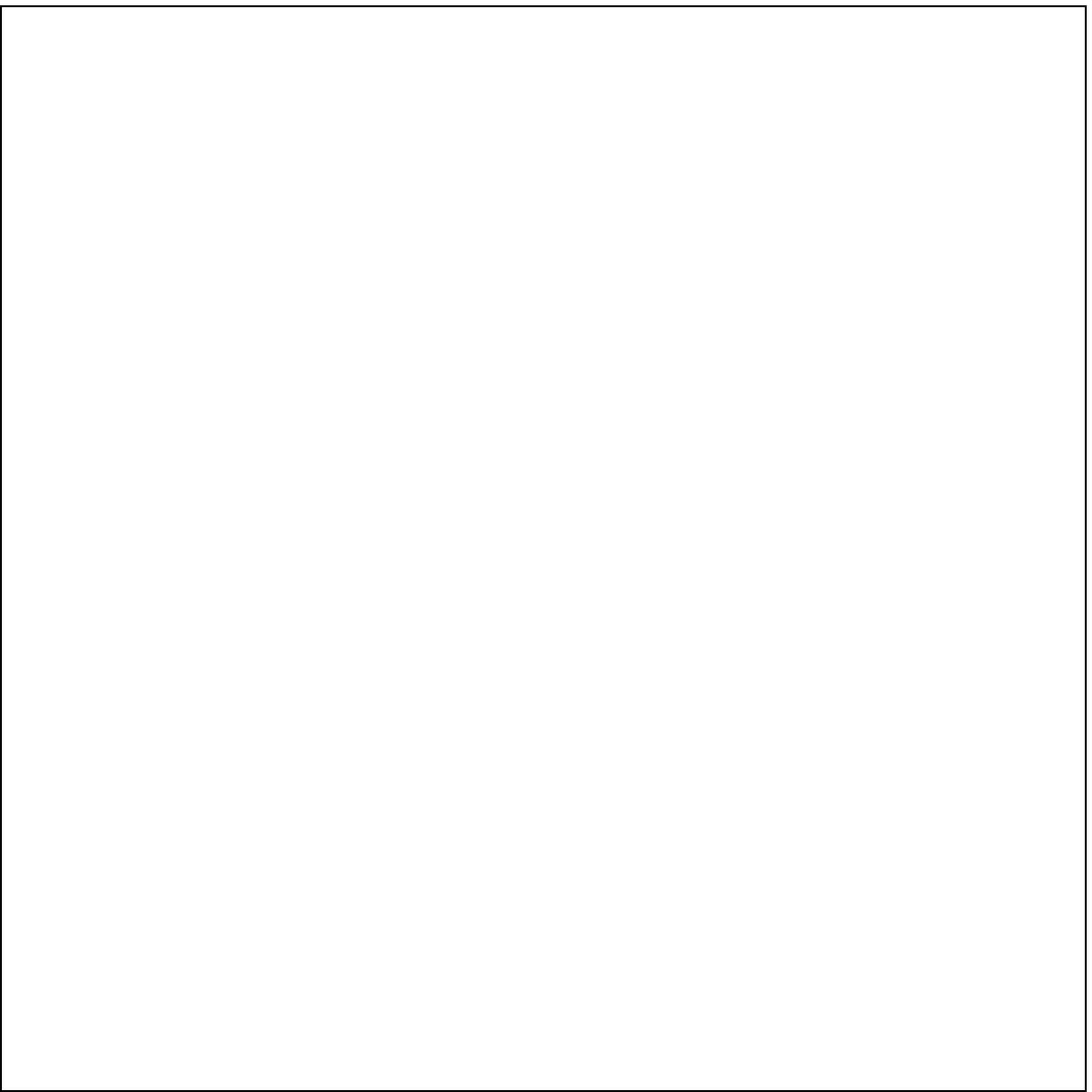}{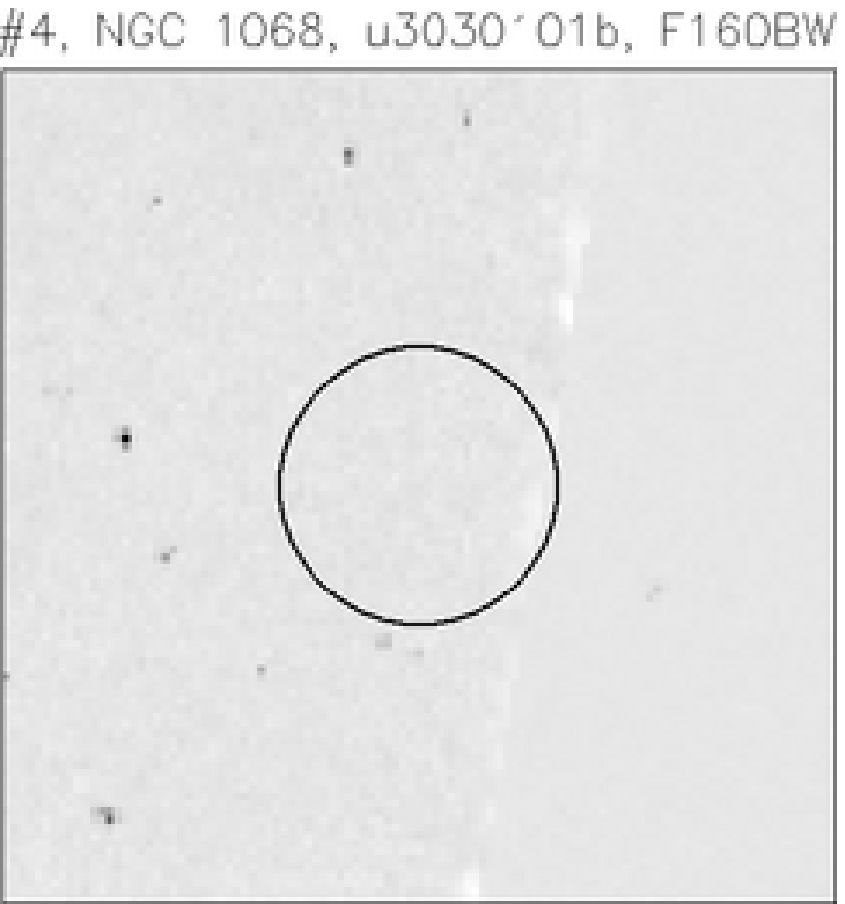}{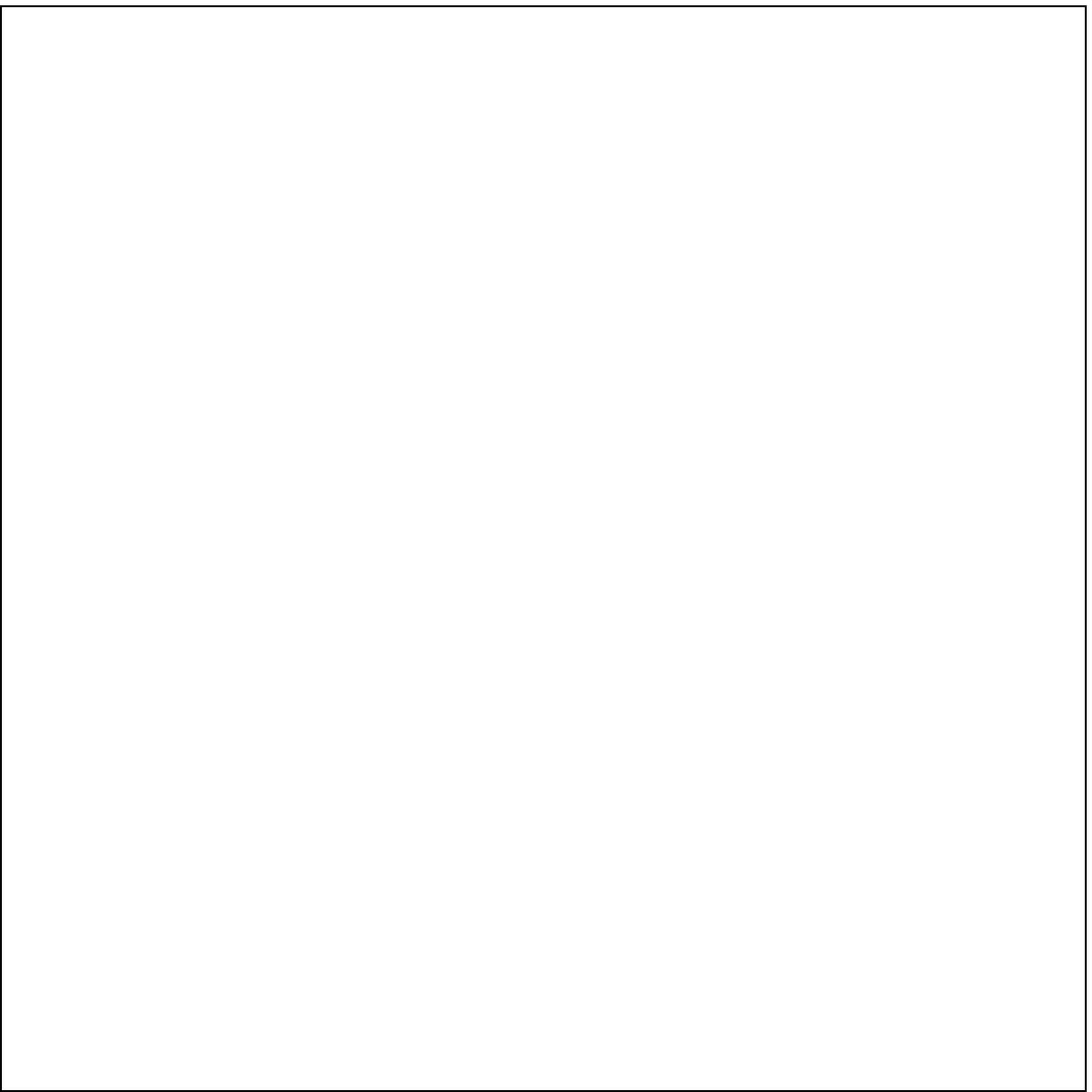}
\setcounter{figure}{1}
\figcaption{\figcapstamps}
\end{figure*}


\section{Photometry}
\label{s:photometry}

We searched all the images for potential ULX optical counterparts. In
doing so we restricted our attention to a search radius in the optical
images that is twice the astrometric registration error for each
ULX. All the ULXs in our sample reside in nearby galaxies, and many
of these galaxies are spiral galaxies.  Such galaxies can have
complicated morphologies, and we therefore commonly found diffuse
structures of various kinds (e.g., parts of spiral arms, dust bands,
etc.) within the search radius around a ULX. Such structures may be
interesting if a connection to the ULX phenomenon could be
demonstrated. However, we decided not to catalog such structures in
the present context. Instead, we decided to restrict our attention to
sources within the search radius that appear point-like, i.e., sources
that are either consistent with the HST point spread function (PSF) or
which are only slightly more extended. Such sources tend to be
associated with star clusters, diffuse regions, or individual stars. All
of these can provide essential insight into the ULX phenomenon if a
direct association with a particular ULX could be demonstrated.

Our method for the identification of point-like sources depended on
the galaxy morphology around the ULX. Based on visual inspection we
discriminated between several different classes of morphologies within
the search radius. We classify as simple morphologies those cases in
which there are one or more easily identifiable, point-like sources on
a relatively smooth or constant background. For these morphologies we
used the standard source detection routine {\tt daofind} in the {\tt
IRAF} software package to find point-like sources. We utilized
4$\sigma$ as the detection threshold. We will refer to datasets in
this category as case S (S for ``simple''). Other ULXs are found in
regions of more complicated morphology. These show diffuse, extended
or dusty structures. In these cases we found that automated routines
for source detection did not provide reliable results. Nonetheless, in
some fraction of cases with these complicated morphologies it was
possible to manually identify point-like sources. We will refer to
these as case M (M for ``manual''). We emphasize the point that the source
detection in these cases is nevertheless somewhat arbitrary,
particularly with regard to whether faint sources are considered
to be real or artifacts.  For the remainder of the
complicated morphologies it was not possible to place any reliable
limits on the presence or absence of point-like sources. We do not
report any results for these cases, which we will refer to as case C
(C for ``complex''). This primarily occurred in the cases of two
galaxies, NGC 1068 and NGC 3034 (M82).  In the case of NGC 1068, while
it appears that in some fields manual source detection would be
possible, taken as a whole the combination of large numbers of sources
within the 2$\sigma$ error circle (3.3'' in radius) and complex
spatial morphology would complicate the interpretation. We
therefore defer detailed analysis in this case for future work when
better astrometry may be available (e.g., by tying the
astrometry of WFPC2 images to ground-based images).  We also classified a
source as ``C'' due to it being on the edge of the chip (a possible
counterpart to ULX 24 in the u3040205b observation).

Another category is formed by those ULXs for
which the optical morphology within the search radius is smooth and
straightforward, but in which there is no sign of detectable
point-like sources. We will refer to these as case N (N for
``nothing''). A final category is formed by those ULXs for which the
optical data was unreliable due to a high fraction of unrealistically
negative pixels within the search radius.  This problem of the data
calibration was discussed in Section~\ref{s:images}. It only affected
a few datasets so significantly that we felt that a quantitative
analysis would not be warranted. These datasets are referred to as
case X (X for ``excluded''). For each ULX and dataset we indicate in
column~(5) of Table~3 to what case it belongs.  Of the
145 unique dataset, ULX, and photometric case combinations, 20 (14\%)
represent case S, 35 (24\%) represent case M, 37 (26\%) represent case
C, 44 (30\%) represent case N and 9 (6\%) represent case X. 

For 9 ULXs we have one or more datasets of case S, and for 20 ULXs we
have one or more datasets of case M. These are the cases for which one
or more point-like sources were either automatically or visually
identified within the search region around the ULX. In total, we found
potential counterparts for 28 of the 44 ULXs in our sample (
there is overlap between S and M cases for ULX 40).  We
assigned  each of the sources thus identified a unique ID number. The
sources are labeled and shown schematically in separate panels of
Figure 2.  Information on the sources is provided in
Table~3.  Columns~(7) and~(8) give the offset of each
source w.r.t.~ULX position (in RA and
Dec.). Column~(9) gives the corresponding 
total offset.  For each source we performed aperture photometry using the
routine {\tt phot} in the {\tt IRAF} software package. The photometry
on the source was performed with an aperture of radius 0.1''
(this is 1 pixel on the Wide Field Camera CCDs and approximately 2
pixels on the Planetary Camera CCD). An appropriate background value
was subtracted from the results. For case S, the local background was
automatically measured in an annulus surrounding the point source. For
case M, the background was kept fixed at the value determined manually
from a representative (but carefully chosen) region near the ULX.

The measured count rates in apertures were converted to magnitudes
using the zeropoints for each filter given in the WFPC2 Data Handbook
(Mobasher \etal 2002). The resulting magnitudes are in the so-called
{\tt VEGAMAG} system. This is a synthetic system in which the
magnitude of the star Vega is defined to be exactly zero in all
filters. In the traditional Johnson-Cousins system the magnitude of
Vega is approximately zero in the UBVRI bands, but not exactly
(Bessel, Castelli, \& Plez 1998). In general, magnitudes on the {\tt
VEGAMAG} system are close to those on the Johnson-Cousins system,
provided the filters have similar central wavelengths. However, to
actually transform the magnitudes onto the Johnson-Cousins system it
is necessary to know the spectral shape, or at least a color, of each
of the identified sources. Since this is not generally known, or known
well, we have not attempted such transformations.

Various corrections are necessary to the inferred magnitudes. First,
an aperture correction is necessary to estimate the true magnitudes
from those that are measured in a small aperture. We used the TinyTim
software (Krist \& Hook 2001) to generate synthetic PSFs for each of
the filters. From aperture photometry on these PSFs we calculated the
aperture corrections between the apertures that we used for the data
analysis and a $0.5''$ radius aperture. These corrections were applied
to the inferred magnitudes. An additional correction of $0.1$ mag was
applied to transform the results to a nominal infinite aperture
(Mobasher \etal 2002). A second correction that must be applied is for
geometric distortion, which results in a change of the surface area
covered by each pixel. This pixel area variation is imprinted on point
source counts during the flat-fielding step (which is defined to make
a source of uniform surface brightness look ``flat'' on the detector).
We corrected for this using the known pixel area variations (Mobasher
\etal 2002). The correction amounts to at most $\sim 0.05$ mag in the 
corners of each chip. The third correction that must be applied is for
imperfect charge-transfer efficiency (CTE), which causes charge to
be lost from the aperture during read-out. We corrected all photometric
results (detections only) using the CTE correction algorithm of
Dolphin (2000), using the equations posted at
http://purcell.as.arizona.edu/wfpc2\_calib/ (December 2004 version). The CTE 
corrections increase with decreasing background. 

The final magnitudes obtained after application of all corrections are
listed in column~(10) of Table~3.  We calculated the total
statistical error as $[cg + 2A(\sigma g)^2]^{1/2}$, where $c$ is the
net number of counts in the aperture, A is the area of the aperture in
pixels, $\sigma$ is the noise level per pixel determined from the
background region, and g is the gain (either 7 or 14 electron/data
unit).  This error is more conservative than the error computed by the
{\tt phot} routine and was on average $0.02$/$0.05$ magnitudes larger for
sources brighter/fainter than 24. It is likely that these are
underestimates of the true 
errors, especially in cases with complicated morphologies. Implied absolute
magnitudes were calculated based on the galaxy distances quoted in
Table~\ref{t:sample} and are listed in column~(12). Signal-to-noise
ratio ($S/N$) estimates were calculated 
for each counterpart in each image and are listed in column~(13),
using the formula $S/N = \frac{c}{\sigma A^{0.5}}$. 
Sources for case~S mostly had $S/N \gtrsim 4$, due to the
detection threshold used in the {\tt daofind} routine. For case~M
sources we found that our manual source identification was also able
to detect sources down to $S/N \approx 4$, although most of the
sources thus identified have much higher $S/N$.

For each ULX and each dataset (except those of case~X for which the
data were suspect and case~C where the local background level is
ambiguous) we estimated the limiting magnitude down to which a 
source might plausibly have been detected. For this we calculate the
noise in a background aperture of the same size as was used for source
photometry. We multiply the result by 4 to get the counts needed for a
$4\sigma$ detection. We transform this result to magnitudes using the
exposure time and filter zeropoints and apply the appropriate aperture
correction. The results are listed in column~(6) of
Table~3. For comparison, we also calculated the
limiting magnitudes using the Exposure Time Calculators (ETCs) for
WFPC2 maintained by STScI. These more idealized calculations generally
give somewhat fainter limits than the ones we calculated. This is as
expected, given that the ETCs do not take into account that we are
searching for sources in nearby galaxies. The galaxy light adds both
noise, fluctuations and complexity. 

\begin{figure}
\plotone{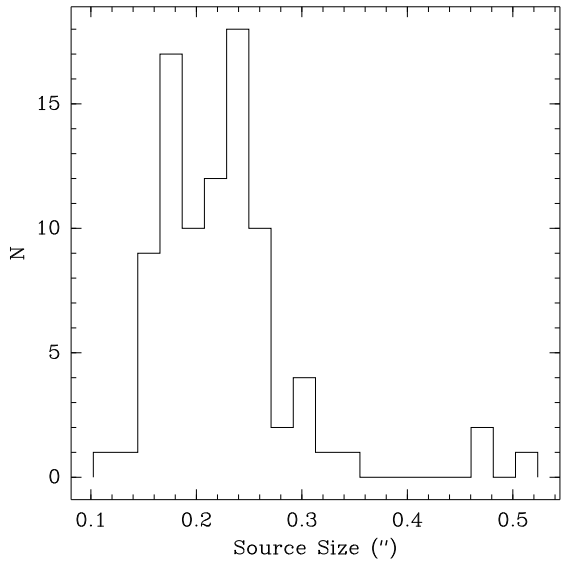}
\caption{The size distribution of the ULX counterparts. The sizes
plotted are radii estimated visually from the
mosaic images. These source sizes are probably representative of point
sources, where the mean radius corresponds to $\sim 85\%$ of the encircled
energy fraction (see text for details). \label{f:size_hist}}
\end{figure}

In the final column we list source radius estimates.  We found that
kernel-fitting routines such as {\tt imexamine} did not converge
consistently, particularly for ``M'' sources. We therefore estimated
the source sizes visually from the mosaic images (with a uniform pixel
size of 0.1'').  While subjective, the intent here is to simply give
an estimate for the spatial extent of the sources and not a detailed
spatial analysis.
We plot in Figure \ref{f:size_hist} a
histogram of the
estimated source sizes.  The majority of sizes were in the range
0.1-0.4'', with a few outliers having larger sizes.  Excluding these
outliers, the mean and standard deviation of the source radii are
0.22'' and 0.05'', respectively.  For comparison, from the WFPC2
Instrument Handbook, this mean radius corresponds to an encircled energy fraction of $\sim
85\%$ at 4000\AA.  In cases where the use of {\tt imexamine} resulted
in a 
reasonable
full-width half-maximum (FWHM) estimate, the mean value was 0.15''
whereas the mean visual source size for these sources was
0.19''. Therefore most sources are likely to be point sources or at
most only marginally resolved, with our visual size estimate being
$\sim 30\%$ larger than the FWHM.  
We also list the linear extent corresponding to the source
size assuming the distances listed in Table 1.

\section{Discussion}
\label{s:discuss}
We have presented a systematic analysis of archival WFPC2 observations
for a sample of ULXs.  The main motivation was to determine the
characteristics of possible optical counterparts to ULXs.  Minimally a
key goal was to determine what is the probability of detecting a
counterpart.  Other goals would be to establish the typical optical
luminosity of counterparts and how the implied X-ray to optical (or
UV) luminosity ratio compares to known source types.  Finally, in
cases in which observations are available in multiple wide-band
filters, the optical colors might also be a useful discriminator of
source type.  Below we address the color information content of the
sample prior to computing optical luminosities since  
any knowledge of the spectral types of the counterparts would benefit
the computation.  Of course, when multiple counterparts are present, obviously
at most only one of the counterparts within the error circle 
can be the true counterpart of the ULX.  Also, when only a single, faint
counterpart is observed, particularly in late-type galaxies, the
association may simply be due to chance coincidence.  Therefore this
discussion should be taken as an assessment of what properties any
potential counterparts would have {\it on average}.

\subsection{ULX WFPC2 Counterpart Frequency}
\begin{figure*}
\plottwo{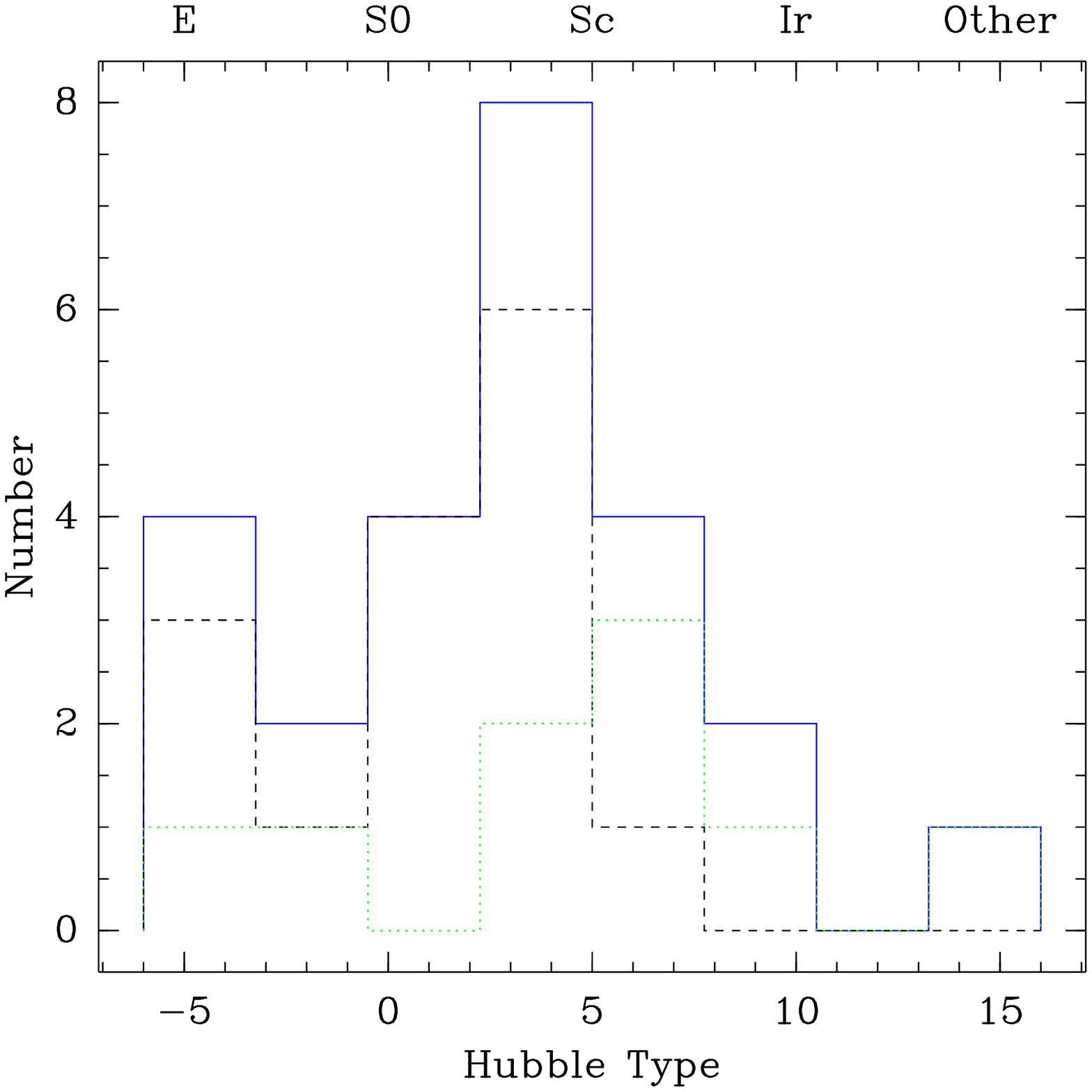}{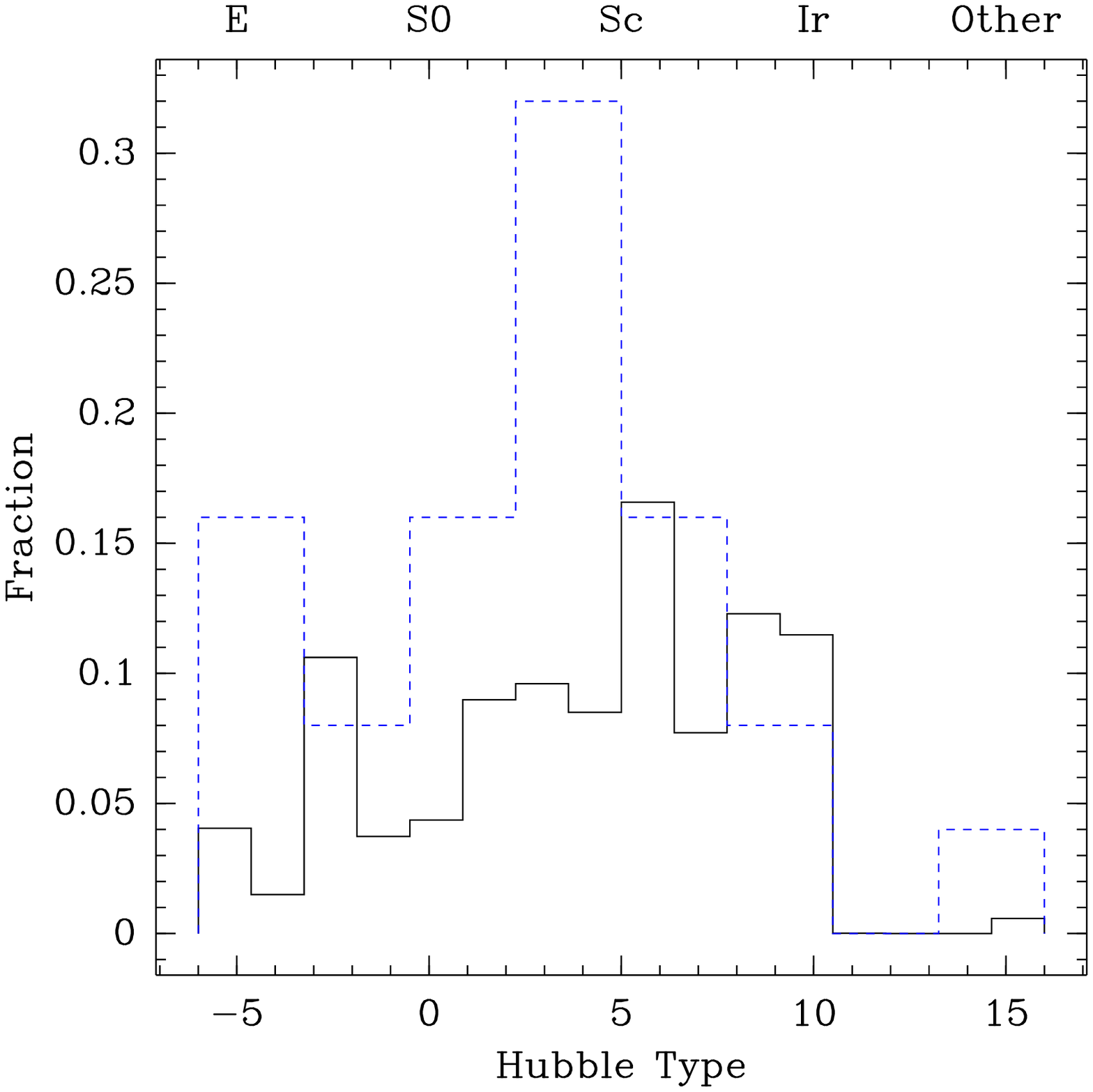}
\caption{The Hubble type distributions of galaxies in the
  Chandra/WFPC2 sample.  (left) The full (blue, solid line) sample
  distribution along with the subset in which at most one counterpart
  was detected (black, dashed line) and the subset in which more than 
  one counterpart was detected and/or an observation was marked as
  complex (green, dotted line).  (right) The fractional type
  distribution of galaxies in the full Chandra/WFPC2 sample (blue,
  dashed line) along with the type distribution of nearby ($cz < 5000
  \rm \ km \ s^{-1}$) RC3 galaxies (black line). \label{f:htypes}} 
\end{figure*}

As is evident from Table 3 and Figure \ref{f:stamps},
variably zero, one or many counterparts are detected within the
(combined X-ray/HST) error circles.   Since the stellar
populations in elliptical galaxies are old (and hence faint), a simple assumption would then be that finding no counterpart would be more common in early-type 
galaxies.  Also, since there is an association between X-ray binaries
and globular clusters in elliptical galaxies (see, e.g., Maccarone,
Kundu, \& Zepf 2003), we might expect to find counterparts to
extra-nuclear X-ray sources in elliptical galaxies that are likely to be
globular clusters.    To test these assumptions, we plot in Figure
\ref{f:htypes} histograms showing 
the Hubble type distribution for the Chandra/WFPC2 galaxy sample (with
25 galaxies total), the subset in which at most one counterpart was
detected (15 galaxies) and the subset in which more than one counterpart
was detected and/or an observation was marked as complex (10
galaxies). As expected, there is a tendency for single or no
counterparts to 
be found in early-type galaxies and multiple counterparts
and/or complex cases to occur in late-type galaxies.  

In Figure \ref{f:htypes} we also
compare the Hubble types of the galaxies in our Chandra/WFPC2 sample
to the type distribution of nearby ($cz < 5000 \rm \ km \ s^{-1}$, or
D $<$ 66 Mpc for $H_0$ = 75 km s$^{-1}$
 Mpc$^{-1}$  and $q_0$ = 0.5) galaxies in RC3, in
order to assess if there is any bias in our galaxy sample.
While the statistics are limited, there is a slight excess of
ellipticals and irregular/peculiars, and a more pronounced excess of
Sb/c galaxies in our sample relative to nearby galaxies.  Note that
the prominent peak in the Chandra/WFPC2 galaxy type distribution at Sb/c
is mainly due to a large number of Sb (type = 3) galaxies (see Table
\ref{t:sample}).  Therefore, in addition to the excess of Sb galaxies,
the main bias in the sample is a lack of early-type (i.e., S0, Sa) and
late type (i.e, Sd) spiral galaxies.

Finally, we computed the number
of background sources expected for the galaxies with Hubble type $<$3
and Hubble type $\geq$ 3.   This was done by taking the larger of the
limiting flux in each Chandra field and the flux associated with a
$L_X = 10^{39} \rm \ ergs \ s^{-1}$ source (i.e., see Ptak \& Colbert
2004), in both cases computed in the 2-10 keV band based on the
full-band count rates.  The number of 
background sources expected in each field was then taken from the 2-10
keV logN-logS plot given in Bauer et al. (2004) derived from the
Chandra Deep Fields.  This resulted in an estimate of 5.5 background
sources in the early-type galaxies (35\% of the total) and 3.8
background sources in the late-type galaxies (14\% of the total).

\subsection{Optical Colors}
\begin{figure*}
\plottwo{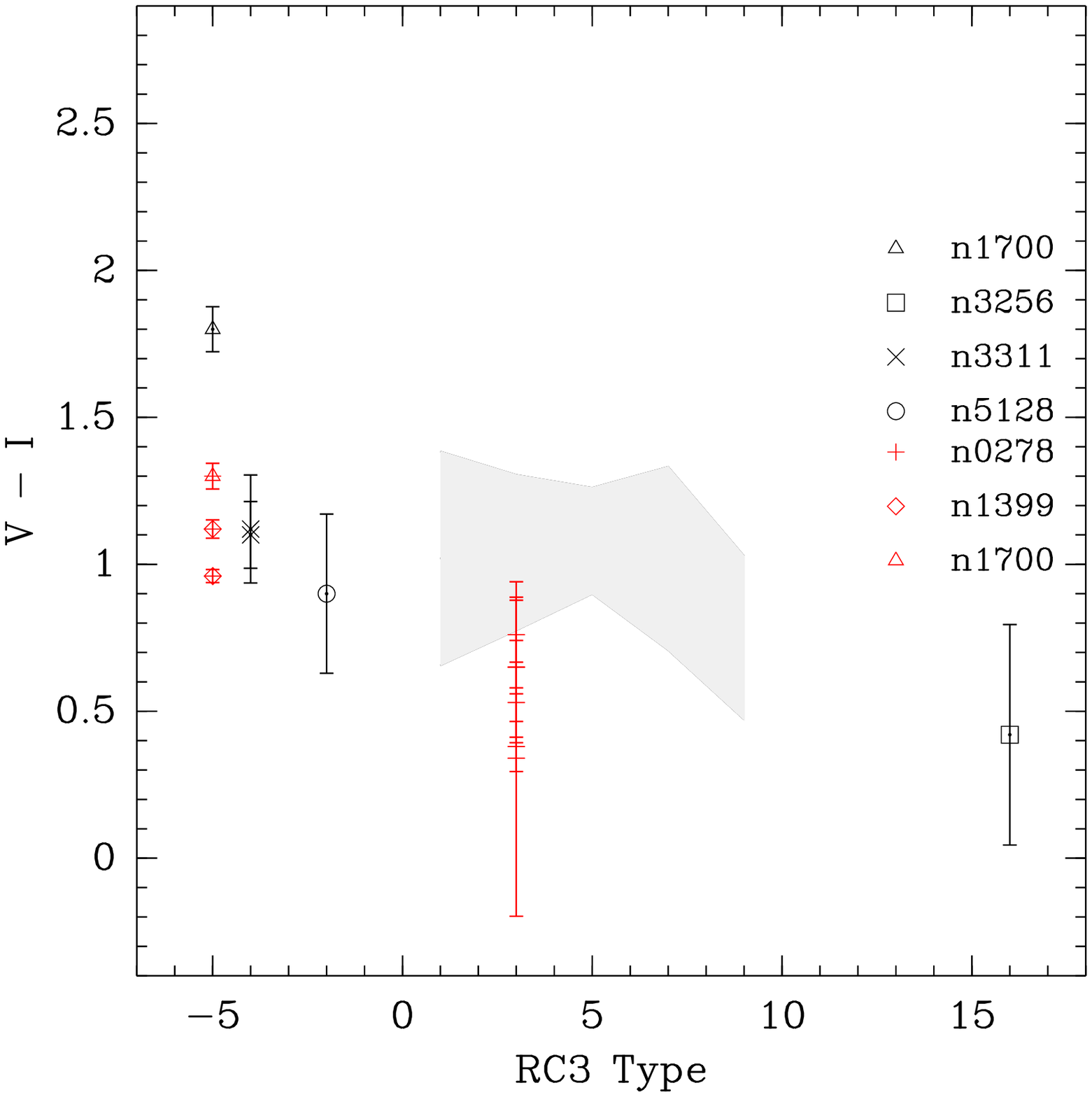}{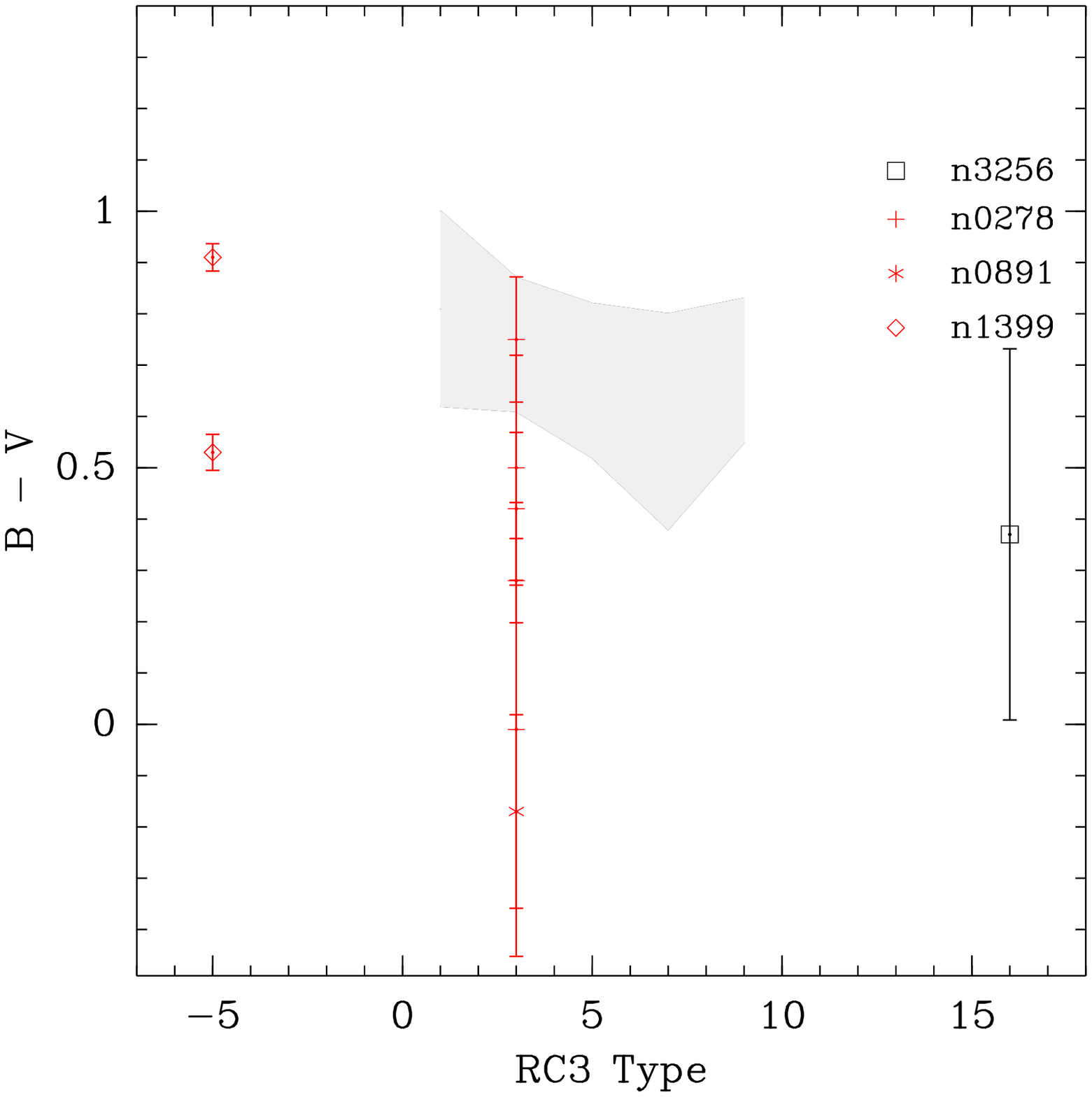}
\caption{V-I (left) and B-V (right) colors for counterparts to ULX
  sources.  F891W and F450W have been used as proxies for I and B,
  while F555W and F606W have been used as a proxy for V (plotted with
  black and red points, respectively).  Note that a large fraction of
  the data points in these plots are due to counterparts for ULX 1
  (see Table 3).  The shaded regions show the 1-$\sigma$
  range in global color determined for spiral galaxies in de Jong (1996).
  For reference, the WFPC2 Exposure-Time Calculator Tool was used to
  compute colors using the Sbc galaxy spectral model (T=4), which resulted in 
  B-V = 0.4 (0.6) and V-I=1.0 (0.8) for V=F555W (F606W), consistent
  with the de Jong values. \label{f:colors}}
\end{figure*}

We turn now to cases in which there are observations of a ULX in more
than one wide-band filter and at least one detection (and hence a
color or color limit can be computed).  Here we restrict our sample to
observations with the filters F336W, F450W, F555W, F606W and F814W
since they map reasonably well to standard UBVI magnitudes. With both
F555W and F606W approximating the V band, hereafter we refer to them as
$V_{555}$ and $V_{606}$.  Only a few ULXs did not
have a (CADC) observation in one of these filters, and
in those cases the observations were entirely in either narrow-band 
(e.g., ULX 41 in Circinus) or UV (e.g., ULXs 43 and 44) filters.  The
colors and color limits are listed in Table \ref{t:colors}.  The
colors B-V and V-I are plotted (as a function of Hubble type of the
host galaxy) in Figure \ref{f:colors}.  Note that 6/9 of the
$V_{606}$-I and 5/8 of the B-$V_{606}$ data points are due to ULX 1 in
NGC 278 (Hubble type 3 = Sb).   The color
F555W - F814W differs from Johnson V-I negligibly for V-I in the
range -1 to 3 (see Holzman et al. 1995
\footnote{\tiny see also http://www.stsci.edu/instruments/wfpc2/Wfpc2\_hand\_current/ch8\_calibration10.html}).
The sense of the correction between $V_{606}$ and $V_{555}$ is such that it 
is negligible for sources with V-I $<$ 0.5.  For red sources, $V_{555} -
V_{606} \sim 0.3-0.6$ (for stellar types G and M0III), i.e., at worst
we are underestimating V-I where V=$V_{606}$.

\subsection{Optical Luminosities}
Excluding the counterparts marked as complex, there are 67
unique counterparts in which one of UBVI was used.  For observations
in these filters we first compute the absolute magnitude and the
luminosity in solar units, e.g., $\log L_V = -0.4(M_V-M_{\odot,V}$), where we
use $M_{\odot, U} = 5.51$, $M_{\odot, B} = 5.41$, $M_{\odot, V}=4.79$,
$M_{\odot, I}=4.03$ for solar absolute magnitudes (Zombeck 1990).  The
results are 
listed in Table \ref{t:lopt}. Since $\sim 50\%$ of the counterparts include
either a $V_{555}$ or $V_{606}$ measurements, and V is a convenient
magnitude for comparison with published photometry and models, we
proceed by computing an effective V-band luminosity. For cases with
color information available, we classified the counterpart as red if
V-I $>$ 0.5, B-V $>$ 0.5, B-I $>$ 1.0, or U-V $>$ 0.7 and blue otherwise.
If no color information is available, then we
assumed sources in galaxies with Hubble type $< 3$ to be red and
blue otherwise. 
 For red (blue) counterparts, we computed $V_{eff}$
assuming a G5 (A0) stellar spectrum.  We then computed $M_{V, eff}$,
the effective absolute V magnitude, and $L_V = \nu L_{\nu}$ (we assumed
an effective wavelength of 5500\AA~ and zero-point flux density of
$3.67 \times 10^{-20} \rm \ ergs \ cm^{-2} \ s^{-1} \ Hz^{-1}$;
Zombeck 1990), also listed in Table \ref{t:lopt}.  $L_V$ is listed in
units of $L_{\odot,V} = 2.9 \times 10^{33} \rm \ ergs \ s^{-1}$, and
these values are in good agreement with those 
computed from the absolute magnitudes in individual filters.  

\begin{figure*}
\plottwo{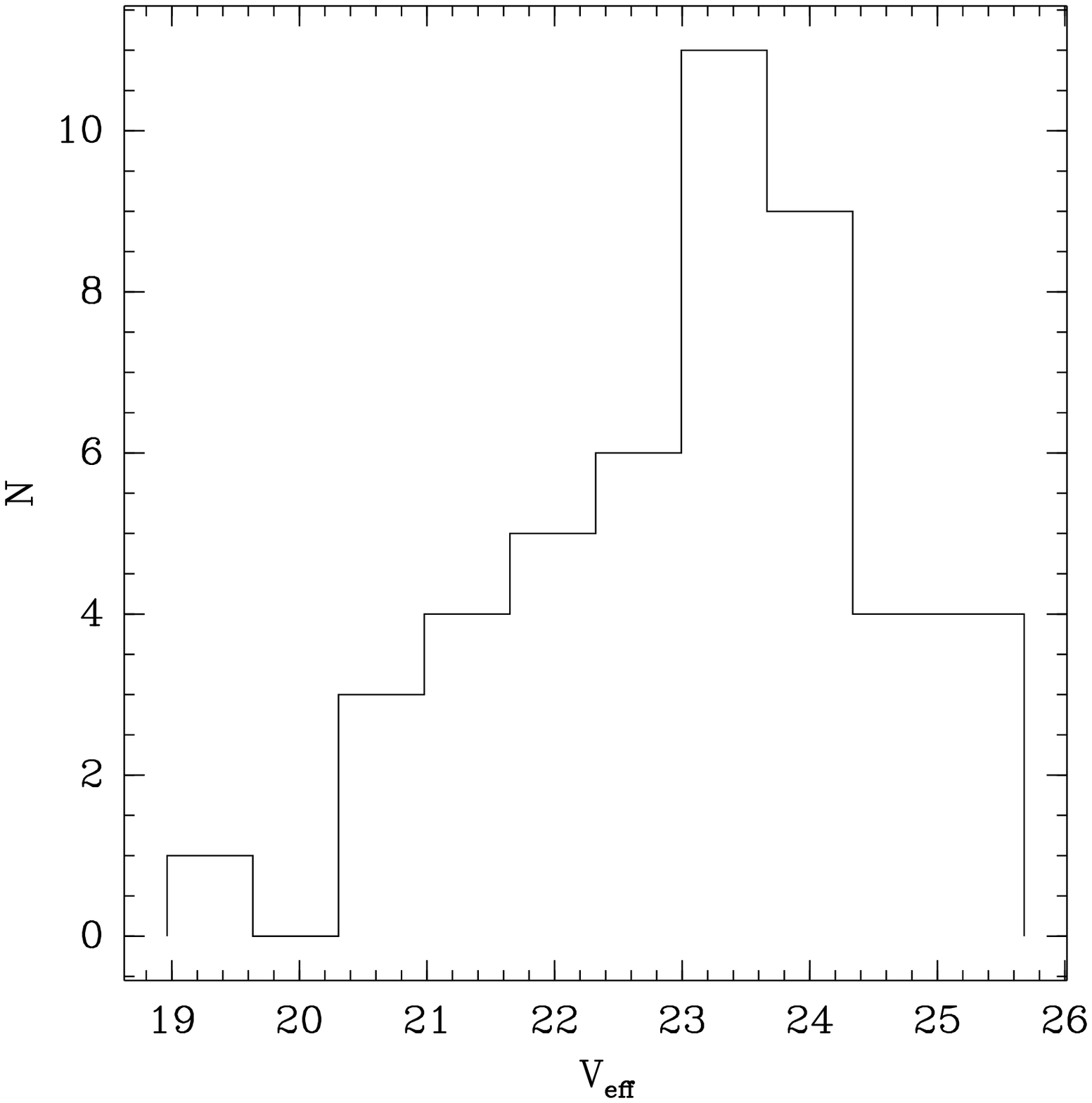}{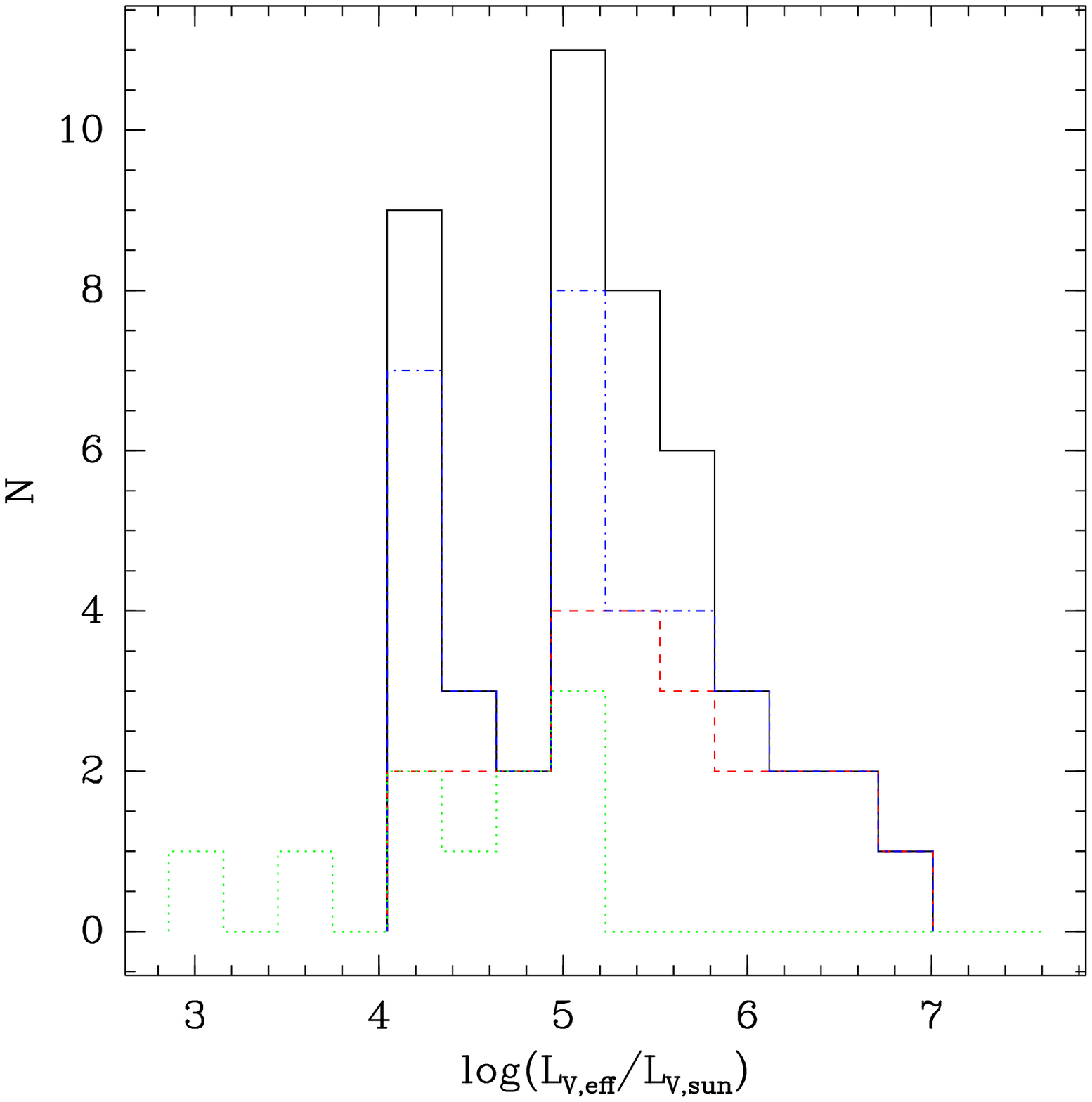}
\caption{Histograms of the computed effective V magnitudes ($V_{eff}$, left)
  and the corresponding luminosities in solar units (right). The
  luminosity plot also includes the cases of only including sources
  brighter than $V_{eff}$ = 23.5 (dashed line) and $V_{eff}$ = 24
  (dot-dashed line), as well as upper-limits (dotted line).  This shows that 
this sample is only approaching completeness for $L_V/L_{\odot, V} > 10^6$.
\label{f:Veffs}}
\end{figure*}

In Figure \ref{f:Veffs} we show the distributions of
$V_{eff}$ and $\log L_V$.  While the
statistics are limited, the
$V_{eff}$ distribution is reminiscent of a power-law distribution with
a cut-off at $V \sim 24$, which is consistent with the average
limiting magnitude of the fields discussed here (24.9).  The $\log
L_V$ distribution shows that we are 
observing sources at least as luminous $10^4 L_{\odot, V}$, with a
mean value of $\sim 10^{5} L_{\odot, V}$.   We also plot the
luminosity distributions only including sources brighter that V=24 and
V=23.5.  The large drop-offs in number counts below $\sim 10^{6} L_{V,
  \odot}$ suggests that 
we are only approaching completeness for counterparts with $L > 10^{6} L_{V,
  \odot}$.  We also plot the effective V band luminosity distribution for
upper-limits (also listed in Table \ref{t:lopt}). For the majority of cases the
limiting luminosity was $10^5$ or higher, suggesting that 
the lack of a counterpart was due at least in part to the lack of sufficient
depth in those fields to detect counterparts of comparable luminosity to
the counterparts in the entire sample.  

\begin{figure}
\plotone{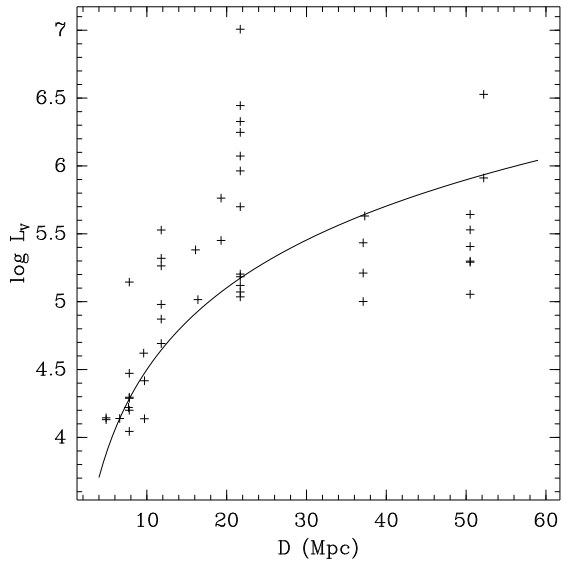}
\caption{Effective V-band luminosity plotted as a function of distance
  to the host galaxies.  The line plotted shows the trend expected for
  a source with constant flux (i.e., $L_V \propto D^2$, where $D$
  is the distance).  \label{f:Veff_dist}}
\end{figure}

An important issue is to what extent there is a trend in counterpart
luminosity with distance since at larger distances only brighter
sources would be detectable (i.e., the Malmquist bias), and source
confusion may also become problematic at larger distances.
In Figure \ref{f:Veff_dist} we
plot the the optical luminosities of the counterparts as a function of
distance, and there is in fact a strong correlation between $\log L_V$ and
distance, although only at distances $\lesssim 15$ Mpc.  We also plot the
luminosity expected for a source at a constant flux level (with arbitrary
normalization) to show the sense of the Malmquist bias.  The observed
optical luminosities increase more rapidly than this trend, although
of course in practice the observations were performed with a range of
filters and exposure times resulting in a large range in limiting
flux. Source confusion may also be contributing, however the lack of a
strong trend at larger distances suggests the neither source confusion
nor Malmquist bias is a factor at distances $\ga 20$ Mpc.  Since our
discussion is largely based on optical luminosities spanning $\sim 2$
dex, the presence of some source confusion is not going to strongly
impact our conclusions.

\subsection{Source Assessment}
Unless an ULX is in reality an interloper (i.e., a foreground star or
background AGN), the most likely physical model is that they are some
sort of X-ray binary (XRB; e.g., see Colbert et al. 2004).  In this
case, the most optimistic scenario is that a given optical counterpart
is the companion star.  However, the ULX counterpart may also
represent a cluster of stars or simply a clump of gas or stars
(particularly in an arm of a spiral galaxy).  We address these in turn
below.  

\subsubsection{Isolated X-ray Binary}
\begin{figure}
\plotone{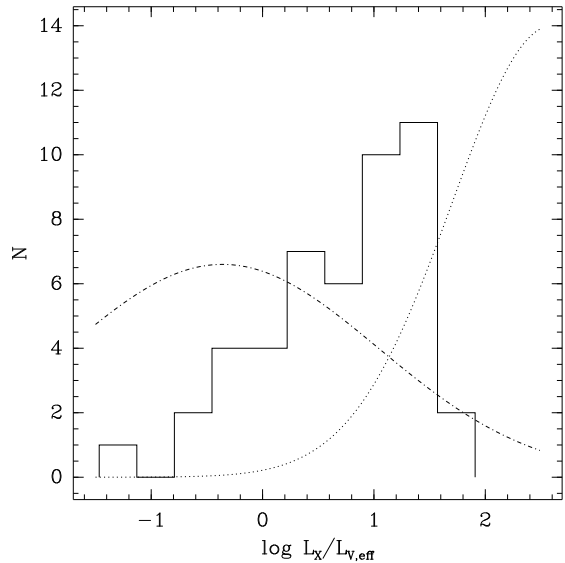}
\caption{The distribution of $\log L_X/L_{V_eff}$ from our sample.
  Also shown are the distributions in $  \log L_{2-11 \rm \
  keV}/L_{3000-7000}$ for low-mass (dotted line) and high-mass
  (dot-dashed line) X-ray binaries in Bradt \& McClintock (1983). The
  correction from $\log L_X/L_{V_eff}$ to 
  $L_{2-11 \rm \ keV}\log L_{3000-7000}$ should be small since the
  peak wavelength of the V band, $\sim 5500 \AA$, is near the midpoint
  of 3000-7000\AA. \label{f:fx_fopt}}
\end{figure}

The absolute magnitudes in Table \ref{t:lopt} are in the range -5 to
-12, corresponding to $\log L_V/L_{\odot,V} \sim 4-6$.
Blue giant stars have $\log L_V/L_{\odot,V} \sim 4-5$ and supergiants can
have $\log L_V/L_{\odot,V} \sim 5-6$.
In the supergiant case the colors
are not restrictive since blue, yellow or red supergiants could be
present.  However in all of these cases the stars would be
very young ($ \lesssim 10^7$ years; Bertelli et al. 1994) and massive
($\gtrsim 10 M_{\odot}$) and therefore the XRB would be a high-mass XRB.    
High-mass XRB typically have relatively low X-ray/optical flux ratios,
in the range of  $-2 < \log L_{2-11 \rm \ keV}/L_{3000-7000} < 1$
(Bradt \& McClintock 1983).  In Figure \ref{f:fx_fopt} we show the distribution
of $\log L_X/L_{V,eff}$ values from our sample along with the observed
values of $\log L_{2-11 \rm \ keV}/L_{3000-7000}$ from Bradt \&
McClintock (1983) for low and high-mass XRB.
In general, $\log L_X/L_{V, eff} \sim  \log L_{2-11 \rm \
keV}/L_{3000-7000}$.  By 
definition ULXs are orders of magnitude brighter than typical
high-mass XRB however the 
stellar companion would not necessarily be more luminous than those
found in Galactic high-mass XRB.  
Accordingly the $\log L_X/L_{V, eff}$ values for our counterparts
would be expected to be 1-2 dex higher than those found high-mass XRB,
and Figure \ref{f:fx_fopt} is consistent with this.
Therefore the high-mass XRB scenario is not ruled out by our data,
although of course such young stars would not be expected in
early-type galaxies and are therefore unlikely in galaxies with Hubble
types less than $\sim 0$ or for ULXs that are clearly bulge or halo
sources, such as ULX 32 in NGC 4565. Conversely a high-mass XRB
association with ULXs would be consistent with the ULX/star formation
connection that has been suggested (e.g., Swartz et al. 2004).

Another possibility is that the ULX is an X-ray binary with an
accretion disk, and that the accretion disk is contributing
significantly to the optical-UV part of the spectrum.  \citet{copper05} discuss this possibility, taking into account the impact of
X-ray irradiation of the accretion disk and companion star. They find
that the expected V-band absolute magnitudes ($M_V$) would be in the
range of -4 to -9, with the companion star often dominating over the
accretion disk at low black hole masses ($10 < M_{BH} < 100
M_{\odot}$) unless irradiation is significant.  This implies optical
counterparts with $3.5 < \log L_V < 5.5$, consistent with many of our
counterparts (see Figure \ref{f:Veffs}). 
Copperwheat et al. also generally predict blue colors for the disk +
star system.  While 
these results were calculated for the case of an isotropic X-ray
luminosity of $10^{40} \rm \ ergs\ s^{-1}$, we find similar results
for higher luminosities (and hence higher accretion rates for a given
black hole mass).   Note 
that the assumption of the X-ray luminosity being $10\%$ of the
bolometric luminosity gives 
$\log L_X/L_{V} \sim 1.0-1.6$,
which is often consistent with our observed values.  
Finally, similar results to those discussed above would be expected in
the case of Bondi accretion onto an intermediate-mass black hole in a 
molecular cloud (Krolik 2004), with the optical emission originating
several hundred gravitational radii from the black hole as in the
thin-disk case (J. Krolik, priv. comm.). 

\subsubsection{Structure in Spiral Disks}
Inspection of Figure 2 suggests that in several cases we are
observing fine structure in the spiral arm of the galaxy.  In this
case the counterparts would likely be local maxima in the surface
brightness and/or holes in the extinction.  The mean source extent of 0.2''
corresponds to a linear size of $\sim 20$ pc for
the mean distance of 20 Mpc for the galaxies in this sample (see Table
3).  In Figure \ref{f:colors} we also plot the 1-$\sigma$ range in B-V and V-I
color for spiral galaxies given in de Jong (1996).  The galaxies 
NGC 278 and NGC 891 are the only spiral galaxies (both Sb galaxies) with color
information for counterparts, and again most of the data points are due to
ULX 1 in NGC 278.  While the data are (very) limited, the colors are
bluer than expected for the integrated emission of an Sb galaxy,
possibly implying regions that are younger than Sb disks are on
average.  This also would be crudely consistent with the ULX/star
formation connection that has been suggested. 

\subsubsection{Clusters}
We next consider the case of the counterparts being stellar clusters.
In the case of early-type galaxies this would 
of course correspond to globular clusters (GCs).  Although the numbers are 
limited, for earlier type galaxies, B-V and V-I tend
to be red ($>$ 0.5), possibly indicating an older stellar population
(e.g., age $\gtrsim 10^{8-9}$ yrs; see Leitherer et al. 1999; Bruzual
\& Charlot 2003).  For comparison, the mean B-V and V-I colors for
GCs in M31
(the Milky Way) are 0.72 (0.71) and 0.96 (0.94) (Barmby et al. 2000), and
the globular clusters in NGC 1399 associated with ULXs 8 and 9 (see
Angelini et al. 2000) have B-V = 0.5, 0.9 and V-I=1.1,
1.0. Chandar, Whitmore, \& Lee (2004) present an HST survey of GCs in
nearby spiral 
galaxies and find that the GC colors similarly lie in the range 0.5
$<$ B-V $<$ 1.2 and 
0.7 $<$ V-I $<$ 1.5, and that the GC luminosities peak at $M_V \sim -6$
to $-8$, consistent with most 
counterparts in our sample.  
The ULX counterparts in NGC 3311 (a lenticular) and ULX 11 in NGC
1700 (an elliptical) have red colors (V-I $>$ 1-2), and ULX 40 in
Cen-A has a counterpart with V-I $\sim$ 0.9.  These are likely to be GCs
also.   The ULX counterpart in NGC 4565 has also been identified as a
globular cluster (albeit with a blue color) by Wu et al. (2002).

Conversely, V-I and B-V $<$ 0.5 would be indicative of a stellar population
that is very young ($< 10^7$ yrs) and/or has low 
metallicity (Leitherer et al. 1999; see also the discussion in Soria
et al. concerning the application of stellar evolutionary models to
counterparts in the vicinity of a ULX in NGC 4559). V-I and/or B-I
colors in the range 0.0-0.5 
are observed for several counterparts in late-type galaxies (ULX 2 in 
NGC 891 and several of the counterparts for ULX 1 in NGC 278). Only
U-V is available (i.e., with our CADC Association data selection criterion) for
the ULXs in the Antennae.  These counterparts are all blue, with
U-V $\lesssim$ 0.5, and the counterparts in the Antennae that
appear more red are probably in regions of higher extinction.  Also note
that U-F336W 
varies from 0 to 0.8 as U-B varies from 0 to -1.5 (Holtzman et
al. 1995), and therefore the very blue U-V colors in Table
\ref{t:colors} are over-estimated (but the true U-V are nevertheless
blue).  Interestingly there are also some counterparts in elliptical
galaxies with blue colors (e.g., ULX 37 in NGC 5018 with U-V =
-1.5). If associated with the galaxies these 
are likely to be young stellar clusters captured or formed after a recent
encounter or merger. 
Finally, there are counterparts in
late-type galaxies with red colors (specifically several of 
the counterparts to ULX 1 in NGC 278 and ULXs 15-16 in NGC 3256)
however this may be due to  extinction in these galaxies.
Alternatively the red 
colors may be due in part to the onset of a red supergiant phase,
although this would require the rather specific circumstances of a
$\sim 10^{7}$ year old cluster resulting from an instantaneous star
formation burst (Leitherer et al. 1999).

Even if the optical counterpart is a cluster, the X-ray emission is
still likely to be due to an accreting X-ray binary. In this case the
X-ray/optical flux ratio would be ``diluted'' by the optical emission
of the stars from the cluster, i.e., the observed X-ray/optical flux
ratio would be lower than that of an isolated X-ray binary, by a
factor of $L_{V,c}/L_{V,b} - 1$, where $L_{V,c}$ is the cluster
V-band luminosity and $L_{V,b}$ is the X-ray binary V-band
luminosity. As discussed above, some of the counterparts in our sample
have $\log L_V > 5.5$, in excess of what might be expected from an
isolated ULX (Copperwheat et al. 2005) by at least 1-2 dex. 
In the case of high-mass X-ray binaries (HMXB), the distribution of the
expected $L_X/L_V$ ratio largely overlaps the observed distribution of
our counterparts, although the peak of the HMXB $L_X/L_V$ distribution is $\sim
2$ dex lower than the peak of the counterpart $L_X/L_V$.  Therefore
the counterpart $L_X/L_V$ distribution is consistent with HMXB
distribution assuming on the order of 1\% of the cluster optical
emission is due to the HMXB.  Conversely, in old clusters ($> 10^{9}$
years), 
we would expect any X-ray binary to be a low-mass X-ray binary (LMXB).  Here
larger amounts of dilution would be consistent with the observed
$L_X/L_{V,eff}$ values since the typical LMXB $L_X/L_{V,eff}$ values
are 2 dex or more larger than observed in our counterparts.  In
summary, our observed $L_V$ and X-ray/optical flux ratios are
generally consistent with a LMXB in old clusters, but probably only
consistent with a HMXB in young clusters if the
optical light  of the X-ray binary is a larger fraction of the optical
light of the entire cluster than is the case with a LMXB in an old cluster.

\begin{figure}
\plotone{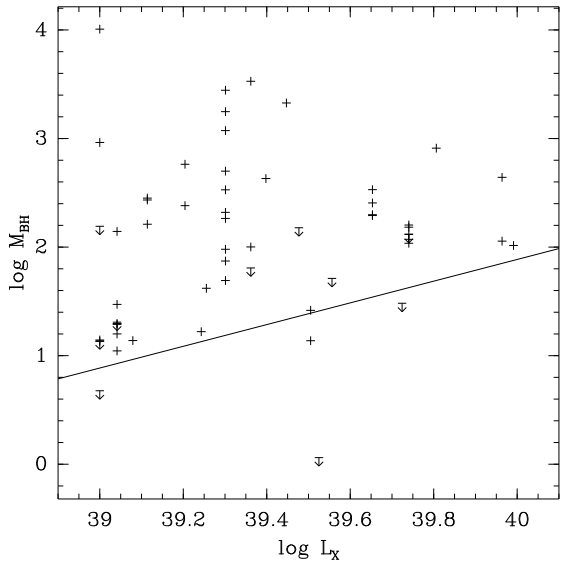}
\caption{Estimated black hole mass assuming the optical counterparts
  are clusters with 0.1\% of their cluster mass in a black hole (and a
  mass-to-light ratio of 1) plotted as a function of the ULX X-ray
  luminosity.  More generally the M/L ratio of a cluster ranges from
  $\sim 0.05$ to $\sim 5$ for ages from $10^7$ to $10^{10}$ years
  \citep{mclaughlin05}, and
  the black hole mass estimate plotted above would scale accordingly.
  The line shows the black hole mass expected if the ULX were
  emitting at the Eddington luminosity in X-rays. Note that in many
  cases there are multiple counterparts for a ULX. \label{f:bhmass}}
\end{figure}

 
Finally, we estimate the black hole mass that would be expected if
a black hole were present with $\sim 0.1\%$ of the cluster mass.  This
is roughly consistent with the well-known bulge mass / black-hole mass
trend, although the slope is highly uncertain at the low mass
end. Laor (2001) finds that $M_{BH}/M_{bulge}$ tends to be lower for
lower-mass systems (i.e., $M_{BH}/M_{bulge} \sim 0.5\%$ in ellipticals
versus $M_{BH}/M_{bulge} \sim 0.05\% $ in late-type spirals). However
recent modeling has shown that clusters may form a black hole with mass
$\sim 0.1\%$ of the cluster mass in a relatively short time ($\lesssim
10$ Myr; Portegies Zwart \& McMillan 2002; Geurkan, Freitag, \& Rasio 2004).  
The results are shown in Figure \ref{f:bhmass} and there is no obvious
trend between $L_V$ (and hence estimated $M_{BH}$) 
and $L_X$. Note that we plotted black hole mass estimates based on a
mass-to-light (M/L) ratio of 1, i.e., $M_{BH} = \log L_V - 3.$, which
is only appropriate for a 
cluster of age several billion years.   From population synthesis
modeling the M/L ratio of a cluster is expected to range from $\sim
0.05$ to $\sim 5$ for clusters of age $10^7$ to $10^{10}$ years
\citep{mclaughlin05}.
Therefore for young clusters the black hole masses would be $\sim 1$
dex lower than plotted here.
We
also plot the Eddington luminosity for a given black hole mass,
$L_{Edd} = 1.3 \times 10^{38} \ 
\frac{M}{M_{\odot}} \ \rm ergs \ s^{-1}$, which is the maximum
expected X-ray luminosity for 
isotropic emission. Most of the X-ray luminosities are well
below the predicted Eddington luminosity given the estimated $L_V$,
although there are some cases where the X-ray luminosity exceeds this
value, suggesting either anisotropic emission or a larger black hole
mass than $\sim 0.1\%$ of the cluster mass (again the $M/L_V$ ratio
may differ significantly from 1 as assumed here).  Also note that,
particularly in the case of spiral galaxies, multiple counterparts are
found for some ULXs, and the corresponding BH masses for each
counterpart are plotted in
Figure \ref{f:bhmass}.  Of course it is unlikely that all counterparts
in those cases are bound systems with an intermediate-mass black hole.

\subsubsection{Interlopers}
We turn now to the case of either a foreground star or a background
AGN.  As stated above, we calculate that $\sim 9.3$ of the ULXs are
likely to be background or foreground X-ray sources ($\sim 21\%$). 
The typical values of X-ray to optical flux ratio expected in
the case of an AGN would be $-1 \lesssim \log L_X/L_{opt} \lesssim 1$
(see, e.g., Norman et al. 2004; Bauer et al. 2004) while in the case
of foreground stars it would be $< -1$ (c.f., Georgakakis et
al. 2004).  From 
Table \ref{t:lopt}, most of the counterparts are consistent with the
background AGN possibility, particularly in the early-type galaxies,
although a few have very low or very high X-ray to optical ratios,
making a foreground star or an X-ray binary in one of the scenarios
above more likely, respectively.

\section{Conclusions}
\label{s:conc}
  We have presented a comprehensive analysis of the available WFPC2
  Association data with coverage of a sample of ULXs.  The relative
  astrometry between the Chandra and HST data was corrected whenever
  possible, in some cases requiring shifts of several arcseconds.  The
  final error circles were 0.3-1.7'' depending on the availability of
  sources useful for the registration.  With the error circles there
  were varyingly no, one or many potential counterparts.  The limiting
  luminosities are often above the faintest observed luminosities
  suggesting that the lack of multiple counterparts in some cases is
  due to insufficient depth. 
  We focused 
  on the ``wide'' filter observations corresponding to UBVI filters
  since the colors could be compared to known source properties.  This
  analysis showed that there was a slight tendency for early-type
  galaxies to have counterparts that were red (i.e., B-V $>$ 0.5, V-I $>$
  0.5) and conversely late-type galaxies had blue counterparts.  In
  general the source properties did not differ significantly from the
  properties of those expected from either structure in spiral disks
  (although in a few cases the colors were blue suggesting a young
  region) or clusters.  The counterparts found in early-type galaxies
  are generally consistent with globular clusters in both their colors
  and luminosities.  The X-ray/optical flux ratios were generally
  consistent with high-mass X-ray binaries, background AGN or
  low-mass X-ray binaries in clusters (i.e., with
  artificially lowered flux ratios).  Therefore for most of the
  sources, any one scenario cannot be ruled out although there are
  several cases (i.e., with very low optical luminosities and/or
  unusual X-ray/optical flux ratios) that may be isolated X-ray
  binaries.  If the counterparts are in fact (bound) clusters, then
  the observed X-ray luminosities are well within the Eddington
  luminosities implied by the predicted central black hole masses
  (i.e., $\sim 0.1\%$ of the 
  cluster mass estimated from the optical luminosity).  We also find
  that the scenario in which the optical flux is dominated by the
  emission of an accretion disk around a black hole results in
  optical and X-ray fluxes consistent with those observed, although
  the black hole mass would have to be $\gtrsim 10^{2} M_{\odot}$
  unless the disk is irradiated. 


\acknowledgments
We thank the anonymous referee for very useful comments that improved
the paper.  We thank Julian Krolik and Richard Mushotzky for useful
discussions concerning this work. 
Support for HST Archival research proposal \#9545 was provided by NASA
through a grant from the Space Telescope Science Institute, which is
operated by the Association of Universities for Research in Astronomy,
Inc., under NASA contract NAS 5-26555.  We also acknowledge support
from CXC grant AR5007X.  A.P. acknowledges support from NASA LTSA
grant NNG04GE13G.



{\bf Appendix}

\section{Notes on Individual ULXs}
\subsection{ULX 1}
ULX 1 is found in disk of NGC 278 , which is a face-on spiral (T=3). 6
potential counterparts are found, mostly in the 1-2$\sigma$ error
annulus although the brightest counterpart is within the 1$\sigma$
error circle.  The B-V and V-I colors are generally blue although some
counterparts have B-V $>$ 0.5.

\subsection{ULX 2}
ULX 2 is found in NGC 891, an edge-on spiral galaxy (T=3), located just
outside of a prominent dust lane.  B-V = -0.2 and V-I $<$ 0.1,
suggesting a very blue counterpart or else the optical source varied
between the different filter observations.

\subsection{ULXs 3-4}
These ULXs are found in NGC 1068, a well-known face-on spiral
galaxy (T=3) with strong starburst and Seyfert activity.  In both
cases the source density is high and in most cases different
counterparts are found in different filters.  However in this case the
astrometry could not be corrected, due in part to the complexity of
the nuclear region in both the X-ray and optical bands and the small
FOV of the WFPC2 images.  Future work may improve the registration by
leveraging ground-based optical images.

\subsection{ULX 5}
ULX is found in the disk of NGC 1097, a face-on spiral galaxy (T=3).
The only Association image was in the UV and no counterparts were
found.

\subsection{ULXs 6-8}
These ULXs are found in NGC 1399, an elliptical galaxy (T=-5).  ULXs 7
and 8 have single counterparts that are consistent with globular clusters
(see also Angelini \etal 2001).  ULX 6 is possibly due to a background
AGN, although its $\log L_X/L_{V,eff}$ value of $>$1.4 is somewhat
large relative to typical Chandra Deep Field sources (see Bauer et
al. 2004).

\subsection{ULXs 9-11}
These ULXs are found in NGC 1700, an elliptical galaxy (T=-5). NGC
1700 is one of the most distant galaxies in our sample at 52 Mpc.  No
counterparts are found for ULX 10. The counterparts for ULX 9 and 11 have
red colors, consistent with a globular cluster interpretation.  

\subsection{ULXs 12-14}
These ULXs are found in M82 (T=90), which is the archetypal, and one of the
nearest, starburst galaxies.  In all cases the optical emission is too
complex for individual sources to be identified, with the source
regions dominated by diffuse emission with non-uniform extinction.

\subsection{ULXs 15-16}
ULXs 15 and 16 are found in NGC 3256, a merger (T=99).  Two 
counterparts are found for each ULX. One of the counterparts for ULX
16 has a red color limit (B-I $>$ 1.0).  This may be due to extinction
however note that this counterpart is only marginally within the
2$\sigma$ error circle and may be unrelated to the ULX.

\subsection{ULXs 17-18}
These ULXs are in NGC 3311, the central galaxy of the Hydra cluster
(T=-4).  The ULXs are found near the nucleus which is dusty.  Multiple
counterparts are found for both, with one being very red, most
likely due to dust extinction.

\subsection{ULX 19}
ULX 19 is found in the edge-on spiral galaxy NGC 3628 (T=3).  The ULX
is in the center of a prominent dust lane, and a single counterpart is
detected with $\log L_X/L_{V_eff}$ = 1.6.  This is consistent with an
isolated X-ray binary scenario however the V-band flux is probably
underestimated due to extinction.

\subsection{ULX 20}
ULX 20 is found in the spiral galaxy (T=3) NGC 3627, located on the
edge of the disk.   Here it should be noted that we were not able to
correct the HST astrometry and the position of the ULX (relative to
nearby stars) differs by $\sim 6$'' between the two V band images.
Therefore the two 3.3'' (2$\sigma$) error circles do not overlap and the
detection of a counterpart (beyond the 1$\sigma$ error radius) in one
image but not the other does not imply variability.

\subsection{ULX 21-28}
These ULXs are found in the Antennae galaxy merger.  As discussed
in Section~\ref{s:intro}, the ULX/HST correlation in the Antennae
has been assessed in Zezas et al. (2002) and Zezas \& Fabbiano
(2002).  They did not correct the HST astrometry and assumed a 2''
error circle, and often found multiple potential counterparts,
typically with blue colors suggesting a young population.  Our results
are consistent, with our (Method 2) 2$\sigma$ error circle being
1.6'' in radius.  ULX 25 was identified as a background AGN by
\citet{clark05}.

\subsection{ULX 29}
ULX 29 is found in NGC 4150, a lenticular galaxy (T=-2).  The only
Association image available was an I band exposure in which no
counterparts were found.

\subsection{ULX 30}
ULX 30 is found in NGC 4490 which is a late-type spiral galaxy (T=7)
with the ULX being located along the disk.  Not surprisingly, multiple
HST sources are found within the 2$\sigma$ error circle.

\subsection{ULX 31}
ULX 31 is found in NGC 4559, also a late-type spiral (T=6). The ULX is
located in the disk near the nucleus.  Two faint point-like
counterparts are found although the region is somewhat complex.
Cropper et al. (2004) found no counterparts in the vicinity of this
ULX (their X10), using Association data.  Cropper et al. (2004) also
use B and I band images which were not 
available at the start of this study.   Similarly the HST data
discussed in Cropper et al. (2004) and Soria et al. (2005) overlapping
another ULX in NGC 4559 (their X7) was not available at the start of
this study.

\subsection{ULX 32}
ULX 32 is found in NGC 4565, an edge-on spiral (T=3).  The ULX is on
the edge of the bulge and has been identified with a globular cluster
in Wu et al. (2002).  

\subsection{ULX 33}
ULX 33 is found in NGC 4594 (the Sombrero galaxy; T=1).  The ULX is
located in the disk near the nucleus (outside the dust lane).  An H$\alpha$ 
counterpart
is found marginally within the 2$\sigma$ error circle (a cosmic ray
not completely removed is noticeable in the 1$\sigma$ error circle).

\subsection{ULX 34}
ULX 34 is found in NGC 4725, an early-type spiral (T=2).  The ULX is
located inside an outer spiral arm.  Diffuse flux is evident at the
ULX position and a (point-like) V-band counterpart is found at the 2$\sigma$
radius.  

\subsection{ULX 35}
ULX 35 is found in NGC 4945, a late-type spiral (T=6).  No counterpart
is found, however the ULX region is located near the nucleus within the
disk and the lack of a counterpart may be due to extinction.

\subsection{ULXs 36-37}
These ULXs are found in the elliptical galaxy NGC 5018 (T=-5), with
ULX 37 being located near the nucleus. The counterpart for ULX 37 has a
very blue color (U-V=-1.6).  This counterpart is therefore likely to
be a background source, with an X-ray/optical flux ratio consistent
with AGN.

\subsection{ULX 38-39}
These ULXs are found in the face-on, late-type spiral galaxy NGC 5033
(T=5).  The ULXs are located close to each other, within the disk.
Only UV Association images were available and several potential
counterparts were detected.

\subsection{ULX 40}
ULX 40 is found in the elliptical galaxy NGC 5128 (Centaurus A;
T=-2).  The ULX is located outside of the dusty central region.  
Two counterparts are found.  Neither of our counterparts corresponds
to the counterparts discussed in \citet{Ghosh06}, based on the same
WFPC2 data.  This is because Ghosh et al. applied astrometric
corrections to the HST data based on a cataloged globular cluster (an
approach not attempted here) and a nearby transient source that was not
detected in the ACIS image used here (but was detected in a subsequent
ACIS observation not public at the start of our study).  The brighter
of the two counterparts listed here is $\sim 1$ magnitude brighter than
the counterpart discussed in Ghosh et al., however we derive a similar
color (V-I $\sim 1$ compared with V-I $\sim 0.7$ in Ghosh et al.)
implying this counterpart is similar in nature to the counterpart
discussed in Ghosh et al.  Since our discussion is based on the
properties of counterparts on average (and often the counterparts to a
given ULX have similar colors), our conclusions are not affected our
(potentially incorrect) counterpart selection here.

\subsection{ULXs 41-42}
These ULXs are found in the Circinus galaxy, a face-on spiral (T=3).
The ULXs are located in the disk of the galaxy.  Only narrow-band
Association images were available that overlapped ULX 41. ULX 41 is CG
X-2 discussed in Bauer et al. (2001) where they identify it as a
supernova based on the detection of an H$\alpha$ counterpart (as found
here) and a radio point source.  ULX 42 is CG X-1 in Bauer et
al. (2001), and they find no counterpart in their analysis of the
archival F606W image.  Their limiting magnitude is 25.3, which is
consistent with our value.  Weisskopf et al. (2004) claim to have
detected a faint, point-like source in the WFPC2 F606W image within
0.25'' of the Chandra position after registering the frames using the
AGN in Circinus (as we have done here).  However they do not give the
significance of the source and consider the optical magnitude, 24.3,
to be an upper-limit.  Also note that in this case the galaxy is
located low in the plane of the Milky Way with an estimated extinction
of $A_V \sim 4.8$ (NED). 

\subsection{ULX 43}
ULX 43 is found in the early-type spiral galaxy NGC 6500 (T=1.7).  The
only (usable) Association image was in the UV where no
counterpart was detected.  

\subsection{ULX 44}
ULX 44 is found in the early-type spiral galaxy NGC 7174 (T=1.8).  NGC
7174 is part of the group HCG 90, and has a highly disturbed
morphology due to interaction with  NGC 7176.  As with ULX 43 the only
Association image was in the UV/U (F300W) and a single counterpart was
detected.






\clearpage
\LongTables

\setcounter{table}{1}
\begin{deluxetable*}{lccccccccccc}
\tablecolumns{9}
\tablewidth{0pc}
\tabletypesize{\scriptsize}
\tablecaption{HST/WFPC2 Datasets}
\tablehead{
\colhead{Dataset} & \colhead{Galaxy} & \colhead{Filter} & \colhead{Num. of} & \colhead{Exp.} & \colhead{Relative} & \colhead{Shift Method} & \colhead{ULX} & \colhead{X} & \colhead{Y} & \colhead{Chip} \\
\colhead{ } & \colhead{ } & \colhead{ } & \colhead{Exposures} & \colhead{Time} & \colhead{Exp. Times} & \colhead{ } & \colhead{ID} & \colhead{ } & \colhead{ } & \colhead{ } & \colhead{ }\\ 
\colhead{(1)} & \colhead{(2)} & \colhead{(3)} & \colhead{(4)} & \colhead{(5)} & \colhead{(6)} & \colhead{(7)} & \colhead{(8)} & \colhead{(9)} & \colhead{(10)} & \colhead{(11)} \\ }
\startdata
u6eae601b & NGC 0278 & F450W & 2 & 320 & 1.00,1.00 & 2(7) & 1 & 121.2 & 197.1 & 2\\
u6eae603b & NGC 0278 & F606W & 2 & 320 & 1.00,1.00 & 2(8) & 1 & 121.5 & 197.2 & 2\\
u6eae605b & NGC 0278 & F814W & 2 & 320 & 1.00,1.00 & 2(8) & 1 & 121.6 & 197.1 & 2\\
u6ea1201b & NGC 0891 & F450W & 2 & 460 & 1.00,1.00 & 2(5) & 2 & 199.9 & 350.3 & 4\\
u29r0701b & NGC 0891 & F606W & 2 & 160 & 1.00,1.00 & 2(6) & 2 & 319.3 & 159.4 & 2\\
u6ea1203b & NGC 0891 & F814W & 2 & 460 & 1.00,1.00 & 2(5) & 2 & 201.5 & 350.4 & 4\\
u3030101b & NGC 1068 & F160BW & 2 & 2400 & 1.00,1.00 & 3 & 3 & 141.9 & 192.0 & 4\\
&  &  &  &  &  &  & 4 & 30.2 & 109.8 & 4\\
u2m3010eb & NGC 1068 & F218W & 2 & 2400 & 0.67,1.33 & 3 & 3 & 97.3 & 50.8 & 4\\
&  &  &  &  &  &  & 4 & 137.2 & 32.4 & 2\\
u2m3010cb & NGC 1068 & F336W & 2 & 900 & 0.67,1.33 & 3 & 3 & 97.3 & 50.8 & 4\\
&  &  &  &  &  &  & 4 & 137.2 & 32.4 & 2\\
u2m3010ab & NGC 1068 & F343N & 2 & 1800 & 0.67,1.33 & 3 & 3 & 97.3 & 50.8 & 4\\
&  &  &  &  &  &  & 4 & 137.2 & 32.4 & 2\\
u2m30108b & NGC 1068 & F375N & 2 & 1800 & 0.67,1.33 & 3 & 3 & 97.3 & 50.8 & 4\\
&  &  &  &  &  &  & 4 & 137.2 & 32.4 & 2\\
u3030103b & NGC 1068 & F487N & 2 & 3400 & 0.65,1.35 & 3 & 3 & 141.9 & 192.0 & 4\\
&  &  &  &  &  &  & 4 & 30.2 & 109.8 & 4\\
u2m30104b & NGC 1068 & F502N & 2 & 900 & 0.67,1.33 & 3 & 3 & 97.3 & 50.8 & 4\\
&  &  &  &  &  &  & 4 & 137.2 & 32.4 & 2\\
u2m30106b & NGC 1068 & F658N & 2 & 900 & 0.67,1.33 & 3 & 3 & 97.3 & 50.8 & 4\\
&  &  &  &  &  &  & 4 & 137.2 & 32.4 & 2\\
u3030105b & NGC 1068 & F673N & 2 & 1800 & 0.67,1.33 & 3 & 3 & 141.9 & 192.0 & 4\\
&  &  &  &  &  &  & 4 & 30.2 & 109.8 & 4\\
u2m3010gb & NGC 1068 & F791W & 2 & 440 & 0.64,1.36 & 3 & 3 & 97.3 & 50.8 & 4\\
&  &  &  &  &  &  & 4 & 137.2 & 32.4 & 2\\
u33z0101b & NGC 1097 & F218W & 2 & 1200 & 1.00,1.00 & 1 & 5 & 689.3 & 507.3 & 3\\
u34m0204b & NGC 1399 & F450W & 4 & 5200 & 1.00,1.00 & 1 & 6 & 316.5 & 32.3 & 2\\
&  &  &  &  &  &  & 7 & 188.1 & 444.2 & 3\\
&  &  &  &  &  &  & 8 & 240.0 & 479.3 & 4\\
u5cv0101b & NGC 1399 & F606W & 8 & 4000 & 1.00,1.00 & 1 & 6 & 262.4 & 50.8 & 2\\
&  &  &  &  &  &  & 7 & 511.5 & 63.1 & 4\\
&  &  &  &  &  &  & 8 & 152.7 & 572.4 & 4\\
u34m0201b & NGC 1399 & F814W & 3 & 1800 & 1.00,1.00 & 1 & 6 & 311.6 & 37.2 & 2\\
&  &  &  &  &  &  & 7 & 193.0 & 449.1 & 3\\
&  &  &  &  &  &  & 8 & 244.9 & 474.3 & 4\\
u2bm0401b & NGC 1700 & F555W(1) & 2 & 1000 & 1.00,1.00 & 1 & 9 & 119.5 & 504.6 & 2\\
&  &  &  &  &  &  & 10 & 298.6 & 436.6 & 1\\
u2bm0403b & NGC 1700 & F814W(1) & 2 & 460 & 1.00,1.00 & 1 & 9 & 119.5 & 504.6 & 2\\
&  &  &  &  &  &  & 10 & 298.7 & 436.7 & 1\\
u3m73201b & NGC 1700 & F555W(2) & 4 & 1600 & 1.00,1.00 & 1 & 10 & 497.6 & 592.1 & 1\\
&  &  &  &  &  &  & 11 & 592.1 & 401.3 & 3\\
u3m73206b & NGC 1700 & F814W(2) & 3 & 1800 & 0.33,1.33 & 1 & 10 & 497.6 & 592.5 & 1\\
&  &  &  &  &  &  & 11 & 592.1 & 401.1 & 3\\
u6fiub02b & NGC 1700 & F606W(1) & 2 & 2500 & 0.80,1.20 & 1 & 11 & 262.9 & 787.1 & 4\\
u6fiud02b & NGC 1700 & F606W(2) & 2 & 2500 & 0.80,1.20 & 1 & 11 & 279.8 & 696.7 & 4\\
u45t0103b & NGC 3034 & F555W(1) & 4 & 3100 & 0.45,2.06 & 2(2) & 12 & 314.8 & 650.0 & 4\\
&  &  &  &  &  &  & 13 & 200.7 & 492.3 & 4\\
&  &  &  &  &  &  & 14 & 179.3 & 449.9 & 4\\
u3jv0101b & NGC 3034 & F656N & 2 & 1000 & 1.00,1.00 & 2[u45t0103b] & 12 & 331.5 & 69.1 & 2\\
&  &  &  &  &  &  & 13 & 299.4 & 413.9 & 1\\
&  &  &  &  &  &  & 14 & 391.8 & 366.5 & 1\\
u3jv0103b & NGC 3034 & F658N & 2 & 1200 & 1.00,1.00 & 2[u45t0103b] & 12 & 331.3 & 69.8 & 2\\
&  &  &  &  &  &  & 13 & 298.6 & 413.2 & 1\\
&  &  &  &  &  &  & 14 & 391.0 & 365.8 & 1\\
u2s04201b & NGC 3034 & F656N(2) & 2 & 600 & 1.00,1.00 & 2[u45t0103b] & 13 & 527.3 & 326.5 & 1\\
&  &  &  &  &  &  & 14 & 436.5 & 376.8 & 1\\
u3jv0201b & NGC 3034 & F656N(3) & 2 & 1000 & 1.00,1.00 & 2[u45t0103b] & 13 & 559.3 & 436.5 & 1\\
&  &  &  &  &  &  & 14 & 482.4 & 506.3 & 1\\
u3jv0203b & NGC 3034 & F658N(4) & 2 & 1200 & 1.00,1.00 & 2[u45t0103b] & 13 & 559.1 & 435.9 & 1\\
&  &  &  &  &  &  & 14 & 482.2 & 505.6 & 1\\
u3jv0206b & NGC 3034 & F502N(5) & 4 & 3600 & 0.67,1.44 & 2[u45t0103b] & 13 & 559.1 & 435.9 & 1\\
&  &  &  &  &  &  & 14 & 482.2 & 505.6 & 1\\
u3jv020bb & NGC 3034 & F631N(6) & 2 & 1200 & 1.00,1.00 & 2[u45t0103b] & 13 & 559.0 & 435.5 & 1\\
&  &  &  &  &  &  & 14 & 482.1 & 505.2 & 1\\
u45t010fb & NGC 3034 & F555W(7) & 4 & 2500 & 0.32,1.60 & 2[u45t0103b] & 13 & 331.2 & 750.7 & 4\\
&  &  &  &  &  &  & 14 & 309.8 & 708.4 & 4\\
u2bl0303b & NGC 3256 & F450W(2) & 2 & 1800 & 1.00,1.00 & 2[u67q4203b] & 15 & 100.9 & 683.4 & 4\\
u67q4101b & NGC 3256 & F555W(1) & 2 & 320 & 1.00,1.00 & 2[u67q4203b] & 15 & 102.8 & 376.1 & 4\\
u67q4201b & NGC 3256 & F555W(2) & 2 & 320 & 1.00,1.00 & 2[u67q4203b] & 15 & 113.5 & 437.6 & 4\\
u67q4103b & NGC 3256 & F814W(2) & 2 & 320 & 1.00,1.00 & 2[u67q4203b] & 15 & 102.8 & 376.1 & 4\\
u67q4203b & NGC 3256 & F814W(3) & 2 & 320 & 1.00,1.00 & 2(21) & 15 & 113.5 & 437.7 & 4\\
u2bl0301b & NGC 3256 & F814W(4) & 2 & 1600 & 1.00,1.00 & 2[u67q4203b] & 15 & 99.6 & 684.0 & 4\\
u6dw0801b & NGC 3256 & F300W & 2 & 600 & 1.00,1.00 & 2[u67q4203b] & 15 & 160.1 & 468.1 & 3\\
&  &  &  &  &  &  & 16 & 61.3 & 412.4 & 4\\
u2bl0203b & NGC 3256 & F450W(1) & 2 & 1800 & 1.00,1.00 & 2[u67q4203b] & 15 & 131.7 & 112.5 & 4\\
&  &  &  &  &  &  & 16 & 47.7 & 745.4 & 4\\
u2bl0201b & NGC 3256 & F814W(1) & 2 & 1600 & 1.00,1.00 & 2[u67q4203b] & 15 & 131.8 & 112.8 & 4\\
&  &  &  &  &  &  & 16 & 47.8 & 745.6 & 4\\
u3vz0304b & NGC 3311 & F555W & 4 & 3700 & 0.97,1.08 & 2(5) & 17 & 520.7 & 439.0 & 1\\
&  &  &  &  &  &  & 18 & 516.4 & 374.2 & 1\\
u3vz0306b & NGC 3311 & F814W & 4 & 3800 & 0.95,1.05 & 2(6) & 17 & 521.9 & 440.3 & 1\\
&  &  &  &  &  &  & 18 & 517.6 & 375.6 & 1\\
u29r1l01b & NGC 3628 & F606W & 2 & 160 & 1.00,1.00 & 2(1) & 19 & 65.2 & 437.1 & 4\\
u29r1k01b & NGC 3627 & F606W(1) & 2 & 160 & 1.00,1.00 & 3 & 20 & 654.2 & 623.8 & 3\\
u67n3402b & NGC 3627 & F606W(2) & 2 & 560 & 0.57,1.43 & 3 & 20 & 538.2 & 523.4 & 3\\
u6a02901b & NGC 3627 & F814W & 2 & 700 & 1.00,1.00 & 3 & 20 & 579.1 & 589.0 & 3\\
u304010lb & NGC 4038 & F336W & 4 & 4500 & 0.89,1.07 & 2[u3040205b] & 21 & 236.4 & 153.6 & 1\\
&  &  &  &  &  &  & 22 & 489.9 & 298.1 & 2\\
&  &  &  &  &  &  & 23 & 85.8 & 420.7 & 2\\
&  &  &  &  &  &  & 24 & 394.5 & 88.9 & 3\\
&  &  &  &  &  &  & 25 & 140.6 & 695.0 & 3\\
&  &  &  &  &  &  & 26 & 529.7 & 224.6 & 3\\
&  &  &  &  &  &  & 27 & 713.0 & 132.1 & 3\\
u3040205b & NGC 4038 & F555W & 2 & 60 & 1.00,1.00 & 2(2) & 21 & 286.4 & 511.8 & 4\\
&  &  &  &  &  &  & 22 & 284.0 & 394.6 & 1\\
&  &  &  &  &  &  & 23 & 48.9 & 293.7 & 3\\
&  &  &  &  &  &  & 24 & 370.3 & 62.2 & 4\\
&  &  &  &  &  &  & 26 & 151.1 & 507.9 & 3\\
&  &  &  &  &  &  & 27 & 334.5 & 415.4 & 3\\
&  &  &  &  &  &  & 28 & 520.7 & 602.1 & 3\\
u3040202b & NGC 4038 & F658N & 4 & 3800 & 0.84,1.16 & 2[u3040205b] & 21 & 268.6 & 162.3 & 1\\
&  &  &  &  &  &  & 22 & 493.9 & 283.4 & 2\\
&  &  &  &  &  &  & 23 & 89.8 & 406.1 & 2\\
&  &  &  &  &  &  & 24 & 379.8 & 84.9 & 3\\
&  &  &  &  &  &  & 25 & 125.9 & 691.0 & 3\\
&  &  &  &  &  &  & 26 & 515.0 & 220.6 & 3\\
&  &  &  &  &  &  & 27 & 698.4 & 128.1 & 3\\
u2tv1501b & NGC 4150 & F814W & 2 & 320 & 1.00,1.00 & 2(5) & 29 & 462.3 & 228.8 & 1\\
u29r2501b & NGC 4490 & F606W & 2 & 160 & 1.00,1.00 & 2(2) & 30 & 68.0 & 638.5 & 2\\
u29r2b01b & NGC 4559 & F606W & 2 & 160 & 1.00,1.00 & 2(4) & 31 & 247.8 & 41.9 & 2\\
u31s010cb & NGC 4565 & F450W(1) & 3 & 630 & 0.95,1.10 & 1 & 32 & 425.1 & 121.1 & 4\\
u3ji030bb & NGC 4565 & F450W(2) & 2 & 460 & 1.00,1.00 & 1 & 32 & 283.4 & 304.3 & 3\\
u3ji0307b & NGC 4565 & F555W & 2 & 320 & 1.00,1.00 & 1 & 32 & 282.7 & 304.4 & 3\\
u31s0107b & NGC 4565 & F814W(1) & 3 & 480 & 1.00,1.00 & 1 & 32 & 420.2 & 125.6 & 4\\
u3ji0309b & NGC 4565 & F814W(2) & 2 & 320 & 1.00,1.00 & 1 & 32 & 282.4 & 304.4 & 3\\
u2uh0601b & NGC 4594 & F658N & 2 & 1600 & 1.00,1.00 & 1 & 33 & 577.4 & 664.1 & 1\\
u67n4602b & NGC 4725 & F606W & 2 & 560 & 0.57,1.43 & 1 & 34 & 290.3 & 711.8 & 2\\
u29r2p01b & NGC 4945 & F606W & 2 & 160 & 1.00,1.00 & 2(4) & 35 & 457.6 & 361.1 & 2\\
u2st0201b & NGC 5018 & F336W & 3 & 1800 & 1.00,1.00 & 1 & 36 & 486.3 & 397.0 & 3\\
&  &  &  &  &  &  & 37 & 724.7 & 522.5 & 1\\
u3m72505b & NGC 5018 & F555W & 6 & 1200 & 0.50,2.00 & 1 & 36 & 509.6 & 394.5 & 3\\
&  &  &  &  &  &  & 37 & 679.4 & 521.9 & 1\\
u2kt0302b & NGC 5033 & F218W & 2 & 4500 & 0.93,1.07 & 1 & 38 & 90.0 & 57.0 & 4\\
&  &  &  &  &  &  & 39 & 140.2 & 93.6 & 4\\
u2kt0303b & NGC 5033 & F300W & 2 & 2000 & 1.00,1.00 & 1 & 38 & 89.0 & 57.6 & 4\\
&  &  &  &  &  &  & 39 & 139.3 & 94.1 & 4\\
u3lba104b & NGC 5128 & F555W & 3 & 180 & 1.00,1.00 & 3 & 40 & 493.9 & 615.3 & 2\\
u3lba101b & NGC 5128 & F814W & 3 & 180 & 1.00,1.00 & 3 & 40 & 493.9 & 615.3 & 2\\
u4im0101b & Circinus & F502N & 2 & 1800 & 1.00,1.00 & 1 & 41 & 101.3 & 75.0 & 4\\
&  &  &  &  &  &  & 42 & 153.0 & 676.5 & 1\\
u4im0104b & Circinus & F656N & 2 & 1600 & 1.00,1.00 & 1 & 41 & 101.4 & 74.1 & 4\\
&  &  &  &  &  &  & 42 & 151.0 & 676.4 & 1\\
u3320801b & Circinus & F606W & 2 & 600 & 0.67,1.33 & 1 & 42 & 777.0 & 403.5 & 1\\
u2ex0i01b & NGC 6500 & F218W & 2 & 2200 & 1.00,1.00 & 1 & 43 & 208.7 & 115.9 & 2\\
u2uh1202b & NGC 6500 & F547M & 2 & 276 & 0.12,1.88 & 1 & 43 & 390.4 & 175.2 & 2\\
u67gb301b & NGC 7174 & F300W & 2 & 1000 & 1.00,1.00 & 1 & 44 & 526.2 & 172.2 & 1\\

\enddata

\tablecomments{The table lists basic properties of the HST/WFPC2 data
used for our study, selected as described in Section~\ref{ss:HSTdata}.
Column~(1) lists the HST/WFPC2 association name. Column~(2) lists the
host galaxy name. Column~(3) lists the filter with which the data were
obtained. When multiple datasets with the same filter are available
for a given galaxy, a unique identifier is added in parentheses.
Column~(4) lists the number of exposures in the WFPC2
association. Column~(5) lists the total exposure time. Column~(6) indicates
the relative exposure times of the different exposures in the association.  The first
number is the shortest exposure time divided by the average exposure time.  The second number
is the longest exosure time divided by the average.  Column~(7) 
indicates the method by which the astrometry was calibrated, as
described in Section~\ref{s:astrocor}.  Method 1 coordinates were
shifted to align the galaxy nucleus with the AGN coordinates in
Chandra data; method 2 coordinates were shifted to align observed
USNO-B1.0 stars with their catalog coordinates; and method 3
coordinates are obtained from the original HST pipeline headers.  When
the number is followed by the name of a dataset in square brackets,
that means that the image was aligned with the listed dataset for the
same galaxy (which itself was aligned using the listed method). For
method 2 the number of USNO-B1.0 stars that was identified is listed
in parentheses. The relative accuracy of the Chandra/HST alignment for
each method is discussed in Section~\ref{s:astrocor}. Column~(8) lists
the ID numbers (from Table~\ref{t:sample}) of the ULXs that fall on
the image. Columns~(9)--(11) list the (X,Y) pixel position in WFPC2
detector coordinates and WFPC2 chip number that correspond to the
position of the ULX given in Table~\ref{t:sample}.\label{t:WFPC2data}}
\end{deluxetable*}

\begin{deluxetable*}{lcccccccccccccc}
\tablecolumns{9}
\tablewidth{0pc}
\tabletypesize{\scriptsize}
\tablecaption{Point-like Optical Sources Coincident with ULXs}
\tablehead{
\colhead{ULX} & \colhead{Galaxy} & \colhead{Src.} & \colhead{Filter} &
\colhead{Phot.} & \colhead{m$_{lim}$} & \colhead{$\delta$ RA } &
\colhead{$\delta$ Dec.} & \colhead{Offset} & \colhead{App.} &
\colhead{Error} & \colhead{Abs.} & \colhead{S/N} & \colhead{Size}\\
\colhead{ID} & \colhead{ } & \colhead{No.} & \colhead{ } &\colhead{Flag} & \colhead{ } & \colhead{ (") } & \colhead{ (") } & \colhead{(")} & \colhead{Mag.} & \colhead{ } & \colhead{Mag.} & \colhead{ } & \colhead{''/pc}\\
\colhead{(1)} & \colhead{(2)} & \colhead{(3)} & \colhead{(4)} & \colhead{(5)} & \colhead{(6)} & \colhead{(7)} & \colhead{(8)} & \colhead{(9)} & \colhead{(10)} & \colhead{(11)} & \colhead{(12)} & \colhead{(13)} & \colhead{(14)}\\ }
\startdata
1 & NGC 0278 &  1 &  F450W &  M &  23.98 &  0.20 & 0.71 & 0.74 &  22.31 & 0.10 & -8.05 &  17.8 & 0.2/13.3\\
 & &  &  F606W &  M &  23.55 &  0.25 & 0.68 & 0.73 &  21.56 & 0.07 & -8.80 &  24.7 & 0.2/14.3\\
 & &  &  F814W &  M &  23.12 &  0.24 & 0.69 & 0.73 &  20.91 & 0.06 & -9.45 &  30.1 & 0.2/12.2\\
 & &  2 &  F450W &  M &  23.98 &  1.20 & 0.43 & 1.27 &  22.91 & 0.17 & -7.45 &  10.1 & 0.2/13.0\\
 & &  &  F606W &  M &  23.55 &  1.23 & 0.31 & 1.27 &  22.41 & 0.14 & -7.95 &  11.3 & 0.3/16.8\\
 & &  &  F814W &  M &  23.12 &  1.24 & 0.31 & 1.28 &  21.65 & 0.11 & -8.71 &  15.1 & 0.2/12.2\\
 & &  3 &  F450W &  M &  23.98 &  -0.60 & 1.17 & 1.32 &  21.62 & 0.06 & -8.74 &  33.9 & 0.2/13.4\\
 & &  &  F606W &  M &  23.55 &  -0.56 & 1.13 & 1.26 &  21.34 & 0.06 & -9.02 &  30.2 & 0.3/14.5\\
 & &  &  F814W &  M &  23.12 &  -0.56 & 1.18 & 1.31 &  20.96 & 0.06 & -9.40 &  28.7 & 0.2/14.3\\
 & &  4 &  F450W &  M &  23.97 &  -1.35 & -0.39 & 1.40 &  23.42 & 0.26 & -6.94 &  6.3 & 0.2/10.4\\
 & &  &  F606W &  M &  23.55 &  -1.33 & -0.43 & 1.40 &  23.43 & 0.36 & -6.93 &  4.4 & 0.2/10.3\\
 & &  &  F814W &  M &  23.12 &  -1.33 & -0.43 & 1.40 &  23.09 & 0.40 & -7.27 &  3.9 & 0.2/11.2\\
 & &  5 &  F450W &  M &  23.98 &  0.14 & 1.50 & 1.51 &  22.42 & 0.11 & -7.94 &  16.1 & 0.2/13.1\\
 & &  &  F606W &  M &  23.55 &  0.08 & 1.44 & 1.44 &  22.00 & 0.10 & -8.36 &  16.5 & 0.3/18.0\\
 & &  &  F814W &  M &  23.12 &  0.03 & 1.39 & 1.39 &  21.47 & 0.10 & -8.89 &  18.0 & 0.2/13.5\\
 & &  6 &  F450W &  N &  24.20 & \\
 & &  &  F606W &  M &  23.55 &  -1.06 & 0.96 & 1.43 &  22.68 & 0.18 & -7.68 &  8.7 & 0.2/13.7\\
 & &  &  F814W &  M &  23.12 &  -1.10 & 0.98 & 1.47 &  22.03 & 0.15 & -8.33 &  10.6 & 0.1/8.5\\
2 & NGC 0891 &  1 &  F450W &  M &  24.88 &  -1.76 & -0.47 & 1.82 &  22.99 & 0.10 & -6.92 &  19.8 & 0.2/10.1\\
 & &  &  F606W &  M &  24.26 &  -1.55 & -0.59 & 1.65 &  23.16 & 0.16 & -6.75 &  11.0 & 0.2/7.3\\
 & &  &  F814W &  N &  23.03 & \\
3 & NGC 1068 &  &  F336W &  C & \\
 & &  &  F791W &  C & \\
 & &  &  F673N &  C & \\
 & &  &  F336W &  C & \\
 & &  &  F487N &  C & \\
 & &  &  F658N &  C & \\
 & &  &  F160BW &  C & \\
 & &  &  F218W &  X & \\
 & &  &  F343N &  X & \\
 & &  &  F375N &  X & \\
 & &  &  F502N &  X & \\
4 & NGC 1068 &  &  F218W &  X & \\
 & &  &  F336W &  C & \\
 & &  &  F343N &  X & \\
 & &  &  F375N &  X & \\
 & &  &  F502N &  X & \\
5 & NGC 1097 &  &  F218W &  N &  20.67 & \\
6 & NGC 1399 &  &  F450W &  N &  25.79 & \\
 & &  &  F606W &  N &  25.20 & \\
 & &  &  F814W &  N &  24.28 & \\
7 & NGC 1399 &  1 &  F450W &  S &  26.74 &  0.16 & -0.12 & 0.20 &  22.83 & 0.03 & -8.60 &  142.5 & 0.2/15.6\\
 & &  &  F606W &  S &  27.04 &  0.16 & 0.01 & 0.16 &  22.30 & 0.02 & -9.13 &  312.5 & 0.2/17.8\\
 & &  &  F814W &  S &  25.35 &  0.16 & -0.13 & 0.21 &  21.18 & 0.02 & -10.25 &  182.1 & 0.2/15.9\\
8 & NGC 1399 &  1 &  F450W &  S &  26.67 &  0.07 & -0.14 & 0.16 &  22.43 & 0.02 & -9.00 &  194.2 & 0.2/14.7\\
 & &  &  F606W &  S &  26.61 &  0.12 & -0.12 & 0.17 &  21.52 & 0.01 & -9.91 &  424.7 & 0.2/21.9\\
 & &  &  F814W &  S &  25.07 &  0.08 & -0.15 & 0.17 &  20.56 & 0.02 & -10.87 &  252.5 & 0.2/17.2\\
9 & NGC 1700 &  1 &  F555W(1) &  S &  25.84 &  0.10 & -0.11 & 0.15 &  22.07 & 0.02 & -11.52 &  127.1 & 0.2/40.0\\
 & &  &  F814W(1) &  S &  24.19 &  0.11 & -0.11 & 0.16 &  19.04 & 0.01 & -14.55 &  451.5 & 0.2/43.8\\
10 & NGC 1700 &  &  F555W(1) &  N &  24.92 & \\
 & &  &  F555W(2) &  N &  25.41 & \\
 & &  &  F814W(1) &  N &  23.21 & \\
 & &  &  F814W(2) &  N &  24.00 & \\
11 & NGC 1700 &  1 &  F555W(2) &  S &  26.64 &  0.12 & -0.18 & 0.22 &  23.85 & 0.07 & -9.74 &  45.4 & 0.2/62.0\\
 & &  &  F606W(1) &  S &  26.91 &  -0.05 & -0.24 & 0.25 &  23.31 & 0.03 & -10.28 &  96.7 & 0.2/54.9\\
 & &  &  F606W(2) &  S &  27.15 &  0.04 & -0.23 & 0.23 &  23.35 & 0.03 & -10.24 &  117.2 & 0.2/45.6\\
 & &  &  F814W(2) &  S &  25.72 &  0.13 & -0.20 & 0.23 &  22.05 & 0.04 & -11.54 &  108.1 & 0.2/60.5\\
12 & NGC 3034 &  &  F555W(1) &  C & \\
 & &  &  F656N &  C & \\
 & &  &  F658N &  C & \\
13 & NGC 3034 &  &  F502N(5) &  C & \\
 & &  &  F555W(1) &  C & \\
 & &  &  F555W(7) &  C & \\
 & &  &  F631N(6) &  C & \\
 & &  &  F656N &  C & \\
 & &  &  F656N(2) &  C & \\
 & &  &  F656N(3) &  C & \\
 & &  &  F658N &  C & \\
 & &  &  F658N(4) &  C & \\
14 & NGC 3034 &  &  F502N(5) &  C & \\
 & &  &  F555W(1) &  C & \\
 & &  &  F555W(7) &  C & \\
 & &  &  F631N(6) &  C & \\
 & &  &  F656N &  C & \\
 & &  &  F656N(2) &  C & \\
 & &  &  F656N(3) &  C & \\
 & &  &  F658N &  C & \\
 & &  &  F658N(4) &  C & \\
15 & NGC 3256 &  1 &  F450W(1) &  M &  25.92 &  -0.09 & -0.00 & 0.09 &  24.99 & 0.18 & -7.86 &  9.4 & 0.3/46.3\\
 & &  &  F450W(2) &  M &  25.95 &  0.16 & -0.15 & 0.22 &  25.50 & 0.28 & -7.35 &  5.9 & 0.2/44.6\\
 & &  &  F555W(1) &  M &  25.21 &  -0.00 & 0.03 & 0.03 &  24.62 & 0.31 & -8.23 &  5.5 & 0.3/46.5\\
 & &  &  F814W(1) &  M &  24.93 &  -0.16 & 0.10 & 0.19 &  24.20 & 0.21 & -8.65 &  7.9 & 0.2/27.8\\
 & &  2 &  F300W &  N &  23.11 & \\
 & &  &  F555W(2) &  N &  24.88 & \\
 & &  &  F814W(1) &  M &  24.93 &  0.42 & 0.13 & 0.44 &  23.95 & 0.17 & -8.90 &  9.9 & 0.2/30.6\\
 & &  &  F814W(2) &  N &  24.15 & \\
 & &  &  F814W(3) &  N &  24.08 & \\
 & &  &  F814W(4) &  N &  24.69 & \\
16 & NGC 3256 & 2  &  F300W &  N &  23.07 & \\
 & &  &  F450W(1) &  N &  25.53 & \\
 & &  &  F814W(1) &  M &  25.37 &  -0.30 & 1.37 & 1.41 &  24.50 & 0.20 & -8.35 &  8.5 & 0.2/42.4\\
17 & NGC 3311 & 1 &  F555W &  N &  26.30 & \\
 & &  &  F814W &  M &  25.16 &  0.89 & 0.07 & 0.90 &  23.85 & 0.13 & -9.67 &  13.0 & 0.5/116.0\\
 & &  2 &  F555W &  M &  26.48 &  0.44 & -0.11 & 0.45 &  24.80 & 0.10 & -8.72 &  18.2 & 0.2/49.4\\
 & &  3 &  F555W &  M &  26.48 &  0.38 & -1.18 & 1.24 &  25.07 & 0.12 & -8.45 &  14.2 & 0.1/31.7\\
 & &  &  F814W &  M &  25.17 &  0.47 & -1.23 & 1.31 &  23.95 & 0.14 & -9.57 &  11.9 & 0.5/115.3\\
 & &  4 &  F555W &  M &  26.48 &  0.84 & -1.14 & 1.41 &  25.09 & 0.12 & -8.43 &  13.9 & 0.1/25.0\\
 & &  &  F814W &  N &  25.01 & \\
18 & NGC 3311 &  1 &  F555W &  M &  26.48 &  -0.35 & 0.73 & 0.81 &  25.68 & 0.20 & -7.84 &  8.0 & 0.2/39.2\\
 & &  &  F814W &  N &  25.02 & \\
 & &  2 &  F555W &  M &  26.28 &  -0.38 & -1.17 & 1.23 &  24.21 & 0.07 & -9.31 &  26.3 & 0.2/39.6\\
 & &  &  F814W &  M &  24.78 &  -0.32 & -1.18 & 1.23 &  23.11 & 0.09 & -10.41 &  18.2 & 0.2/51.8\\
19 & NGC 3628 &  1 &  F606W &  M &  24.81 &  -0.35 & 0.29 & 0.45 &  23.68 & 0.18 & -5.75 &  10.5 & 0.2/6.5\\
20 & NGC 3627 & 1 &  F606W(1) &  N &  24.91 & \\
 & &  &  F606W(2) &  M &  25.36 &  -3.05 & 0.72 & 3.13 &  23.55 & 0.10 & -5.55 &  17.6 & 0.3/9.5\\
 & &  &  F814W &  N &  24.66 & \\
21 & NGC 4038 &  1 &  F336W &  M &  24.69 &  -0.19 & -0.22 & 0.29 &  22.15 & 0.06 & -9.53 &  40.0 & 0.2/17.7\\
 & &  &  F555W &  N &  21.81 & \\
 & &  &  F658N &  M &  23.34 &  0.14 & -0.11 & 0.18 &  19.54 & 0.03 & -12.14 &  129.0 & 0.3/29.4\\
 & &  2 &  F336W &  M &  24.69 &  -0.49 & 0.31 & 0.58 &  19.86 & 0.02 & -11.82 &  335.9 & 0.3/32.6\\
 & &  &  F555W &  M &  23.93 &  -0.28 & 0.23 & 0.36 &  20.37 & 0.05 & -11.31 &  96.0 & 0.2/24.8\\
 & &  &  F658N &  M &  23.34 &  -0.17 & 0.53 & 0.55 &  19.69 & 0.04 & -11.99 &  111.4 & 0.2/24.8\\
 & &  3 &  F336W &  M &  24.69 &  1.11 & 0.94 & 1.46 &  21.22 & 0.04 & -10.46 &  95.1 & 0.3/35.1\\
 & &  &  F555W &  M &  23.92 &  1.28 & 0.84 & 1.53 &  21.30 & 0.08 & -10.38 &  39.3 & 0.3/29.5\\
 & &  &  F658N &  M &  23.34 &  0.85 & 0.94 & 1.26 &  19.64 & 0.04 & -12.04 &  116.6 & 0.5/55.1\\
 & &  4 &  F336W &  M &  24.69 &  -0.59 & 0.15 & 0.61 &  20.78 & 0.03 & -10.90 &  142.7 & 0.3/32.6\\
 & &  &  F555W &  N &  21.81 & \\
22 & NGC 4038 &  1 &  F336W &  M &  24.03 &  -0.16 & -0.31 & 0.35 &  23.39 & 0.25 & -8.29 &  6.8 & 0.2/15.8\\
 & &  &  F555W &  N &  23.66 & \\
 & &  2 &  F336W &  M &  24.03 &  -0.26 & 0.22 & 0.34 &  23.60 & 0.30 & -8.08 &  5.5 & 0.2/18.9\\
 & &  &  F555W &  N &  23.66 & \\
 & &  3 &  F336W &  M &  24.03 &  0.89 & -0.14 & 0.90 &  23.72 & 0.34 & -7.96 &  5.0 & 0.2/21.0\\
 & &  &  F555W &  N &  23.66 & \\
 & &  4 &  F336W &  M &  24.03 &  0.96 & 0.78 & 1.24 &  23.44 & 0.26 & -8.24 &  6.5 & 0.2/21.0\\
 & &  &  F555W &  N &  23.66 & \\
 & &  5 &  F336W &  M &  24.03 &  0.67 & -1.32 & 1.48 &  23.81 & 0.36 & -7.87 &  4.5 & 0.2/17.9\\
 & &  &  F555W &  N &  23.66 & \\
 & &  6 &  F555W &  N &  23.66 & \\
 & &  &  F658N &  M &  23.35 &  -0.52 & 0.40 & 0.65 &  21.84 & 0.15 & -9.84 &  15.0 & 0.2/22.3\\
23 & NGC 4038 &  1 &  F336W &  M &  24.43 &  1.19 & -0.77 & 1.42 &  21.49 & 0.05 & -10.19 &  57.4 & 0.2/18.9\\
 & &  &  F555W &  N &  23.51 & \\
 & &  &  F658N &  N &  22.77 & \\
24 & NGC 4038 &  1 &  F336W &  M &  24.35 &  0.84 & 1.12 & 1.40 &  18.88 & 0.01 & -12.80 &  611.6 & 0.2/26.2\\
 & &  &  F555W &  C & \\
 & &  2 &  F658N &  M &  22.44 &  1.04 & 0.25 & 1.07 &  20.15 & 0.07 & -11.53 &  32.7 & 0.3/27.5\\
25 & NGC 4038 &  1 &  F336W &  M &  24.67 &  -0.39 & 0.17 & 0.43 &  20.58 & 0.04 & -11.10 &  157.0 & 0.3/27.4\\
 & &  &  F658N &  M &  23.28 &  -0.07 & 0.46 & 0.47 &  20.42 & 0.07 & -11.26 &  49.2 & 0.2/24.4\\
26 & NGC 4038 &  &  F336W &  C & \\
 & &  &  F555W &  C & \\
 & &  &  F658N &  C & \\
27 & NGC 4038 &  &  F336W &  N &  24.62 & \\
 & &  &  F555W &  N &  23.47 & \\
 & &  &  F658N &  N &  23.44 & \\
28 & NGC 4038 &  &  F555W &  N &  23.47 & \\
29 & NGC 4150 &  &  F814W &  N &  23.85 & \\
30 & NGC 4490 &  1 &  F606W &  M &  24.44 &  -0.57 & -1.00 & 1.15 &  24.15 & 0.34 & -5.31 &  4.9 & 0.2/9.5\\
 & &  2 &  F606W &  M &  24.44 &  0.98 & -0.12 & 0.99 &  23.08 & 0.14 & -6.38 &  13.4 & 0.2/8.2\\
 & &  3 &  F606W &  M &  24.44 &  0.56 & -1.11 & 1.24 &  23.52 & 0.20 & -5.94 &  8.9 & 0.2/7.0\\
 & &  4 &  F606W &  M &  24.44 &  0.02 & -1.50 & 1.50 &  23.76 & 0.24 & -5.70 &  7.1 & 0.2/5.7\\
 & &  5 &  F606W &  M &  24.44 &  -0.23 & 1.28 & 1.30 &  23.54 & 0.20 & -5.92 &  8.7 & 0.2/6.3\\
 & &  6 &  F606W &  M &  24.44 &  1.11 & -1.30 & 1.71 &  21.40 & 0.04 & -8.06 &  64.3 & 0.2/7.3\\
31 & NGC 4559 &  1 &  F606W &  M &  24.50 &  -0.62 & 0.19 & 0.65 &  24.39 & 0.37 & -5.54 &  4.4 & 0.2/8.6\\
 & &  2 &  F606W &  M &  24.51 &  0.88 & -0.64 & 1.09 &  23.69 & 0.21 & -6.24 &  8.4 & 0.2/10.2\\
32 & NGC 4565 &  1 &  F450W(1) &  S &  25.64 &  0.20 & 0.12 & 0.23 &  23.81 & 0.12 & -7.26 &  20.7 & 0.2/18.3\\
 & &  &  F450W(2) &  N &  25.22 & \\
 & &  &  F555W &  N &  25.22 & \\
 & &  &  F814W(1) &  S &  24.78 &  0.19 & 0.13 & 0.23 &  22.55 & 0.09 & -8.52 &  30.7 & 0.2/18.2\\
 & &  &  F814W(2) &  N &  24.25 & \\
33 & NGC 4594 &  1 &  F658N &  S &  22.07 &  0.53 & 0.29 & 0.60 &  19.09 & 0.04 & -11.72 &  58.7 & 0.3/18.0\\
34 & NGC 4725 &  1 &  F606W &  M &  25.41 &  0.15 & -0.59 & 0.61 &  22.08 & 0.03 & -8.95 &  76.2 & 0.2/16.9\\
35 & NGC 4945 &  &  F606W &  N &  24.17 & \\
36 & NGC 5018 &  &  F336W &  N &  23.30 & \\
 & &  &  F555W &  N &  26.42 & \\
37 & NGC 5018 &  1 &  F336W &  M &  23.77 &  0.01 & -0.05 & 0.05 &  22.00 & 0.11 & -10.86 &  17.9 & 0.2/35.1\\
 & &  &  F555W &  M &  25.67 &  -0.05 & -0.02 & 0.05 &  23.58 & 0.09 & -9.28 &  22.8 & 0.2/33.1\\
38 & NGC 5033 & 1 &  F218W &  N &  21.86 & \\
 & &  &  F300W &  S &  24.18 &  -0.22 & -0.07 & 0.24 &  23.91 & 0.34 & -7.45 &  5.1 & 0.2/18.3\\
39 & NGC 5033 &  1 &  F300W &  S &  24.19 &  -0.23 & 0.07 & 0.24 &  21.91 & 0.08 & -9.45 &  32.5 & 0.2/19.5\\
 & &  2 &  F218W &  S &  22.06 &  0.17 & -0.19 & 0.26 &  20.65 & 0.14 & -10.71 &  14.4 & 0.2/17.2\\
40 & NGC 5128 &  1 &  F555W &  S &  24.40 &  -0.15 & -0.87 & 0.88 &  22.89 & 0.18 & -5.56 &  11.0 & 0.3/6.4\\
 & &  &  F814W &  S &  23.26 &  -0.16 & -0.85 & 0.87 &  21.99 & 0.21 & -6.46 &  8.7 & 0.2/5.7\\
 & &  2 &  F555W &  N &  24.32 & \\
 & &  &  F814W &  M &  23.26 &  0.73 & -1.04 & 1.27 &  22.28 & 0.27 & -6.17 &  6.5 & 0.2/5.0\\
41 & Circinus &  1 &  F502N &  M &  22.67 &  0.50 & 0.17 & 0.53 &  20.70 & 0.10 & -7.14 &  23.5 & 0.2/3.5\\
 & &  &  F656N &  M &  21.92 &  0.42 & 0.25 & 0.49 &  17.93 & 0.03 & -9.91 &  154.5 & 0.2/3.2\\
42 & Circinus &  &  F502N &  N &  22.30 & \\
 & &  &  F606W &  N &  24.97 & \\
 & &  &  F656N &  N &  21.67 & \\
43 & NGC 6500 &  &  F218W &  N &  21.53 & \\
 & &  &  F547M &  X & \\
44 & NGC 7174 &  1 &  F300W &  M &  23.28 &  0.39 & 0.14 & 0.42 &  22.54 & 0.25 & -10.30 &  6.6 & 0.2/42.7\\

\enddata

\tablecomments{Column~(1) lists the ULX ID number from Table~\ref{t:sample}.
Column~(2) lists the host galaxy name. Column~(3) lists a running ID
number of the optical sources spatially coincident with the ULX given
in column~(1). These optical source ID numbers are also used to label
the sources in Figure 2. Column~(4) lists the filter used
for the HST observation, as in Table~2; the
corresponding HST dataset name can be obtained from that
table. Column~(5) indicates the optical
morphology of the region surrounding the ULX, as defined in
Section~\ref{s:photometry}. 
Column~(6)
lists the $4\sigma$ limit for source detection determined as described
in Section~\ref{s:photometry}. 
Columns~(7)
and~(8) lists the offset in RA and Dec. of the optical sources to the
ULX in arcseconds.  Column~(9) lists the
total offset of the optical sources to the ULX in arcseconds.
Column~(10) lists the 
apparent magnitude of the optical source in the {\tt VEGAMAG}
system. Column~(11) lists the formal error in the apparent
magnitude. Column~(12) lists the absolute magnitude of source,
calculated using the galaxy distance in Table~\ref{t:sample}.
Column~(13) lists the $S/N$ level of the optical source
detection.Column~(14) lists an estimate of the source extent in
arcseconds and pc assuming the distances listed in Table 1 (in most
cases the sources are unresolved in which case this estimate is
intended to give an upper-limit to the physical source
size).\label{t:PhotInfo}} 
\end{deluxetable*}











\begin{deluxetable}{cccccc}
\tabletypesize{\scriptsize}
\tablecaption{WFPC2 Counterpart Colors \label{t:colors}}
\tablehead{
\colhead{ULX} & \colhead{Galaxy} & \colhead{T-type} & \colhead{Src.} &
\colhead{Color} & \colhead{Value} \\
\colhead{ID} & & \colhead{RC3} & \colhead{No.} \\
\colhead{(1)} & \colhead{(2)} & \colhead{(3)} & \colhead{(4)} &
\colhead{(5)} & \colhead{(6)}
}
\startdata
1 & n0278 & 3.0 & 1 & B - V$_{606}$ & $0.75 \pm 0.12$ \\ 
1 & n0278 & 3.0 & 1 & B - I & $1.40 \pm 0.12$ \\ 
1 & n0278 & 3.0 & 1 & V$_{606}$ - I & $0.65 \pm 0.09$ \\ 
1 & n0278 & 3.0 & 2 & B - V$_{606}$ & $0.50 \pm 0.22$ \\ 
1 & n0278 & 3.0 & 2 & B - I & $1.26 \pm 0.20$ \\ 
1 & n0278 & 3.0 & 2 & V$_{606}$ - I & $0.76 \pm 0.18$ \\ 
1 & n0278 & 3.0 & 3 & B - V$_{606}$ & $0.28 \pm 0.08$ \\ 
1 & n0278 & 3.0 & 3 & B - I & $0.66 \pm 0.09$ \\ 
1 & n0278 & 3.0 & 3 & V$_{606}$ - I & $0.38 \pm 0.09$ \\ 
1 & n0278 & 3.0 & 4 & B - V$_{606}$ & $-0.01 \pm 0.44$ \\ 
1 & n0278 & 3.0 & 4 & B - I & $0.33 \pm 0.48$ \\ 
1 & n0278 & 3.0 & 4 & V$_{606}$ - I & $0.34 \pm 0.54$ \\ 
1 & n0278 & 3.0 & 5 & B - V$_{606}$ & $0.42 \pm 0.15$ \\ 
1 & n0278 & 3.0 & 5 & B - I & $0.95 \pm 0.15$ \\ 
1 & n0278 & 3.0 & 5 & V$_{606}$ - I & $0.53 \pm 0.14$ \\ 
1 & n0278 & 3.0 & 6 & V$_{606}$ - I & $0.65 \pm 0.24$ \\ 
1 & n0278 & 3.0 & 6 & B - V$_{606}$ & $> 1.52$ \\ 
1 & n0278 & 3.0 & 6 & B - I & $> 2.17$ \\ 
2 & n0891 & 3.0 & 1 & V$_{606}$ - I & $< 0.13$ \\ 
2 & n0891 & 3.0 & 1 & B - V$_{606}$ & $-0.17 \pm 0.19$ \\ 
2 & n0891 & 3.0 & 1 & B - I & $< -0.04$ \\ 
7 & n1399 & -5.0 & 1 & B - I & $1.65 \pm 0.04$ \\ 
7 & n1399 & -5.0 & 1 & B - V$_{606}$ & $0.53 \pm 0.04$ \\ 
7 & n1399 & -5.0 & 1 & V$_{606}$ - I & $1.12 \pm 0.03$ \\ 
8 & n1399 & -5.0 & 1 & B - I & $1.87 \pm 0.03$ \\ 
8 & n1399 & -5.0 & 1 & B - V$_{606}$ & $0.91 \pm 0.03$ \\ 
8 & n1399 & -5.0 & 1 & V$_{606}$ - I & $0.96 \pm 0.02$ \\ 
9 & n1700 & -5.0 & 1 & V$_{555}$ - I & $3.03 \pm 0.03$ \\ 
11 & n1700 & -5.0 & 1 & V$_{555}$ - I & $1.80 \pm 0.08$ \\ 
11 & n1700 & -5.0 & 1 & V$_{555}$ - V$_{606}$ & $0.50 \pm 0.07$ \\ 
11 & n1700 & -5.0 & 1 & V$_{606}$ - I & $1.30 \pm 0.04$ \\ 
15 & n3256 & 99.0 & 1 & B - I & $0.79 \pm 0.28$ \\ 
15 & n3256 & 99.0 & 1 & B - V$_{555}$ & $0.37 \pm 0.36$ \\ 
15 & n3256 & 99.0 & 1 & V$_{555}$ - I & $0.42 \pm 0.37$ \\ 
16 & n3256 & 99.0 & 2 & B - I & $> 1.03$ \\ 
17 & n3311 & -4.0 & 1 & V$_{555}$ - I & $> 2.45$ \\ 
17 & n3311 & -4.0 & 3 & V$_{555}$ - I & $1.12 \pm 0.18$ \\ 
17 & n3311 & -4.0 & 4 & V$_{555}$ - I & $< 0.08$ \\ 
18 & n3311 & -4.0 & 1 & V$_{555}$ - I & $< 0.66$ \\ 
18 & n3311 & -4.0 & 2 & V$_{555}$ - I & $1.10 \pm 0.11$ \\ 
21 & n4038/9 & 9.0 & 1 & U - V$_{555}$ & $< 0.34$ \\ 
21 & n4038/9 & 9.0 & 2 & U - V$_{555}$ & $-0.51 \pm 0.05$ \\ 
21 & n4038/9 & 9.0 & 3 & U - V$_{555}$ & $-0.08 \pm 0.09$ \\ 
21 & n4038/9 & 9.0 & 4 & U - V$_{555}$ & $< -1.03$ \\ 
22 & n4038/9 & 9.0 & 1 & U - V$_{555}$ & $< -0.27$ \\ 
22 & n4038/9 & 9.0 & 2 & U - V$_{555}$ & $< -0.06$ \\ 
22 & n4038/9 & 9.0 & 3 & U - V$_{555}$ & $< 0.06$ \\ 
22 & n4038/9 & 9.0 & 4 & U - V$_{555}$ & $< -0.22$ \\ 
22 & n4038/9 & 9.0 & 5 & U - V$_{555}$ & $< 0.15$ \\ 
23 & n4038/9 & 9.0 & 1 & U - V$_{555}$ & $< -2.02$ \\ 
32 & n4565 & 3.0 & 1 & V$_{555}$ - I & $> 2.67$ \\ 
32 & n4565 & 3.0 & 1 & B - I & $1.26 \pm 0.15$ \\ 
32 & n4565 & 3.0 & 1 & B - V$_{555}$ & $< -1.41$ \\ 
37 & n5018 & -5.0 & 1 & U - V$_{555}$ & $-1.58 \pm 0.14$ \\ 
40 & n5128 & -2.0 & 1 & V$_{555}$ - I & $0.90 \pm 0.27$ \\ 
40 & n5128 & -2.0 & 2 & V$_{555}$ - I & $> 2.04$ \\ 

\enddata
\tablecomments{Column~(1) lists the ULX ID number from
Table~\ref{t:sample}. Column~(2) lists the host galaxy
name. Column~(3) lists the galaxy morphology T-type from the Third
Reference Catalog of Bright Galaxies (RC3, de Vaucouleurs 
\etal 1991).  Column~(4) lists the id number of the source from Table
3.  Column~(5) lists the color being computed, with 
the filters designations U, B, V$_{555}$, V$_{606}$ and I refer to HST
WFPC2 filters F336W, F450W, F555W, F606W and F814W.  Column~(6) lists
the color or color limit.}
\end{deluxetable}

\begin{deluxetable}{ccccccc}
\tabletypesize{\scriptsize}
\tablecaption{WFPC2 Counterpart Optical Luminosities \label{t:lopt}}
\tablehead{
\colhead{ULX} & \colhead{Src.} &
\colhead{$\log L_{\lambda}/L_{\odot, \lambda}$} & 
\colhead{$V_{eff}$} &
\colhead{$M_{V, eff}$} &
\colhead{$\log L_{V,eff}/L_{\odot,V}$} &
\colhead{$\log L_X/L_{V,eff}$} \\
\colhead{ID} & \colhead{No.} \\
\colhead{(1)} & \colhead{(2)} & \colhead{(3)} & \colhead{(4)} & \colhead{(5)} & \colhead{(6)} & \colhead{(7)}
}
\startdata
1 & 1 &5.4(B) , 5.4(V) , 5.4(I)  & 21.9 & -8.5 & 5.3 & 0.5 \\ 
1 & 2 &5.1(B) , 5.1(V) , 5.1(I)  & 22.7 & -7.6 & 5.0 & 0.9 \\ 
1 & 3 &5.7(B) , 5.5(V) , 5.4(I)  & 21.3 & -9.0 & 5.5 & 0.3 \\ 
1 & 4 &4.9(B) , 4.7(V) , 4.5(I)  & 23.4 & -6.9 & 4.7 & 1.2 \\ 
1 & 5 &5.3(B) , 5.3(V) , 5.2(I)  & 22.0 & -8.4 & 5.3 & 0.6 \\ 
1 & 6 &5.0(V) , 4.9(I) , $<$4.6 (B)  & 23.0 & -7.4 & 4.9 & 1.0 \\ 
2 & 1 &4.6(V) , 4.9(B) , $<$4.4 (I)  & 23.2 & -6.8 & 4.6 & 1.2 \\ 
6 & 1 &$<$4.5 (I) , $<$4.4 (B) , $<$4.4 (V)  & $>$25.9 & $>$-5.5 & $<$4.1 & $>$1.4 \\ 
7 & 1 &5.7(I) , 5.6(B) , 5.6(V)  & 22.6 & -8.8 & 5.5 & 0.2 \\ 
8 & 1 &6.0(I) , 5.8(B) , 5.9(V)  & 21.8 & -9.6 & 5.8 & -0.0 \\ 
9 & 1 &6.5(V) , 7.4(I)  & 22.1 & -11.5 & 6.5 & -0.6 \\ 
10 & 1 &$<$5.2 (V) , $<$5.4 (I) , $<$5.4 (V) , $<$5.8 (I)  & $>$25.4 & $>$-8.1 & $<$5.2 & $>$0.8 \\ 
11 & 1 &5.8(V) , 6.2(I) , 6.0(V) , 6.0(V)  & 23.6 & -10.0 & 5.9 & 0.4 \\ 
15 & 1 &5.3(B) , 5.1(I) , 5.1(B) , 5.2(V)  & 24.6 & -8.2 & 5.2 & 0.4 \\ 
15 & 2 &5.2(I) , $<$4.9 (I) , $<$5.1 (I) , $<$5.1 (V) , $<$5.1 (I)  & 24.1 & -8.8 & 5.4 & 0.2 \\ 
16 & 1 &$<$5.1 (B) , $<$4.9 (I)  & $>$25.6 & $>$-7.2 & $<$4.8 & $>$1.1 \\ 
16 & 2 &5.0(I) , $<$5.1 (B)  & 25.1 & -7.7 & 5.0 & 0.9 \\ 
17 & 1 &5.5(I) , $<$4.8 (V)  & 24.5 & -9.0 & 5.5 & 0.7 \\ 
17 & 2 &5.4(V)  & 24.8 & -8.7 & 5.4 & 0.8 \\ 
17 & 3 &5.3(V) , 5.4(I)  & 25.1 & -8.4 & 5.3 & 0.9 \\ 
17 & 4 &5.3(V) , $<$5.0 (I)  & 25.1 & -8.4 & 5.3 & 0.9 \\ 
18 & 1 &5.1(V) , $<$5.0 (I)  & 25.7 & -7.8 & 5.1 & 1.5 \\ 
18 & 2 &5.6(V) , 5.8(I)  & 24.2 & -9.3 & 5.6 & 0.9 \\ 
19 & 1 &4.2(V)  & 23.7 & -5.8 & 4.2 & 1.6 \\ 
20 & 1 &4.1(V) , $<$3.6 (V) , $<$3.4 (I)  & 23.6 & -5.5 & 4.1 & 1.5 \\ 
21 & 1 &6.0(U) , $<$5.9 (V)  & 22.2 & -9.4 & 5.7 & 0.1 \\ 
21 & 2 &6.9(U) , 6.4(V)  & 20.4 & -11.3 & 6.4 & -0.6 \\ 
21 & 3 &6.4(U) , 6.1(V)  & 21.3 & -10.4 & 6.1 & -0.2 \\ 
21 & 4 &6.6(U) , $<$5.9 (V)  & 20.9 & -10.8 & 6.2 & -0.4 \\ 
22 & 1 &5.5(U) , $<$5.1 (V)  & 23.5 & -8.2 & 5.2 & 1.1 \\ 
22 & 2 &5.4(U) , $<$5.1 (V)  & 23.7 & -8.0 & 5.1 & 1.2 \\ 
22 & 3 &5.4(U) , $<$5.1 (V)  & 23.8 & -7.9 & 5.1 & 1.2 \\ 
22 & 4 &5.5(U) , $<$5.1 (V)  & 23.5 & -8.2 & 5.2 & 1.1 \\ 
22 & 5 &5.4(U) , $<$5.1 (V)  & 23.9 & -7.8 & 5.0 & 1.2 \\ 
22 & 6 &$<$5.1 (V)  & $>$23.7 & $>$-8.0 & $<$5.1 & $>$1.2 \\ 
23 & 1 &6.3(U) , $<$5.2 (V)  & 21.6 & -10.1 & 6.0 & -0.4 \\ 
24 & 1 &7.3(U)  & 19.0 & -12.7 & 7.0 & -1.5 \\ 
25 & 1 &6.6(U)  & 20.7 & -11.0 & 6.3 & -0.3 \\ 
27 & 1 &$<$5.0 (U) , $<$5.2 (V)  & $>$24.7 & $>$-7.0 & $<$4.7 & $>$1.4 \\ 
28 & 1 &$<$5.2 (V)  & $>$23.5 & $>$-8.2 & $<$5.2 & $>$0.3 \\ 
29 & 1 &$<$4.0 (I)  & $>$24.0 & $>$-6.0 & $<$4.3 & $>$1.3 \\ 
30 & 1 &4.0(V)  & 24.1 & -5.3 & 4.0 & 1.5 \\ 
30 & 2 &4.5(V)  & 23.1 & -6.4 & 4.5 & 1.1 \\ 
30 & 3 &4.3(V)  & 23.5 & -5.9 & 4.3 & 1.3 \\ 
30 & 4 &4.2(V)  & 23.8 & -5.7 & 4.2 & 1.4 \\ 
30 & 5 &4.3(V)  & 23.5 & -5.9 & 4.3 & 1.3 \\ 
30 & 6 &5.1(V)  & 21.4 & -8.1 & 5.1 & 0.4 \\ 
31 & 1 &4.1(V)  & 24.4 & -5.5 & 4.1 & 1.9 \\ 
31 & 2 &4.4(V)  & 23.7 & -6.2 & 4.4 & 1.6 \\ 
32 & 1 &5.0(I) , 5.1(B) , $<$4.3 (V) , $<$4.3 (I) , $<$4.5 (B)  & 23.3 & -7.7 & 5.0 & 1.5 \\ 
34 & 1 &5.5(V)  & 22.4 & -8.7 & 5.4 & 0.4 \\ 
35 & 1 &$<$3.7 (V)  & $>$24.2 & $>$-4.4 & $<$3.7 & $>$1.9 \\ 
36 & 1 &$<$6.0 (U) , $<$4.5 (V)  & $>$26.5 & $>$-6.4 & $<$4.5 & $>$1.8 \\ 
37 & 1 &6.5(U) , 5.6(V)  & 23.6 & -9.3 & 5.6 & 0.3 \\ 
40 & 1 &4.2(I) , 4.1(V)  & 22.9 & -5.6 & 4.1 & 1.4 \\ 
40 & 2 &4.1(I) , $<$3.6 (V)  & 22.9 & -5.5 & 4.1 & 1.4 \\ 
42 & 1 &$<$3.1 (V)  & $>$25.0 & $>$-2.9 & $<$3.1 & $>$3.0 \\ 

\enddata
\tablecomments{Column~(1) lists the ULX ID number from
Table~\ref{t:sample}.  Column~(2) lists ID number of the optical
source from Table 3.  Column~(3) lists the optical
luminosity computed for the band(s) in which photometric data was
available.  Column~(4) lists $V_{eff}$, the effective V band magnitude
computed based on the observed or assumed color of the 
source. Column~(5) lists the absolute magnitude computed using
$V_{eff}$ and the distance listed in Table {t:sample}.  
Column~(6) lists $\log L_{V,eff}/L_{\odot,V}$, 
the optical luminosity of the source based on the $V_{eff}$ estimate,
in solar luminosities. Column~(7) lists the ratio of X-ray to optical
luminosity.} 

\end{deluxetable}



\end{document}